\documentclass[12pt]{article}
\pdfoutput=1
\usepackage{ae} 
\usepackage[T1]{fontenc}
\usepackage[ansinew]{inputenc}
\usepackage{amsmath}
\usepackage{amssymb}
\usepackage{mathtools}
\usepackage{graphicx}
\usepackage{bbm}
\usepackage{slashed}
\usepackage{booktabs}
\usepackage{color}
\usepackage{hyperref}
\usepackage{epsfig}
\usepackage{graphicx}
\usepackage{subcaption}
\usepackage{tikz}
\usetikzlibrary{trees}
\usetikzlibrary{calc}
\usetikzlibrary{decorations.pathmorphing}
\usetikzlibrary{decorations.markings}
\usetikzlibrary{decorations.text} 
\usetikzlibrary{intersections}
\usepackage{cite}

\newcommand{\p}{\partial}

\newcommand{\Tr}{{\rm Tr}}

\newcommand{\cO}{{\cal O}}

\def\eps{{\epsilon}}

\def\e{\epsilon}

\def\s{\sigma}

\def\g{\gamma}

\def\su2{{SU(2)}}

\def\i2{\frac{i}{2}}

\def \Xcolor {red!90!yellow}
\def \Zcolor {black!70!blue}

\tikzset{
photon/.style={decorate, decoration={snake}},
particle/.style={postaction={decorate},
    decoration={markings,mark=at position .5 with {\arrow{>}}}},
antiparticle/.style={postaction={decorate},
    decoration={markings,mark=at position .5 with {\arrow{<}}}},
gluon/.style={decorate, decoration={coil,amplitude=2pt, segment length=4pt},color=purple},
wilson/.style={color=blue, thick},
scalarZ/.style={thick, \Zcolor, postaction={decorate},decoration={markings, mark=at position .5 with{\arrow[scale=1]{stealth}}}},
scalarZf/.style={thick, \Zcolor, postaction={decorate},decoration={markings, mark=at position 1 with{\arrow[thick,\Zcolor,xshift=2pt,>=latex,scale=0.7]{>}}}},
scalarZfaint/.style={thick, \Zcolor, opacity = 0.2, postaction={decorate},decoration={markings, mark=at position .5 with{\arrow[scale=1]{stealth}}}},
scalarX/.style={thick, \Xcolor, postaction={decorate}, dashed, dash pattern = on 4pt off 2pt, dash phase = 2pt, decoration={markings, mark=at position .53 with{\arrow[scale=1]{stealth}}}},
scalarXnoarrow/.style={thick, \Xcolor, postaction={decorate}, dashed, dash pattern = on 4pt off 2pt, dash phase = 2pt},
scalarXf/.style={thick, \Xcolor, postaction={decorate}, dashed, dash pattern = on 4pt off 2pt, dash phase = 2pt, decoration={markings, mark=at position 1 with{\arrow[thick,\Xcolor,xshift=2pt,>=latex,scale=0.7]{>}}}},
scalarXfaint/.style={thick, \Xcolor, opacity = 0.2, postaction={decorate}, dashed, dash pattern = on 4pt off 2pt, dash phase = 2pt, decoration={markings, mark=at position .6 with{\arrow[scale=1]{stealth}}}},
scalarZw/.style={thick, \Zcolor, postaction={decorate},decoration={markings, mark=at position .7 with{\arrow[scale=1]{stealth}}}},
scalarZy/.style={thick, \Zcolor, postaction={decorate},decoration={markings, mark=at position .6 with{\arrow[scale=1]{stealth}}}},
scalarZx/.style={thick, \Zcolor, postaction={decorate},decoration={markings, mark=at position .55 with{\arrow[scale=1]{stealth}}}},
scalarXw/.style={thick, \Xcolor, postaction={decorate}, dashed, dash pattern = on 4pt off 2pt, dash phase = 2pt, decoration={markings, mark=at position .6 with{\arrow[scale=1]{stealth}}}},
scalarXy/.style={thick, \Xcolor, postaction={decorate}, dashed, dash pattern = on 4pt off 2pt, dash phase = 2pt, decoration={markings, mark=at position .65 with{\arrow[scale=1]{stealth}}}},
  adjointa/.style =  {
  \Zcolor,
    preaction={
      draw=\Zcolor,
      double distance between line centers=3pt,
      line width=1pt,
      shorten >=0pt,
      shorten <=0pt,
    },
    draw=white,
    line width=3pt,
    shorten >=-.1pt,
    shorten <=-.1pt,
    postaction={
      decorate,
    },
    decoration={
      markings,
      mark=at position 0.53  with {
        \arrow[thick,\Zcolor, yshift= 1.8pt,xshift=.8pt,>=latex,scale=0.7]{>},
        \arrow[thick,\Zcolor, yshift=-1.8pt,xshift=+.3pt,>=latex,scale=0.7]{<}
      },
    },
  },
  adjointb/.style =  {
  \Xcolor,
    preaction={
      draw=\Xcolor,
      dashed,
      double distance between line centers=3pt,
      line width=1pt,
      shorten >=0pt,
      shorten <=0pt,
    },
    draw=white,
    line width=3pt,
    shorten >=-.1pt,
    shorten <=-.1pt,
    postaction={
      decorate,
    },
    decoration={
      markings,
      mark=at position 0.53  with {
        \arrow[thick,\Xcolor, yshift= 1.8pt,xshift=-.4pt,>=latex,scale=0.7]{>},
        \arrow[thick,\Xcolor, yshift=-1.8pt,xshift=+.4pt,>=latex,scale=0.7]{<}
      },
    },
    },
     adjointaALT/.style =  {
  \Zcolor,
    preaction={
      draw=\Zcolor,
      double distance between line centers=3pt,
      line width=1pt,
      shorten >=0pt,
      shorten <=0pt,
    },
    draw=white,
    line width=3pt,
    shorten >=-.1pt,
    shorten <=-.1pt,
    postaction={
      decorate,
    },
    decoration={
      markings,
      mark=at position 0.44  with {
        \arrow[thick,\Zcolor, yshift= 1.8pt,xshift=.8pt,>=latex,scale=0.7]{>},
        \arrow[thick,\Zcolor, yshift=-1.8pt,xshift=+.3pt,>=latex,scale=0.7]{<}
      },
    },
  },
  adjointbALT/.style =  {
  \Xcolor,
    preaction={
      draw=\Xcolor,
      dashed,
      double distance between line centers=3pt,
      line width=1pt,
      shorten >=0pt,
      shorten <=0pt,
    },
    draw=white,
    line width=3pt,
    shorten >=-.1pt,
    shorten <=-.1pt,
    postaction={
      decorate,
    },
    decoration={
      markings,
      mark=at position 0.58  with {
        \arrow[thick,\Xcolor, yshift= 1.8pt,xshift=-.4pt,>=latex,scale=0.7]{>},
        \arrow[thick,\Xcolor, yshift=-1.8pt,xshift=+.4pt,>=latex,scale=0.7]{<}
      },
    },
  }
}

\newcommand{\dgab}{
  \begin{minipage}[c][20px][c]{50px}
    \centering
    \begin{tikzpicture}[xscale=1/4, yscale = 1/2]
      \draw[\Xcolor, rounded corners = 2pt] (1,0)  -- (1,1);
      \draw[\Xcolor, rounded corners = 2pt] (2,0) -- (2,0.25) -- (3,0.75) -- (3,1);
      \draw[\Zcolor, rounded corners = 2pt] (3,0) -- (3,0.25) -- (2,0.75) -- (2,1);
      \draw[\Zcolor, rounded corners = 2pt] (4,0) -- (4,1);
      \draw[\Zcolor, rounded corners = 2pt] (5,0) -- (5,1);
      \draw[\Zcolor, rounded corners = 2pt] (6,0) -- (6,1);      
      \draw[\Zcolor, rounded corners = 2pt] (7,0) -- (7,1);      
\end{tikzpicture}
\end{minipage}
}

\newcommand{\dgba}{
  \begin{minipage}[c][20px][c]{50px}
    \centering
  \begin{tikzpicture}[xscale=1/4, yscale = 1/2]
  \draw[\Xcolor, rounded corners = 2pt] (1,0) -- (1,0.25) -- (2,0.75) -- (2,1);
  \draw[\Zcolor, rounded corners = 2pt] (2,0) -- (2,0.25) -- (1,0.75) -- (1,1);
  \draw[\Xcolor, rounded corners = 2pt] (3,0)  -- (3,1);
  \draw[\Zcolor, rounded corners = 2pt] (4,0) -- (4,1);
  \draw[\Zcolor, rounded corners = 2pt] (5,0) -- (5,1);
  \draw[\Zcolor, rounded corners = 2pt] (6,0) -- (6,1);
  \draw[\Zcolor, rounded corners = 2pt] (7,0) -- (7,1);
\end{tikzpicture}
\end{minipage}
}

\newcommand{\dgbc}{
  \begin{minipage}[c][20px][c]{50px}
    \centering
    \begin{tikzpicture}[xscale=1/4, yscale = 1/2]
      \draw[\Xcolor, rounded corners = 2pt] (1,0)  -- (1,1);
      \draw[\Zcolor, rounded corners = 2pt] (2,0) -- (2,1);
      \draw[\Zcolor, rounded corners = 2pt] (3,0) -- (3,1);
      \draw[\Zcolor, rounded corners = 2pt] (4,0) -- (4,1);
      \draw[\Xcolor, rounded corners = 2pt] (5,0) -- (5,0.25) -- (6,0.75) -- (6,1);
      \draw[\Zcolor, rounded corners = 2pt] (6,0) -- (6,0.25) -- (5,0.75) -- (5,1);
      \draw[\Zcolor, rounded corners = 2pt] (7,0) -- (7,1);
    \end{tikzpicture}
  \end{minipage}
}

\newcommand{\dgcb}{
  \begin{minipage}[c][20px][c]{50px}
    \centering
    \begin{tikzpicture}[xscale=1/4, yscale = 1/2]
      \draw[\Xcolor, rounded corners = 2pt] (1,0) -- (1,0.25) -- (2,0.75) -- (2,1);
      \draw[\Zcolor, rounded corners = 2pt] (2,0) -- (2,0.25) -- (1,0.75) -- (1,1);
      \draw[\Zcolor, rounded corners = 2pt] (3,0) -- (3,1);
      \draw[\Xcolor, rounded corners = 2pt] (4,0) -- (4,1);
      \draw[\Zcolor, rounded corners = 2pt] (5,0) -- (5,1);
      \draw[\Zcolor, rounded corners = 2pt] (6,0) -- (6,1);      
      \draw[\Zcolor, rounded corners = 2pt] (7,0) -- (7,1);      
\end{tikzpicture}
\end{minipage}
}

\newcommand{\dgcc}{
  \begin{minipage}[c][20px][c]{50px}
    \centering
    \begin{tikzpicture}[xscale=1/4, yscale = 1/2]
      \draw[\Xcolor, rounded corners = 2pt] (1,0) -- (1,1);
      \draw[\Zcolor, rounded corners = 2pt] (2,0) -- (2,1);      
      \draw[\Zcolor, rounded corners = 2pt] (3,0) -- (3,1);      
      
      \draw[\Xcolor, rounded corners = 2pt] (4,0) -- (4,0.25) -- (5,0.75) -- (5,1);
      \draw[\Zcolor, rounded corners = 2pt] (5,0) -- (5,0.25) -- (4,0.75) -- (4,1);
      \draw[\Zcolor, rounded corners = 2pt] (6,0) -- (6,1);
      \draw[\Zcolor, rounded corners = 2pt] (7,0) -- (7,1);

\end{tikzpicture}
\end{minipage}
}

\newcommand{\dggaa}{
  \begin{minipage}[c][20px][c]{50px}
    \centering
    \begin{tikzpicture}[xscale=1/4, yscale = 1/4]
      \draw[\Zcolor, rounded corners = 2pt] (3,0) -- (3,0.5) -- (1,1.5) -- (1,2);
      \draw[\Xcolor, rounded corners = 2pt] (1,0) -- (1,0.5) -- (2,1.5) -- (2,2);
      \draw[\Xcolor, rounded corners = 2pt] (2,0) -- (2,0.5) -- (3,1.5) -- (3,2);
      \draw[\Zcolor, rounded corners = 2pt] (4,0) -- (4,2);
      \draw[\Zcolor, rounded corners = 2pt] (5,0) -- (5,2);
      \draw[\Zcolor, rounded corners = 2pt] (6,0) -- (6,2);
      \draw[\Zcolor, rounded corners = 2pt] (7,0) -- (7,2);
    \end{tikzpicture}
  \end{minipage}
}

\newcommand{\dggac}{
  \begin{minipage}[c][20px][c]{50px}
    \centering
    \begin{tikzpicture}[xscale=1/4, yscale = 1/4]
      \draw[\Xcolor, rounded corners = 2pt] (1,0) -- (1,1) -- (1,2);      
      \draw[\Zcolor, rounded corners = 2pt] (4,0) -- (4,0.5) --
      (3,1.5) -- (3,2);
      \draw[\Zcolor, rounded corners = 2pt] (3,0) -- (3,0.5) --
      (2,1.5) -- (2,2);
      \draw[\Xcolor, rounded corners = 2pt] (2,0) --
      (2,0.5) -- (4,1.5) -- (4,2);
      \draw[\Zcolor, rounded corners = 2pt] (5,0) -- (5,1) -- (5,2);
      \draw[\Zcolor, rounded corners = 2pt] (6,0) -- (6,1) -- (6,2);
      \draw[\Zcolor, rounded corners = 2pt] (7,0) -- (7,1) -- (7,2);
    \end{tikzpicture}
  \end{minipage}
}

\newcommand{\dggbb}{
  \begin{minipage}[c][20px][c]{50px}
    \centering
    \begin{tikzpicture}[xscale=1/4, yscale = 1/4]
      \draw[\Zcolor, rounded corners = 2pt] (2,0) -- (2,0.5) --
      (1,1.5) -- (1,2);
      \draw[\Xcolor, rounded corners = 2pt] (1,0) --
      (1,0.5) -- (2,1.5) -- (2,2);
      \draw[\Xcolor, rounded corners =
      2pt] (3,0) -- (3,0.5) -- (4,1.5) -- (4,2);
      \draw[\Zcolor, rounded corners = 2pt] (4,0) --
      (4,0.5) -- (3,1.5) -- (3,2);
      \draw[\Zcolor, rounded corners = 2pt] (5,0) -- (5,2);
      \draw[\Zcolor, rounded corners = 2pt] (6,0) -- (6,2);
      \draw[\Zcolor, rounded corners = 2pt] (7,0) -- (7,2);
    \end{tikzpicture}
  \end{minipage}
}

\newcommand{\dggca}{
  \begin{minipage}[c][20px][c]{50px}
    \centering
    \begin{tikzpicture}[xscale=1/4, yscale = 1/4]
      \draw[\Zcolor, rounded corners = 2pt] (3,0) -- (3,0.5) --
      (2,1.5) -- (2,2);
      \draw[\Zcolor, rounded corners = 2pt] (2,0) -- (2,0.5) --
      (1,1.5) -- (1,2);
      \draw[\Xcolor, rounded corners = 2pt] (1,0) --
      (1,0.5) -- (3,1.5) -- (3,2);
      \draw[\Xcolor, rounded corners = 2pt] (4,0) -- (4,1) -- (4,2);      
      \draw[\Zcolor, rounded corners = 2pt] (5,0) -- (5,1) -- (5,2);
      \draw[\Zcolor, rounded corners = 2pt] (6,0) -- (6,1) -- (6,2);
      \draw[\Zcolor, rounded corners = 2pt] (7,0) -- (7,1) -- (7,2);
    \end{tikzpicture}
  \end{minipage}
}

\newcommand{\dggbc}{
  \begin{minipage}[c][20px][c]{50px}
    \centering
    \begin{tikzpicture}[xscale=1/4, yscale = 1/4]
      \draw[\Zcolor, rounded corners = 2pt] (3,0) -- (3,0.5) --
      (2,1.5) -- (2,2);
      \draw[\Zcolor, rounded corners = 2pt] (2,0) -- (2,0.5) --
      (1,1.5) -- (1,2);
      \draw[\Xcolor, rounded corners = 2pt] (1,0) --
      (1,0.5) -- (3,1.5) -- (3,2);
      \draw[\Zcolor, rounded corners = 2pt] (4,0) -- (4,1) -- (4,2);      
      \draw[\Zcolor, rounded corners = 2pt] (5,0) -- (5,1) -- (5,2);
      \draw[\Xcolor, rounded corners = 2pt] (6,0) -- (6,1) -- (6,2);
      \draw[\Zcolor, rounded corners = 2pt] (7,0) -- (7,1) -- (7,2);
    \end{tikzpicture}
    \end{minipage}
}

\newcommand{\dggcb}{
  \begin{minipage}[c][20px][c]{50px}
    \centering
    \begin{tikzpicture}[xscale=1/4, yscale = 1/4]
      \draw[\Zcolor, rounded corners = 2pt] (3,0) -- (3,0.5) --
      (2,1.5) -- (2,2);
      \draw[\Zcolor, rounded corners = 2pt] (2,0) -- (2,0.5) --
      (1,1.5) -- (1,2);
      \draw[\Xcolor, rounded corners = 2pt] (1,0) --
      (1,0.5) -- (3,1.5) -- (3,2);
      \draw[\Zcolor, rounded corners = 2pt] (4,0) -- (4,1) -- (4,2);      
      \draw[\Xcolor, rounded corners = 2pt] (5,0) -- (5,1) -- (5,2);
      \draw[\Zcolor, rounded corners = 2pt] (6,0) -- (6,1) -- (6,2);
      \draw[\Zcolor, rounded corners = 2pt] (7,0) -- (7,1) -- (7,2);
    \end{tikzpicture}
    \end{minipage}
}

\newcommand{\dgggab}{
  \begin{minipage}[c][20px][c]{50px}
    \centering
    \begin{tikzpicture}[xscale=1/4, yscale = 1/4]
      \draw[\Xcolor, rounded corners = 2pt] (1,0) -- (1,0.5) -- (2,1.5) -- (2,2);
      \draw[\Xcolor, rounded corners = 2pt] (2,0) -- (2,0.5) -- (4,1.5) -- (4,2);
      \draw[\Zcolor, rounded corners = 2pt] (3,0) -- (3,0.5) -- (1,1.5) -- (1,2);
      \draw[\Zcolor, rounded corners = 2pt] (4,0) -- (4,0.5) -- (3,1.5) -- (3,2);
      \draw[\Zcolor, rounded corners = 2pt] (5,0) -- (5,2);      
      \draw[\Zcolor, rounded corners = 2pt] (6,0) -- (6,2);
      \draw[\Zcolor, rounded corners = 2pt] (7,0) -- (7,2);      
    \end{tikzpicture}
  \end{minipage}
}

\newcommand{\dgggac}{
  \begin{minipage}[c][20px][c]{50px}
    \centering
    \begin{tikzpicture}[xscale=1/4, yscale = 1/4]
      \draw[\Xcolor, rounded corners = 2pt] (1,0) -- (1,2);      
      \draw[\Xcolor, rounded corners = 2pt] (2,0) -- (2,0.5) -- (5,1.5) -- (5,2);
      \draw[\Zcolor, rounded corners = 2pt] (3,0) -- (3,0.5) -- (2,1.5) -- (2,2);
      \draw[\Zcolor, rounded corners = 2pt] (4,0) -- (4,0.5) -- (3,1.5) -- (3,2);
      \draw[\Zcolor, rounded corners = 2pt] (5,0) -- (5,0.5) -- (4,1.5) -- (4,2);
      \draw[\Zcolor, rounded corners = 2pt] (6,0) -- (6,2);      
      \draw[\Zcolor, rounded corners = 2pt] (7,0) -- (7,2);      
    \end{tikzpicture}
  \end{minipage}
}

\newcommand{\dgggba}{
  \begin{minipage}[c][20px][c]{50px}
    \centering
    \begin{tikzpicture}[xscale=1/4, yscale = 1/4]
      \draw[\Xcolor, rounded corners = 2pt] (1,0) -- (1,0.5) -- (3,1.5) -- (3,2);      
      \draw[\Zcolor, rounded corners = 2pt] (4,0) -- (4,0.5) --
      (2,1.5) -- (2,2);
      \draw[\Zcolor, rounded corners = 2pt] (2,0) -- (2,0.5) --
      (1,1.5) -- (1,2);
      \draw[\Zcolor, rounded corners = 2pt] (5,0) -- (5,2);
      \draw[\Zcolor, rounded corners = 2pt] (6,0) -- (6,2);
      \draw[\Zcolor, rounded corners = 2pt] (7,0) -- (7,2);
      \draw[\Xcolor, rounded corners = 2pt] (3,0) --
      (3,0.5) -- (4,1.5) -- (4,2);
    \end{tikzpicture}
  \end{minipage}
}

\newcommand{\dgggbb}{
  \begin{minipage}[c][20px][c]{50px}
    \centering
    \begin{tikzpicture}[xscale=1/4, yscale = 1/4]
      \draw[\Xcolor, rounded corners = 2pt] (1,0) -- (1,2);
      \draw[\Zcolor, rounded corners = 2pt] (2,0) -- (2,2);      
      \draw[\Xcolor, rounded corners = 2pt] (3,0) -- (3,0.5) -- (6,1.5) -- (6,2);
      \draw[\Zcolor, rounded corners = 2pt] (4,0) -- (4,0.5) -- (3,1.5) -- (3,2);
      \draw[\Zcolor, rounded corners = 2pt] (5,0) -- (5,0.5) -- (4,1.5) -- (4,2);
      \draw[\Zcolor, rounded corners = 2pt] (6,0) -- (6,0.5) -- (5,1.5) -- (5,2);
      \draw[\Zcolor, rounded corners = 2pt] (7,0) -- (7,2);      
    \end{tikzpicture}
  \end{minipage}
}

\newcommand{\dgggbc}{
  \begin{minipage}[c][20px][c]{50px}
    \centering
    \begin{tikzpicture}[xscale=1/4, yscale = 1/4]
      \draw[\Zcolor, rounded corners = 2pt] (5,0) -- (5,0.5) --
      (4,1.5) -- (4,2);
      \draw[\Zcolor, rounded corners = 2pt] (4,0) -- (4,0.5) --
      (3,1.5) -- (3,2);
      \draw[\Xcolor, rounded corners = 2pt] (3,0) --
      (3,0.5) -- (5,1.5) -- (5,2);
      \draw[\Xcolor, rounded corners = 2pt] (1,0) -- (1,0.5) -- (2,1.5) -- (2,2);
      \draw[\Zcolor, rounded corners = 2pt] (2,0) -- (2,0.5) -- (1,1.5) -- (1,2);
      \draw[\Zcolor, rounded corners = 2pt] (6,0) -- (6,2);      
      \draw[\Zcolor, rounded corners = 2pt] (7,0) -- (7,2);      
    \end{tikzpicture}
  \end{minipage}
}

\newcommand{\dgggca}{
  \begin{minipage}[c][20px][c]{50px}
    \centering
    \begin{tikzpicture}[xscale=1/4, yscale = 1/4]
      \draw[\Xcolor, rounded corners = 2pt] (1,0) -- (1,2);
      \draw[\Zcolor, rounded corners = 2pt] (2,0) -- (2,2);      
      \draw[\Zcolor, rounded corners = 2pt] (3,0) -- (3,2);      
      \draw[\Xcolor, rounded corners = 2pt] (4,0) -- (4,0.5) -- (7,1.5) -- (7,2);
      \draw[\Zcolor, rounded corners = 2pt] (5,0) -- (5,0.5) -- (4,1.5) -- (4,2);
      \draw[\Zcolor, rounded corners = 2pt] (6,0) -- (6,0.5) -- (5,1.5) -- (5,2);
      \draw[\Zcolor, rounded corners = 2pt] (7,0) -- (7,0.5) -- (6,1.5) -- (6,2);
    \end{tikzpicture}
  \end{minipage}
}

\newcommand{\dgggcb}{
  \begin{minipage}[c][20px][c]{50px}
    \centering
    \begin{tikzpicture}[xscale=1/4, yscale = 1/4]
      \draw[\Zcolor, rounded corners = 2pt] (3,0) -- (3,0.5) --
      (2,1.5) -- (2,2);
      \draw[\Zcolor, rounded corners = 2pt] (2,0) -- (2,0.5) --
      (1,1.5) -- (1,2);
      \draw[\Xcolor, rounded corners = 2pt] (1,0) --
      (1,0.5) -- (3,1.5) -- (3,2);
      \draw[\Xcolor, rounded corners = 2pt] (4,0) -- (4,0.5) -- (5,1.5) -- (5,2);
      \draw[\Zcolor, rounded corners = 2pt] (5,0) -- (5,0.5) -- (4,1.5) -- (4,2);
      \draw[\Zcolor, rounded corners = 2pt] (6,0) -- (6,2);
      \draw[\Zcolor, rounded corners = 2pt] (7,0) -- (7,2);      
    \end{tikzpicture}
  \end{minipage}
}

\newcommand{\dgggcc}{
  \begin{minipage}[c][20px][c]{50px}
    \centering
    \begin{tikzpicture}[xscale=1/4, yscale = 1/4]
      \draw[\Zcolor, rounded corners = 2pt] (3,0) -- (3,0.5) --
      (2,1.5) -- (2,2);
      \draw[\Zcolor, rounded corners = 2pt] (2,0) -- (2,0.5) --
      (1,1.5) -- (1,2);
      \draw[\Xcolor, rounded corners = 2pt] (1,0) --
      (1,0.5) -- (3,1.5) -- (3,2);
      \draw[\Zcolor, rounded corners = 2pt] (4,0) -- (4,2);
      \draw[\Xcolor, rounded corners = 2pt] (5,0) -- (5,0.5) -- (6,1.5) -- (6,2);
      \draw[\Zcolor, rounded corners = 2pt] (6,0) -- (6,0.5) -- (5,1.5) -- (5,2);
      \draw[\Zcolor, rounded corners = 2pt] (7,0) -- (7,2);      
    \end{tikzpicture}
  \end{minipage}
}

\newcommand{\dggggaa}{
  \begin{minipage}[c][20px][c]{50px}
    \centering
    \begin{tikzpicture}[xscale=1/4, yscale = 1/8]
      \draw[\Xcolor, rounded corners = 2pt] (1,0) -- (1,0.5)  -- (3,3.5) -- (3,4);
      \draw[\Xcolor, rounded corners = 2pt] (2,0) -- (2,0.5)  -- (4,3.5) -- (4,4);
      \draw[\Zcolor, rounded corners = 2pt] (4,0) -- (4,0.5)  -- (2,3.5) -- (2,4);
      \draw[\Zcolor, rounded corners = 2pt] (3,0) -- (3,0.5)  -- (1,3.5) -- (1,4);
      \draw[\Zcolor, rounded corners = 2pt] (5,0) -- (5,4);
      \draw[\Zcolor, rounded corners = 2pt] (6,0) -- (6,4);
      \draw[\Zcolor, rounded corners = 2pt] (7,0) -- (7,4);
    \end{tikzpicture}
  \end{minipage}
}

\newcommand{\dggggab}{
  \begin{minipage}[c][20px][c]{50px}
    \centering
    \begin{tikzpicture}[xscale=1/4, yscale = 1/4]
      \draw[\Xcolor, rounded corners = 2pt] (1,0) -- (1,2);
      \draw[\Xcolor, rounded corners = 2pt] (2,0) -- (2,0.5)  -- (6,1.5) -- (6,2);
      \draw[\Zcolor, rounded corners = 2pt] (3,0) -- (3,0.5)  -- (2,1.5) -- (2,2);
      \draw[\Zcolor, rounded corners = 2pt] (4,0) -- (4,0.5)  -- (3,1.5) -- (3,2);
      \draw[\Zcolor, rounded corners = 2pt] (5,0) -- (5,0.5)  -- (4,1.5) -- (4,2);
      \draw[\Zcolor, rounded corners = 2pt] (6,0) -- (6,0.5)  -- (5,1.5) -- (5,2);
      \draw[\Zcolor, rounded corners = 2pt] (7,0) -- (7,2);
    \end{tikzpicture}
  \end{minipage}
}

\newcommand{\dggggba}{
  \begin{minipage}[c][20px][c]{50px}
    \centering
    \begin{tikzpicture}[xscale=1/4, yscale = 1/4]
      \draw[\Xcolor, rounded corners = 2pt] (1,0) -- (1,2);
      \draw[\Zcolor, rounded corners = 2pt] (2,0) -- (2,2);
      \draw[\Xcolor, rounded corners = 2pt] (3,0) -- (3,0.5)  -- (7,1.5) -- (7,2);
      \draw[\Zcolor, rounded corners = 2pt] (4,0) -- (4,0.5)  -- (3,1.5) -- (3,2);
      \draw[\Zcolor, rounded corners = 2pt] (5,0) -- (5,0.5)  -- (4,1.5) -- (4,2);
      \draw[\Zcolor, rounded corners = 2pt] (6,0) -- (6,0.5)  -- (5,1.5) -- (5,2);
      \draw[\Zcolor, rounded corners = 2pt] (7,0) -- (7,0.5)  -- (6,1.5) -- (6,2);

    \end{tikzpicture}
  \end{minipage}
}

\newcommand{\dggggac}{
  \begin{minipage}[c][20px][c]{50px}
    \centering
    \begin{tikzpicture}[xscale=1/4, yscale = 1/8]
      \draw[\Xcolor, rounded corners = 2pt] (1,0) -- (1,1) -- (2,3) -- (2,4);
      \draw[\Zcolor, rounded corners = 2pt] (3,0) -- (3,1) -- (1,3) -- (1,4);
      \draw[\Zcolor, rounded corners = 2pt] (4,0) -- (4,1) -- (3,3) -- (3,4);            
      \draw[\Xcolor, rounded corners = 2pt] (2,0) -- (2,1) -- (5,3) -- (5,4);
      \draw[\Zcolor, rounded corners = 2pt] (5,0) -- (5,1) -- (4,3) -- (4,4);
      \draw[\Zcolor, rounded corners = 2pt] (6,0) -- (6,4);
      \draw[\Zcolor, rounded corners = 2pt] (7,0) -- (7,4);      
    \end{tikzpicture}
  \end{minipage}
}

\newcommand{\dggggca}{
  \begin{minipage}[c][20px][c]{50px}
    \centering
    \begin{tikzpicture}[xscale=1/4, yscale = 1/8]
      \draw[\Xcolor, rounded corners = 2pt] (1,0) -- (1,1) -- (4,3) -- (4,4);
      \draw[\Zcolor, rounded corners = 2pt] (2,0) -- (2,1) -- (1,3) -- (1,4);
      \draw[\Zcolor, rounded corners = 2pt] (3,0) -- (3,1) -- (2,3) -- (2,4);            
      \draw[\Xcolor, rounded corners = 2pt] (4,0) -- (4,1) -- (5,3) -- (5,4);
      \draw[\Zcolor, rounded corners = 2pt] (5,0) -- (5,1) -- (3,3) -- (3,4);
      \draw[\Zcolor, rounded corners = 2pt] (6,0) -- (6,4);
      \draw[\Zcolor, rounded corners = 2pt] (7,0) -- (7,4);
    \end{tikzpicture}
  \end{minipage}
}

\newcommand{\dggggcb}{
  \begin{minipage}[c][20px][c]{50px}
    \centering
    \begin{tikzpicture}[xscale=1/4, yscale = 1/4]
      \draw[\Zcolor, rounded corners = 2pt] (4,0) -- (4,0.5) --
      (3,1.5) -- (3,2);
      \draw[\Zcolor, rounded corners = 2pt] (3,0) -- (3,0.5) --
      (2,1.5) -- (2,2);
      \draw[\Zcolor, rounded corners = 2pt] (2,0) -- (2,0.5) --
      (1,1.5) -- (1,2);
      \draw[\Xcolor, rounded corners = 2pt] (1,0) -- (1,0.5) -- (4,1.5) -- (4,2);
      \draw[\Xcolor, rounded corners = 2pt] (5,0) -- (5,0.5) -- (6,1.5) -- (6,2);
      \draw[\Zcolor, rounded corners = 2pt] (6,0) -- (6,0.5) -- (5,1.5) -- (5,2);
      \draw[\Zcolor, rounded corners = 2pt] (7,0) -- (7,2);      
    \end{tikzpicture}
  \end{minipage}
}

\newcommand{\dggggbb}{
  \begin{minipage}[c][20px][c]{50px}
    \centering
    \begin{tikzpicture}[xscale=1/4, yscale = 1/8]
      \draw[\Xcolor, rounded corners = 2pt] (1,0) --(1,1)  -- (3,3) -- (3,4);
      \draw[\Zcolor, rounded corners = 2pt] (2,0) -- (2,1) --  (1,3) -- (1,4);
      \draw[\Zcolor, rounded corners = 2pt] (4,0) -- (4,1)  -- (2,3) -- (2,4);            
      \draw[\Xcolor, rounded corners = 2pt] (3,0) -- (3,1)  -- (5,3) -- (5,4);
      \draw[\Zcolor, rounded corners = 2pt] (5,0) -- (5,1) --  (4,3) -- (4,4);
      \draw[\Zcolor, rounded corners = 2pt] (6,0) -- (6,4);
      \draw[\Zcolor, rounded corners = 2pt] (7,0) -- (7,4);

    \end{tikzpicture}
  \end{minipage}
}
\newcommand{\dggggbc}{
  \begin{minipage}[c][20px][c]{50px}
    \centering
    \begin{tikzpicture}[xscale=1/4, yscale = 1/4]
      \draw[\Zcolor, rounded corners = 2pt] (5,0) -- (5,0.5) --
      (4,1.5) -- (4,2);
      \draw[\Zcolor, rounded corners = 2pt] (6,0) -- (6,0.5) --
      (5,1.5) -- (5,2);
      \draw[\Zcolor, rounded corners = 2pt] (4,0) -- (4,0.5) --
      (3,1.5) -- (3,2);
      \draw[\Xcolor, rounded corners = 2pt] (3,0) --
      (3,0.5) -- (6,1.5) -- (6,2);
      \draw[\Xcolor, rounded corners = 2pt] (1,0) -- (1,0.5) -- (2,1.5) -- (2,2);
      \draw[\Zcolor, rounded corners = 2pt] (2,0) -- (2,0.5) -- (1,1.5) -- (1,2);
      \draw[\Zcolor, rounded corners = 2pt] (7,0) -- (7,2);      
    \end{tikzpicture}
  \end{minipage}
}

\newcommand{\dggggcc}{
  \begin{minipage}[c][20px][c]{50px}
    \centering
    \begin{tikzpicture}[xscale=1/4, yscale = 1/4]
      \draw[\Zcolor, rounded corners = 2pt] (2,0) -- (2,0.5) --
      (1,1.5) -- (1,2);
      \draw[\Zcolor, rounded corners = 2pt] (3,0) -- (3,0.5) --
      (2,1.5) -- (2,2);
      \draw[\Xcolor, rounded corners = 2pt] (1,0) --
      (1,0.5) -- (3,1.5) -- (3,2);
      \draw[\Zcolor, rounded corners = 2pt] (5,0) -- (5,0.5) --
      (4,1.5) -- (4,2);
      \draw[\Zcolor, rounded corners = 2pt] (6,0) -- (6,0.5) --
      (5,1.5) -- (5,2);
      \draw[\Xcolor, rounded corners = 2pt] (4,0) --
      (4,0.5) -- (6,1.5) -- (6,2);
      \draw[\Zcolor, rounded corners = 2pt] (7,0) -- (7,2);      
    \end{tikzpicture}
  \end{minipage}
}

\newcommand{\dgggggaa}{
  \begin{minipage}[c][20px][c]{50px}
    \centering
    \begin{tikzpicture}[xscale=1/4, yscale = 1/8]
      \draw[\Xcolor, rounded corners = 2pt] (1,0) -- (1,4);
      \draw[\Xcolor, rounded corners = 2pt] (2,0) -- (2,1)  -- (7,3) -- (7,4);
      \draw[\Zcolor, rounded corners = 2pt] (3,0) -- (3,1)  -- (2,3) -- (2,4);
      \draw[\Zcolor, rounded corners = 2pt] (4,0) -- (4,1)  -- (3,3) -- (3,4);
      \draw[\Zcolor, rounded corners = 2pt] (5,0) -- (5,1)  -- (4,3) -- (4,4);
      \draw[\Zcolor, rounded corners = 2pt] (6,0) -- (6,1)  -- (5,3) -- (5,4);
      \draw[\Zcolor, rounded corners = 2pt] (7,0) -- (7,1)  -- (6,3) -- (6,4);
    \end{tikzpicture}
  \end{minipage}
}

\newcommand{\dgggggab}{
  \begin{minipage}[c][20px][c]{50px}
    \centering
    \begin{tikzpicture}[xscale=1/4, yscale = 1/8]
      \draw[\Xcolor, rounded corners = 2pt] (1,0) -- (1,0.5)  -- (3,3.5) -- (3,4);
      \draw[\Xcolor, rounded corners = 2pt] (2,0) -- (2,0.5)  -- (5,3.5) -- (5,4);
      \draw[\Zcolor, rounded corners = 2pt] (3,0) -- (3,0.5)  -- (1,3.5) -- (1,4);
      \draw[\Zcolor, rounded corners = 2pt] (4,0) -- (4,0.5)  -- (2,3.5) -- (2,4);
      \draw[\Zcolor, rounded corners = 2pt] (5,0) -- (5,0.5)  -- (4,3.5) -- (4,4);
      \draw[\Zcolor, rounded corners = 2pt] (6,0) -- (6,4);
      \draw[\Zcolor, rounded corners = 2pt] (7,0) -- (7,4);
    \end{tikzpicture}
  \end{minipage}
}

\newcommand{\dgggggac}{
  \begin{minipage}[c][20px][c]{50px}
    \centering
    \begin{tikzpicture}[xscale=1/4, yscale = 1/8]
      \draw[\Xcolor, rounded corners = 2pt] (1,0) -- (1,1) -- (2,3) -- (2,4);
      \draw[\Zcolor, rounded corners = 2pt] (3,0) -- (3,1) -- (1,3) -- (1,4);
      \draw[\Zcolor, rounded corners = 2pt] (4,0) -- (4,1) -- (3,3) -- (3,4);            
      \draw[\Xcolor, rounded corners = 2pt] (2,0) -- (2,1) -- (6,3) -- (6,4);
      \draw[\Zcolor, rounded corners = 2pt] (5,0) -- (5,1) -- (4,3) -- (4,4);
      \draw[\Zcolor, rounded corners = 2pt] (6,0) -- (6,1) -- (5,3) -- (5,4);
      \draw[\Zcolor, rounded corners = 2pt] (7,0) -- (7,4);      
    \end{tikzpicture}
  \end{minipage}
}

\newcommand{\dgggggba}{
  \begin{minipage}[c][20px][c]{50px}
    \centering
    \begin{tikzpicture}[xscale=1/4, yscale = 1/8]
      \draw[\Xcolor, rounded corners = 2pt] (1,0) -- (1,0.5)  -- (4,3.5) -- (4,4);
      \draw[\Xcolor, rounded corners = 2pt] (3,0) -- (3,0.5)  -- (5,3.5) -- (5,4);
      \draw[\Zcolor, rounded corners = 2pt] (2,0) -- (2,0.5)  -- (1,3.5) -- (1,4);
      \draw[\Zcolor, rounded corners = 2pt] (4,0) -- (4,0.5)  -- (2,3.5) -- (2,4);
      \draw[\Zcolor, rounded corners = 2pt] (5,0) -- (5,0.5)  -- (3,3.5) -- (3,4);
      \draw[\Zcolor, rounded corners = 2pt] (6,0) -- (6,4);
      \draw[\Zcolor, rounded corners = 2pt] (7,0) -- (7,4);
    \end{tikzpicture}
  \end{minipage}
}

\newcommand{\dgggggbb}{
  \begin{minipage}[c][20px][c]{50px}
    \centering
    \begin{tikzpicture}[xscale=1/4, yscale = 1/8]
      \draw[\Xcolor, rounded corners = 2pt] (1,0) -- (1,1)  -- (5,3) -- (5,4);
      \draw[\Zcolor, rounded corners = 2pt] (2,0) -- (2,1)  -- (1,3) -- (1,4);
      \draw[\Zcolor, rounded corners = 2pt] (3,0) -- (3,1)  -- (2,3) -- (2,4);
      \draw[\Zcolor, rounded corners = 2pt] (4,0) -- (4,1)  -- (3,3) -- (3,4);
      \draw[\Zcolor, rounded corners = 2pt] (5,0) -- (5,1)  -- (4,3) -- (4,4);
      \draw[\Xcolor, rounded corners = 2pt] (6,0) -- (6,1)  -- (7,3) -- (7,4);
      \draw[\Zcolor, rounded corners = 2pt] (7,0) -- (7,1)  -- (6,3) -- (6,4);
    \end{tikzpicture}
  \end{minipage}
}

\newcommand{\dgggggbc}{
  \begin{minipage}[c][20px][c]{50px}
    \centering
    \begin{tikzpicture}[xscale=1/4, yscale = 1/8]
      \draw[\Xcolor, rounded corners = 2pt] (1,0) -- (1,1) -- (3,3) -- (3,4);
      \draw[\Zcolor, rounded corners = 2pt] (2,0) -- (2,1) -- (1,3) -- (1,4);
      \draw[\Zcolor, rounded corners = 2pt] (4,0) -- (4,1) -- (2,3) -- (2,4);            
      \draw[\Xcolor, rounded corners = 2pt] (3,0) -- (3,1) -- (6,3) -- (6,4);
      \draw[\Zcolor, rounded corners = 2pt] (5,0) -- (5,1) -- (4,3) -- (4,4);
      \draw[\Zcolor, rounded corners = 2pt] (6,0) -- (6,1) -- (5,3) -- (5,4);
      \draw[\Zcolor, rounded corners = 2pt] (7,0) -- (7,4);      
    \end{tikzpicture}
  \end{minipage}
}

\newcommand{\dgggggca}{
  \begin{minipage}[c][20px][c]{50px}
    \centering
    \begin{tikzpicture}[xscale=1/4, yscale = 1/8]
      \draw[\Xcolor, rounded corners = 2pt] (1,0) -- (1,1) -- (5,3) -- (5,4);
      \draw[\Zcolor, rounded corners = 2pt] (2,0) -- (2,1) -- (1,3) -- (1,4);
      \draw[\Zcolor, rounded corners = 2pt] (3,0) -- (3,1) -- (2,3) -- (2,4);            
      \draw[\Xcolor, rounded corners = 2pt] (5,0) -- (5,1) -- (6,3) -- (6,4);
      \draw[\Zcolor, rounded corners = 2pt] (4,0) -- (4,1) -- (3,3) -- (3,4);
      \draw[\Zcolor, rounded corners = 2pt] (6,0) -- (6,1) -- (4,3) -- (4,4);
      \draw[\Zcolor, rounded corners = 2pt] (7,0) -- (7,4);
    \end{tikzpicture}
  \end{minipage}
}

\newcommand{\dgggggcb}{
  \begin{minipage}[c][20px][c]{50px}
    \centering
    \begin{tikzpicture}[xscale=1/4, yscale = 1/8]
      \draw[\Xcolor, rounded corners = 2pt] (1,0) -- (1,1) -- (4,3) -- (4,4);
      \draw[\Zcolor, rounded corners = 2pt] (2,0) -- (2,1) -- (1,3) -- (1,4);
      \draw[\Zcolor, rounded corners = 2pt] (3,0) -- (3,1) -- (2,3) -- (2,4);            
      \draw[\Xcolor, rounded corners = 2pt] (4,0) -- (4,1) -- (6,3) -- (6,4);
      \draw[\Zcolor, rounded corners = 2pt] (5,0) -- (5,1) -- (3,3) -- (3,4);
      \draw[\Zcolor, rounded corners = 2pt] (6,0) -- (6,1) -- (5,3) -- (5,4);
      \draw[\Zcolor, rounded corners = 2pt] (7,0) -- (7,4);      
    \end{tikzpicture}
  \end{minipage}
}

\newcommand{\dgggggcc}{
  \begin{minipage}[c][20px][c]{50px}
    \centering
    \begin{tikzpicture}[xscale=1/4, yscale = 1/4]
      \draw[\Zcolor, rounded corners = 2pt] (2,0) -- (2,0.5) --
      (1,1.5) -- (1,2);
      \draw[\Zcolor, rounded corners = 2pt] (3,0) -- (3,0.5) --
      (2,1.5) -- (2,2);
      \draw[\Zcolor, rounded corners = 2pt] (4,0) -- (4,0.5) --
      (3,1.5) -- (3,2);
      \draw[\Xcolor, rounded corners = 2pt] (1,0) --
      (1,0.5) -- (4,1.5) -- (4,2);
      \draw[\Zcolor, rounded corners = 2pt] (6,0) -- (6,0.5) --
      (5,1.5) -- (5,2);
      \draw[\Xcolor, rounded corners = 2pt] (5,0) --
      (5,0.5) -- (7,1.5) -- (7,2);
      \draw[\Zcolor, rounded corners = 2pt] (7,0) -- (7,0.5) --
      (6,1.5) -- (6,2);
    \end{tikzpicture}
  \end{minipage}
}

\begin{document}

\setlength{\jot}{0.3cm}
\renewcommand{\thefootnote}{\fnsymbol{footnote}}
\setcounter{footnote}{0}

\thispagestyle{empty}
\begin{flushright}
LPTENS-16/08\\

\end{flushright}
\vspace{0.6cm} \setcounter{footnote}{0}
\begin{center}
{\Large{\bf Chiral  limit of $\mathcal{N}=4$ SYM and ABJM\\   and integrable Feynman graphs   }\\
}\vspace{8mm}
{ Jo\~ao Caetano$^a$, \hspace{1cm} \"Omer G\"urdo\u gan$^{a,b}$, \hspace{1cm}  Vladimir~Kazakov$^{a}$\\[7mm]
\large\it\small 
$^a$Laboratoire de Physique Th\'eorique\\
de l'Ecole Normale Sup\'erieure et l'Universit\'e Paris-VI,\\
24 rue Lhomond, Paris  CEDEX 75231, France\\
$^b$School of Physics \& Astronomy, University of Southampton,\\
Highfield, Southampton, SO17 1BJ, United Kingdom.\footnote{ \scriptsize \tt\noindent
Email:  \indent  joao.caetano@lpt.ens.fr, \indent  o.c.gurdogan@soton.ac.uk,\indent   kazakov@physique.ens.fr}}

\vspace{4mm}
\includegraphics[height=3cm]{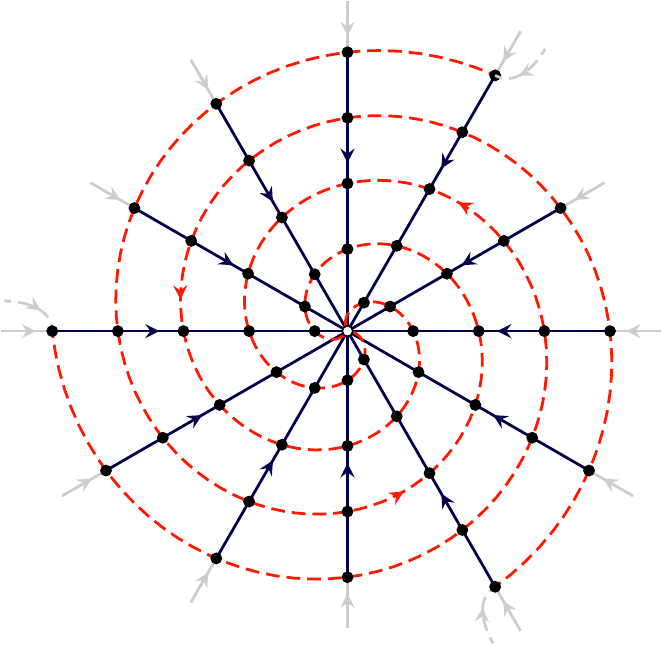}
\hspace{6cm}
\includegraphics[height=3cm]{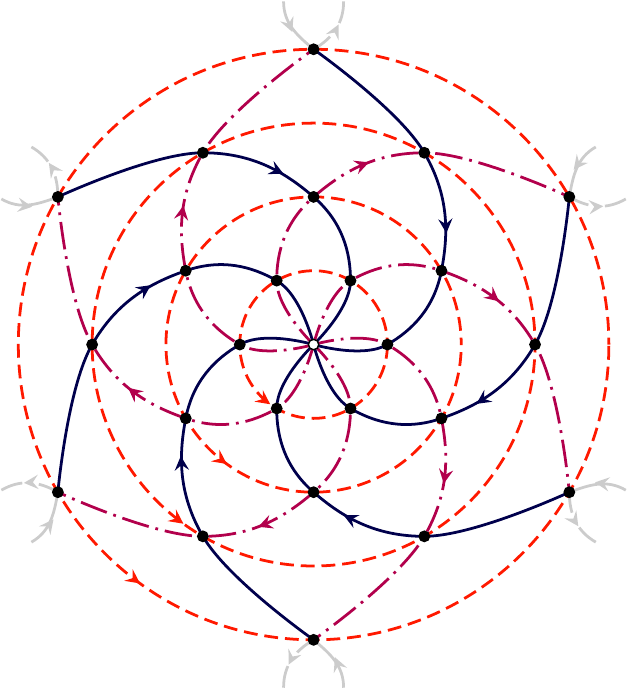}

\end{center}
\noindent\\[-2cm]

\begin{center}
{\sc Abstract}\\[2mm]
\end{center}

We consider a special double scaling limit, recently introduced by two of the authors, combining weak coupling and large imaginary twist,  for the \(\gamma\)-twisted $\mathcal{N}=4$ SYM theory. We also establish the analogous limit for ABJM theory.  The resulting non-gauge chiral 4D and 3D theories of interacting scalars and fermions are integrable in the planar limit.  In spite of the breakdown of conformality by double-trace interactions,  most of the  correlators for local operators of these theories are conformal, with non-trivial anomalous dimensions defined by specific, integrable Feynman diagrams. We discuss the details of this diagrammatics.  We construct the doubly-scaled asymptotic Bethe ansatz (ABA) equations  for multi-magnon states in these theories.   Each entry of the mixing matrix of local conformal operators in the simplest of these theories - the bi-scalar model in 4D and tri-scalar model in 3D -  is given by a single Feynman diagram at  any given loop order. The related diagrams are in principle computable,  up to a few scheme dependent constants,  by integrability methods (quantum spectral curve or ABA). These constants should be fixed from direct computations of a few simplest  graphs. 
This integrability-based method is advocated to be able to provide information about some high loop order graphs which are hardly computable by other known methods.  We exemplify our approach 
with specific five-loop graphs.             

\newpage
\newpage
\renewcommand{\baselinestretch}{0.80}\normalsize
\tableofcontents
\renewcommand{\baselinestretch}{1.04}\normalsize
\newpage

\renewcommand{\thefootnote}{\arabic{footnote}}
\setcounter{footnote}{0}


\section{Introduction}

The examples of solvable, or integrable QFTs in more than two
dimensions are very rare. In fact only two such theories with
non-trivial integrable dynamics are known\footnote{apart from
  topological sectors of some $\mathcal{N}=1$ and $\mathcal{N}=2$ SYM
  theories where typical physical quantities are dominated by
  instanton effects rather than by Feynman perturbation series. }:
four-dimensional $\mathcal{N}=4$ SYM theory and three-dimensional ABJM
model in the planar, or 't~Hooft limit \cite{Beisert:2010jr}.  Gauge
symmetry and a large amount of supersymmetry, as well as the existence
of AdS/CFT string duals for these theories have been long believed to
be a prerequisite for their quantum integrability.

However, in the last years various deformations of these theories have
been considered which seem to preserve planar integrability. 
Examples include the so-called \(\beta\)- and \(\gamma\)-deformations
which break the R-symmetry
\cite{Beisert:2005if,Lunin:2005jy,Leigh1995}, or even more general
deformations \cite{Beisert:2005if,Gromov:2007ky} of $\mathcal{N}=4$
SYM, which break partially or completely the global superconformal
$\mathfrak{psu}(2,2|4)$ symmetry, together with a part of
supersymmetry or even all of it. A similar twisting is possible for
the ABJM model \cite{Chen:2016geo,Imeroni:2008cr}.
It seems that the consequences of these generalizations have not been
yet completely explored.

Recently, two of the current authors proposed in
\cite{Gurdogan:2015csr} a special, double scaling limit of
\(\gamma\)-twisted $\mathcal{N}=4$ SYM theory \cite{Leigh1995}, such
that the three twisting parameters \(q_j=e^{-\frac{i}{2}\g_j},\,\,
j=1,2,3 \), approach infinity (or zero) and the 't~Hooft coupling
\(g=\sqrt{N_c}\, g_{YM}\) is sent to zero, while their product \(\xi_i=gq_j\)
(or ratio) is kept fixed.  Since the twist parameters \(\gamma_j\)
have to be taken imaginary for this limit, the resulting action (see
equations \eqref{chiFT4} and \eqref{chiFT4int} in the following
section) is not real and the theory is not unitary. The interaction
vertices impose a specific clockwise (or anti-clockwise) orientation
on planar Feynman graphs reflecting the chirality property of the
theory. For this reason, slightly abusing the common terminology, we
will call it the ``chiral field theory'' and use the abbreviation
\(\chi\)FT.  This theory, represented by the action
eqs. \eqref{chiFT4}-\eqref{chiFT4int}, contains only three complex
scalars and three complex fermions which are interacting through a few
quartic scalar and Yukawa couplings, all oriented in the same way on
planar graphs. The gauge fields and the gluino decouple in this limit. 
The supersymmetry is completely lost apart from a
particular choice called the \(\beta\)-deformation when all three
couplings are equal (with the action given by equation
\eqref{chiFT4beta}).  Nevertheless, these new theories have to be
integrable in the planar limit and they show indeed multiple signs of
this integrability \cite{Gurdogan:2015csr}. In the simplest case of a
single non-zero coupling the theory reduces just to two interacting
massless complex scalar fields, equation \eqref{bi-scalarL}. We will call
this model the bi-scalar \(\chi\)FT.

One of the remarkable features of these \(\chi\)FTs is a great
simplification of Feynman diagrammatics compared to their
``mother'' theory -- the $\mathcal{N}=4$ SYM.  For the bi-scalar
\(\chi\)FT, one can roughly say that a generic multi-point correlation
function of single-trace local operators
\(\Tr(\chi_{1}\chi_{2}\dots\chi_{L})\), where
\(\chi_j\in\{\phi^1,\phi_1^{\dagger},\phi^2,\phi_2^{\dagger}\}\), can
have at most one Feynman graph at each order of perturbation theory in
the planar limit. The bulk of a higher-loop planar Feynman graph 
looks like a regular quadratic lattice. It was noticed a long time ago that
such regular ``fishnet'' graphs define, due to the star-triangle
relations, an integrable quantum spin chain model
\cite{Zamolodchikov:1980mb} with four-dimensional conformal
$\mathfrak{su}(2,2)$ symmetry. The quantum spins of this spin chain are
four-dimensional coordinates of the physical space and they live in the
principal series representation of 
$\mathfrak{su}(2,2)$. More specifically, the
only graphs that renormalise the simplest operator in the bi-scalar
\(\chi\)FT, such as the ``BMN vacuum'' \(\Tr((\phi^1)^L)\), are
globe-like graphs of the
figure \ref{twopointfunction}, which resemble parallels and meridians
drawn on a globe  \cite{Gurdogan:2015csr}. For the purpose of computing their anomalous
dimensions, the \(L\) propagators ending up at the north (or south)
pole can be amputated giving simpler ``wheel'' graphs with \(L\)
spokes and \(M\) frames, which are of \(ML\)-loop order in the
coupling \(\xi^2\) (see figure \ref{doublewheel}). Replacing a few
\(\phi^1\) fields by \(\phi^2\) fields (``magnons'') inside the trace
will change the boundary conditions of the bulk lattice resulting in
more involved graphs than just wheels, for example those shown in
figures \ref{one_m_spiral}-\ref{two_m_spiral}. However these two-point
functions, or equivalently graphs, will nevertheless be
integrable.

The analysis of possible graph configurations of the bi-scalar \(\chi\)FT
and their computations using integrability will be one of the goals of
this paper.
  
We will also apply the double scaling limit to the three-dimensional
\(\g\)-twisted ABJM model where we will show that it reduces in the simplest limit to a
model of three complex scalars interacting through a single sextic vertex.  It
has also a limited set of graphs for similar physical quantities,
looking like a regular triangular lattice in their bulk. The
\(\beta\)-deformed ABJM model (with only one non-zero twist parameter)
can also contain fermions in a slightly different double-scaling limit.
  
The tools of quantum AdS/CFT integrability, such as asymptotic
twisted Bethe ansatz (ABA) \cite{Beisert:2005if,Beisert:2006ez} for
dimensions of long operators\footnote{or for the exact perturbative
  expansion of these dimensions up to the order preceeding the
  ``wrapping'': roughly, to all orders smaller than the length of the
  operator } or twisted quantum spectral curve (QSC) equations for
dimensions of any single-trace
operator \cite{Gromov:2015dfa,Gromov:2014caa,Kazakov:2015efa}, as well
as the recently developed methods of computation of structure
constants \cite{Basso:2015zoa,Basso:2015eqa,Fleury:2016ykk}, should be
applicable to these new \(\chi\)FT theories after their appropriate
double scaling has been found. In the present paper, we show how to
compute the scaling dimensions of operators with magnons using the
doubly scaled twisted ABA equations in a particular ``broken''
$\mathfrak{su}(3)$ sector. Similar ABA equations for the ABJM models will be
also deduced here. We show the efficiency of this doubly scaled ABA
for computations of multi-magnon anomalous dimensions in these
\(\chi\)FTs to high orders.

Moreover, we will use these integrability tools to relate particularly
complicated five-loop Feynman graphs to simpler ones of the same loop
order. The method is based on constructing the mixing matrix for
multi-magnon operators up to a given order of perturbation theory,
where each entry is given by a constant represented by a single scalar
graph. Then we find as many relations between these constants as
possible by comparing the spectrum of this mixing matrix with the one
given from ABA. Due to the freedom related to similarity
transformations of the mixing matrix, this procedure predicts the
values of Feynman graphs only up to a few scheme dependent parameters
which can be then fixed by direct computation of a few simple
graphs. We demonstrate the method by 
fixing two linear combinations of three relevant five-loop graphs
(up to \(1/\epsilon\) terms in dimensional regularization).

The \(\g\)-deformed $\mathcal{N}=4$ SYM and ABJM theories are not
conformal in a strict sense, even in the planar limit
\cite{Fokken:2013aea,Jin:2013baa}. The reason for this is the presence of
double-trace interactions of the type
\(\Tr(\chi_1\chi_2)\Tr(\chi_3\chi_4)\) that renormalise the Lagrangian
already at one-loop order and that the beta functions for the
couplings of these interactions are not zero (they were computed at
one loop in \cite{Sieg:2016vap}). One of the manifestations of this
non-conformality is the diverging anomalous dimension of operators of
length 2, such as \(\Tr(\phi^1\phi^2)\) or
\(\Tr(\phi^1\phi_1^\dagger)\).  But most of the correlators of local
operators still obey the conformal properties in planar
approximation. On the string side of the AdS/CFT duality
\cite{Beisert:2005if,Frolov:2005dj,Frolov:2005ty,Frolov:2005iq} it was
noticed in \cite{Pomoni2009} that these singularities are due to a
tachyon state which unavoidably appears in the deformed string theory.

Although it is not clear whether our \(\chi\)FTs have a well-defined
string dual, since the weak coupling limit in our double scaling corresponds
to an infinitely-strongly-coupled string theory, the singularity
appearing for the shortest operators of \(\chi\)FTs looks similar to
the tachyonic string singularity. For non-planar contributions, when
we sum over all states, the appearance of such singular \(L=2\) states
is unavoidable: since we sum up all the states around closed cycles
there will be always a ``tachyonic'' state \(L=2\) propagating around
non-trivial cycles of graphs, so the conformal symmetry is broken down
for all physical quantities already at the first \(1/N_c^2\)
correction. Similarly to the original \(\g\)-deformed $\mathcal{N}=4$
SYM and ABJM theories in the 't~Hooft limit, the majority of
correlation functions of \(\chi\)FTs considered here have a conformal
form due to the fact that the couplings \(\xi_j\) are not running at
\(N_c\to\infty\). The only condition for conformality of a correlation
function is that all the involved operators should have the length
\(L>2\) and there should be no intermediate states in the OPE of length \(L=2\)
as well.

This paper is organised as follows: in the first two sections we will
introduce the \(\chi\)FT\(_4\) and \(\chi\)FT\(_3\) theories emerging
from double scaling limits of $\gamma$-deformed \({\cal N}=4 \) SYM
and for ABJM theories, repectively, and describe their planar
diagrammatics for the two-point functions of single-trace
operators. We will then formulate the doubly scaled asymptotic Bethe
Ans\"atze in particular symmetry sectors of these theories. In section
five we will show how to compute particular relations for
dimensionally regulated five-loop Feynman graphs related to the mixing
matrix of multi-magnon operators, with an important help provided by
the integrability in the form of ABA. Finally, in the last part of the
paper we reproduce the values of a class of Feynman integrals using
the strong-deformation limit of the wrapping corrections to the
anomalous dimensions in $\beta$-deformed \(\mathcal{N}=4\) SYM which
were computed in \cite{Fiamberti2009}.


\section{Integrable  chiral field theories  ($\chi$FTs)  from DS limit  of $\gamma$-twisted $\mathcal{N}=4$ SYM }

In this section, we will first remind the definition of the double
scaling limit of \(\gamma\)-twisted $\mathcal{N}=4$ SYM and the
resulting action of \(\chi\mathrm{FT}_4\) with three couplings
proposed in \cite{Gurdogan:2015csr}, as well as its particular cases
with one and two non-zero couplings.  We will give an overview of
possible operators and correlation functions of \(\chi\mathrm{FT}_4\)
and describe their Feynman diagrammatics and the reasons for its
drastic simplification in the planar limit. In particular, in addition to
the globe graphs \cite{Gurdogan:2015csr} for two-point correlators of
BMN-type operators (or their ``amputated'' version -- the wheel
graphs), we will encounter, for (multi)-magnon operators, the
(multi)-spiral types graphs of the type depicted in figure
\ref{one_m_spiral}.  We will
also comment in this section on the reasons of breakdown of
conformality by double-trace interactions in these \(\chi\)FTs and of
the persistence of conformal properties of large majority of
correlators.

\subsection{$\chi\mathrm{FT}_4$ as DS limit of $\mathcal{N}=4$ SYM }

The \(\gamma\)-deformed action of \(\mathcal{N}=4\) SYM is presented in Appendix~\ref{app:actionssym}. It was proposed in~\cite{Gurdogan:2015csr} the following  DS limit of the \(\gamma\)-deformed Lagrangian \eqref{N=4SYMlagrangian} combining weak coupling together with large imaginary gamma parameters: 
 \begin{equation}
 \begin{aligned}\label{chiral}
 & q_3 \sim q_2 \sim q_1 \to\infty, \qquad g\to 0,\\
   & \xi_1\equiv q_1 g,\quad \xi_2\equiv q_2 g, \quad \xi_3\equiv q_3 g \quad
   \text{fixed},
   \end{aligned}
   \end{equation} where the large \(q_i\) limit
 corresponds to sending \(\gamma_i\to i \infty\).  In this DS limit   only certain Yukawa and
 4-scalar interactions survive in \(\mathcal{L}_{\mathrm{int}}\) and we arrived in \cite{Gurdogan:2015csr} 
 at the following  \(\chi\)FT\(_4\) of complex scalars and fermions (no gauge fields!):
\begin{equation}\label{chiFT4}
  {\cal L}_{\phi\psi}=N_c\Tr\left(-\frac{1}{2}\p^\mu\phi^\dagger_j\p_\mu\phi^j
  +i\bar\psi^{\dot\alpha}_{ j}
  (\tilde\sigma^{\mu})^\alpha_{\dot\alpha}\p_\mu \psi^j_{\alpha }\right)
+{\cal L}_{\rm int}
\end{equation}
where
the sum  is taken with respect to doubly repeated \(j=1,2,3\) and 
\begin{equation}
\begin{aligned}
     \mathcal{L}_{\rm int} &={}N_c
\,\Tr\Bigl(\xi_1^2\,\phi_2^\dagger \phi_3^\dagger 
\phi^2\phi^3+\xi_2^2\,\phi_3^\dagger \phi_1^\dagger 
\phi^3\phi^1+\xi_3^2\,\phi_1^\dagger \phi_2^\dagger \phi^1\phi^2\hfill \label{chiFT4int} \\
 &+i\sqrt{\xi_2\xi_3}(\psi^3 \phi^1 \psi^{ 2}+ \bar\psi_{ 3} \phi^\dagger_1 \bar\psi_2 )
 +i\sqrt{\xi_1\xi_3}(\psi^1 \phi^2 \psi^{ 3}+ \bar\psi_{ 1} \phi^\dagger_2 \bar\psi_3 )
 +i\sqrt{\xi_1\xi_2}(\psi^2 \phi^3 \psi^{ 1}+ \bar\psi_{ 2} \phi^\dagger_3 \bar\psi_1 )\,\Bigr).  \\
\end{aligned}
\end{equation}
We suppressed in the last equation the spinorial indices assuming the scalar product of both fermions in each term. 
An interesting particular case emerges from this action in the limit \(\xi_3\to 0\),
\begin{equation}
     \mathcal{L}_{\rm int} = N_c
\,\Tr\Bigl(\xi_1^2\,\phi_2^\dagger \phi_3^\dagger 
\phi^2\phi^3+\xi_2^2\,\phi_3^\dagger \phi_1^\dagger 
\phi^3\phi^1
 +i\sqrt{\xi_1\xi_2}(\psi^2 \phi^3 \psi^{ 1}+ \bar\psi_{ 2} \phi^\dagger_3 \bar\psi_1 )\,\Bigr) \label{chiFT4two}
 \end{equation} when only two fermions are left.  
 
If we send in \eqref{chiFT4int} \(\xi_1\to 0,\,\,\xi_2\to 0\) then the fermions and one of the scalars decouple and we get an even  simpler bi-scalar action~\cite{Gurdogan:2015csr}      
\begin{equation}
    \label{bi-scalarL}
    {\cal L}_{\phi}= \frac{N_c}{2}\Tr
    \left(\p^\mu\phi^\dagger_1 \p_\mu\phi^1+\p^\mu\phi^\dagger_2 \p_\mu\phi^2+2\xi^2\,\phi_1^\dagger \phi_2^\dagger \phi^1\phi^2\right)\,,
  \end{equation}
where we denoted \(\xi\equiv\xi_3=q_3 g\).

A particular case of all equal couplings \(\xi_1=\xi_2=\xi_3=\xi\) represents the so-called \(\beta\)-deformation, with one supersymmetry left intact,  \begin{equation}
\begin{aligned}
\label{chiFT4beta}
     \mathcal{L}_{\rm int} &={}N_c
\,\xi\,^2\Tr\left(\phi_2^\dagger \phi_3^\dagger 
\phi^2\phi^3+\,\phi_3^\dagger \phi_1^\dagger 
\phi^3\phi^1+\,\phi_1^\dagger \phi_2^\dagger \phi^1\phi^2\right)+\\
 &+N_c
\,i\,\xi\,\,\Tr\left(\psi^3 \phi^1 \psi^{ 2}+ \bar\psi_{ 3} \phi^\dagger_1 \bar\psi_2 
 +\psi^1 \phi^2 \psi^{ 3}+ \bar\psi_{ 1} \phi^\dagger_2 \bar\psi_3 +\psi^2 \phi^3 \psi^{ 1}+ \bar\psi_{ 2} \phi^\dagger_3 \bar\psi_1 \right)\,.
 \end{aligned}
 \end{equation}
According to the observations of \cite{Fokken2014c},  if the gauge symmetry is \(U(N_c)\), the \(\beta\)-deformed   \(\mathcal{N}=4\) SYM theory is not conformal because of  the double-trace counterterms generated in RG due to the \(U(1)\) degrees of freedom of scalar fields. The same is true about its doubly scaled version \eqref{chiFT4beta}. However,  for the \(SU(N_c)\) gauge symmetry the \(\beta\)-deformed \(\mathcal{N}=4\) SYM is conformal and its doubly scaled version \eqref{chiFT4beta} as well!

These theories are chiral, in the sense that the actions are not invariant w.r.t. complex conjugation. The missing complex conjugated terms (with opposite chirality) can be retrieved in the opposite, physically equivalent, DS limit
\begin{equation}
\begin{aligned}
\label{antichiral}& q_3 \sim q_2 \sim q_1 \to0, \qquad g\to 0,\\
   & \xi_1\equiv q_1 /g,\quad \xi_2\equiv q_2 /g, \quad \xi_3 \equiv q_3 /g \quad
   \text{fixed},
   \end{aligned}
   \end{equation} where the large \(q_i\) limit
 corresponds to sending \(\gamma_i\to -i \infty\). Thus the chiral and anti-chiral interactions completely decouple in our DS limit.

Let us end this section with a comment on further double scaling limits. Besides the examples we have worked out above, we can scale $q_i$ differently for each $i$ and generate many more Lagrangians. In particular, with the following particular choice
 \begin{equation}
 \begin{aligned}\label{chiral}
 & q_1, q_2 \to\infty, \qquad q_3, g\to 0,\\
   & \xi_1\equiv q_1 \sqrt{g},\quad \xi_2\equiv q_2 g, \quad \xi_3\equiv g/q_3 \quad
   \text{fixed},
   \end{aligned}
   \end{equation} 
the gluino $\psi^4$ and its conjugate can survive and we get the following interacting theory
\begin{equation}
 \mathcal{L}_{\rm int} ={}N_c \Tr \left( \xi_2^2\, \phi^{\dagger}_3 \phi^{\dagger}_1 \phi^3 \phi^1+\xi_3^2\,\phi^{\dagger}_2 \phi^{\dagger}_1 \phi^2 \phi^1 + \sqrt{\xi_2 \xi_3} \left(\bar{\psi }_1 \phi^1 \bar{\psi }_4-\psi ^1 \phi^{\dagger}_1 \psi ^4\right) \right)\,.
\end{equation}
In the following we will explore the diagrammatics and integrability of the uniformly double scaled theories presented above but an analogous analysis applies to this other class of models as well.

\subsection{Planar diagrammatics of  $\chi\mathrm{FT}_4$ and the breakdown of conformality}

\begin{figure}[t]
\begin{center}
  \includegraphics[]{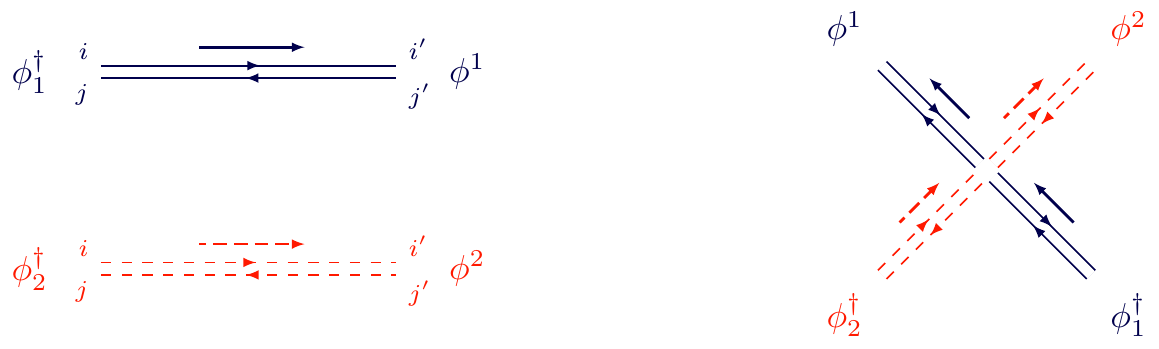}
\end{center}
\caption{Feynman rules for the bi-scalar theory. Double lines represent the colour dependence of the fields. There are two types of propagators:  solid lines correspond to \(\phi^1\) and dashed lines to \(\phi^2\) fields. The arrows besides the double lines indicate the flavour flow of complex scalars. The only interaction vertex is a particular quartic one and its orientation implies a sense of chirality for the graphs it enters. The vertex with opposite orientation (chirality) is absent. }
\label{Feynman_rules}
\end{figure}

We will describe the planar diagrams and the breakdown of conformal properties of  \(\chi\)FT\(_4\)   on the example of the simplest  bi-scalar theory \eqref{bi-scalarL}.  We have there two types of propagators and one interacting vertex, all of them presented on figure \ref{Feynman_rules}. Due to this limited set of building blocks, the theory looks ``almost'' conformal in the 't~Hooft limit.
  
\begin{figure}[t]
   \begin{center}
     \centering
    \begin{minipage}{7cm}
      \centering
      \includegraphics[scale=1]{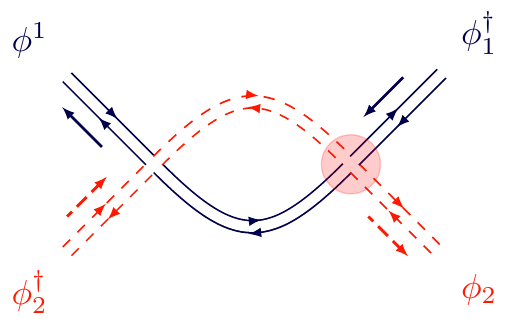}
    \end{minipage}
    \hspace{2cm}
    \begin{minipage}{5cm}
      \centering
      \includegraphics[scale=1]{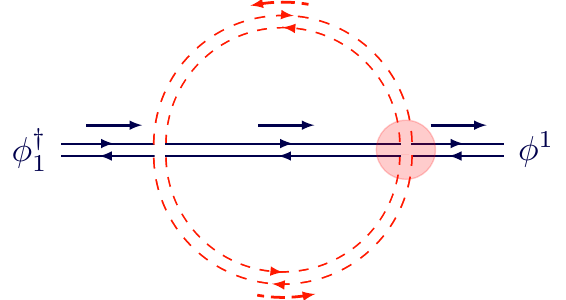}
    \end{minipage}
\end{center}
\caption{The one-loop
  planar Feynman  graphs which could contribute to the renormalization of the single coupling \(\xi\) (on the left picture) or generate the mass (on the right picture). But on each graph, only one of two vertices is present in  perturbation theory and the other vertex, indicated with red, has a wrong ordering of the fields, so that these diagrams do not contribute. This argument can be generalized to any loop order. Therefore the mass is not generated and the coupling \(\xi\)   is not running in the planar limit. }
\label{renormalisation}
\end{figure}

 Indeed, consider for example the  planar diagrams on figure \ref{renormalisation} which could renormalize, respectively, the coupling \(\xi\) and the mass.  However, it is easy to see that, due to the fixed orientation of the single vertex of the bi-scalar model, both diagrams are absent since each of them has vertices with opposite orientations (chiralities). This argument can be generalized to higher loops.

\begin{figure}[t]
  \begin{center}
    \begin{minipage}{7cm}
      \centering
      \includegraphics[scale=1]{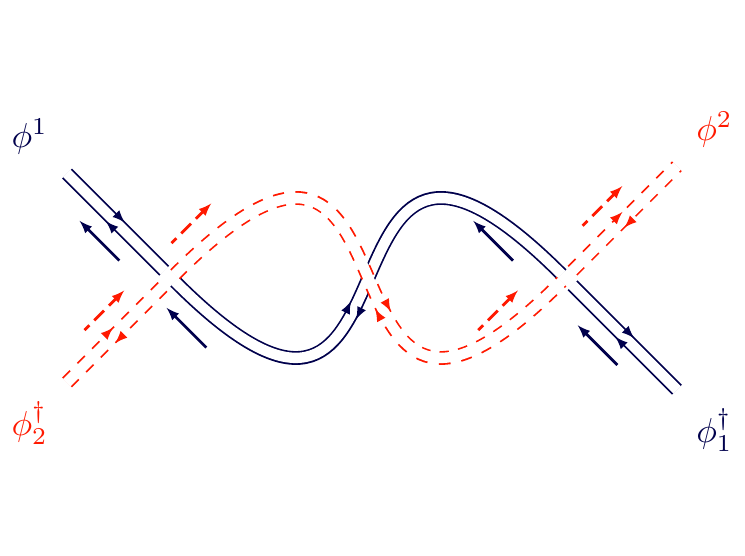}
    \end{minipage}
    \hspace{2cm}
    \begin{minipage}{5cm}
      \centering
      \includegraphics[scale=1]{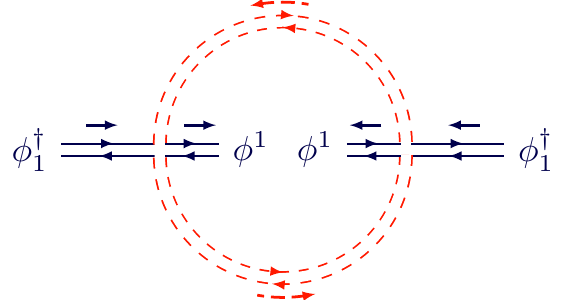}
    \end{minipage}
  \end{center}
\caption{Non-planar diagrams that generate couplings of the form \(\Tr(\phi^1\phi_2^\dagger)\,\Tr(\phi^2\phi_1^\dagger)\) and \(\bigl[\Tr(\phi^1\phi_1^\dagger)\bigr]^2\). }
\label{doubletracecouplings}
\end{figure}

However, the graphs on figure \ref{doubletracecouplings} lead to the new double-trace type vertices which have non-zero beta functions (computed at one loop in \cite{Sieg:2016vap}) already in the planar limit \cite{Fokken:2013aea}.  Hence the theory is not conformal even in the planar limit. On the other hand, since the coupling \(\xi\) is not running in the planar limit, most of the correlators of the theory will have a conformal form in this case. The only correlators which have a non-conformal behaviour are those which  contain  the operators of length two, such as  \((\Tr\phi^i\phi^\dagger_j)^2\) or \((\Tr\phi^i\phi^j)(\Tr \phi_k^\dagger\phi_l^\dagger)\), or similar    intermediate states of length two.

In the next sections, we will consider various examples of conformal operators, leaving so far aside the discussion of length two states breaking the conformality of certain correlators.

\subsection{Planar diagrams for correlation functions of  $\chi\mathrm{FT}_4$}

Let us discuss possible local operators and  the Feynman graphs for various correlation functions of the bi-scalar theory     \eqref{bi-scalarL}. Later we will comment on diagrammatics for the more general action \eqref{chiFT4int}.  Let us stress that, unless it is specified otherwise, we consider here only the  planar graphs for the leading large \(N_c\) order for each quantity. Planar diagrammatics  for \(\chi\)FTs is particularly simple and nice. Moreover, we do not expect the integrability to survive at finite \(N_c\). 

A general local operator is a linear combination of single-trace monomial operators of the type \begin{equation}
\cO_{\chi_1\chi_2\dots\chi_L}(x)= \Tr[\chi_{1}(x)\chi_{2}(x)\dots\chi_{L}(x)]\,,\quad \text{where}\quad \chi_j\in\{\phi^1,\phi_1^{\dagger},\phi^2,\phi_2^{\dagger}\} \, .
\label{general_operators}\end{equation}

The simplest quantity to compute is a two-point correlator of two such operators in different spacetime points
\begin{equation}
K_{\{\chi\};\{\chi'\}}(x^2)=\left<\cO_{\chi_1\chi_2,\dots,\chi_L}(x)\,\cO_{\chi'_1\chi'_2,\dots,\chi'_L}(0)\right>\,.
\end{equation}

\begin{figure}[t]
  \begin{center}    
  \includegraphics[height=5cm]{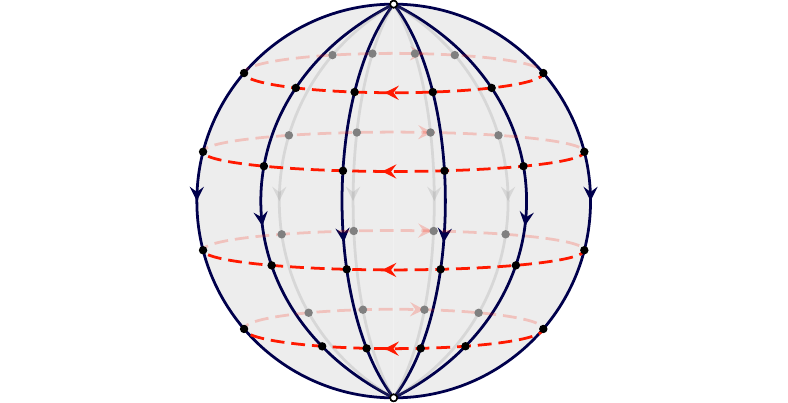}
\end{center}
\caption{Loop corrections to the two-point function of the BMN vacuum
  operator in the bi-scalar theory. It is clearly seen that in the bulk
  these graphs have the regular ``fishnet'' structure.
    }
\label{twopointfunction}
\end{figure}

\begin{figure}[t]
  \begin{center}    
  \includegraphics[height=5cm]{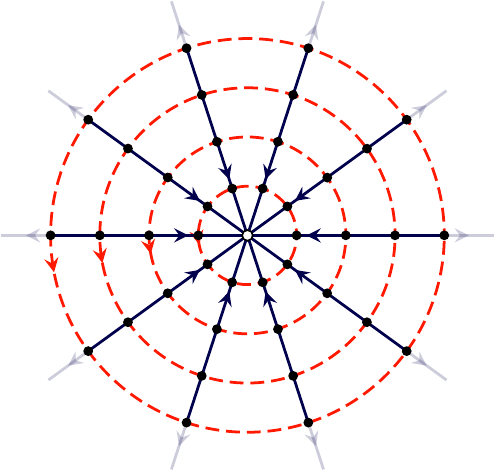}
\end{center}
\caption{Amputated graphs for pair correlators of BMN-vacuum -- wheel graphs.} 
\label{doublewheel}
\end{figure}

\paragraph{``Globe'' and ``wheel'' graphs for BMN vacuum} Let us  describe  the  planar diagrams for such correlators in the  bi-scalar \(\chi\)FT\(_4\).
The  simplest operator is  \begin{equation}
\cO_L(x)= \Tr[(\phi^1)^L]\,.
\end{equation}  It is usually  called BMN vacuum since  in the  undeformed $\mathcal{N}=4$ SYM its bare dimension is protected by supersymmetry and we will continue employing this name. The  corresponding two-point correlator  is conformal,
\begin{equation}
K_L(x^2)=\left<\cO_L(x)\cO_L(0)\right>=\frac{C}{(x^{2})^{\Delta(\xi)}}\,.
\end{equation}
  Unlike in the undeformed case this operator is not protected in the bi-scalar theory: its planar graphs  are nontrivial due to wrapping effects. They are represented on figure \ref{twopointfunction}.
   It is obvious that due to the fixed chirality of the only vertex of this theory no other graphs are possible~\cite{Gurdogan:2015csr}.
If we are not interested in the normalization constant \(C\) and wish to compute only the anomalous dimension we can amputate the propagators converging at the north pole (or, alternatively, at the south pole) on figure \ref{twopointfunction}, thus reducing the computation to the summation of  ``wheel'' graphs of figure \ref{doublewheel}. The integrability approach in Y-system/TBA form allowed to compute this anomalous dimension up to two wrappings (two ``frames'' on the wheel), i.e. up to \(\xi^{4L}\) order \cite{Ahn2011} in terms of infinite sums and double integrals. This result was brought in \cite{Gurdogan:2015csr} to  the form of finite sums of multi-\(\zeta\) numbers. It should be possible to construct an algorithmic expansion in higher wrapping  once the double scaling limit of twisted quantum spectral curve (QSC) \cite{Gromov2014a,Gromov:2014caa,Kazakov:2015efa} will be understood. 

As it was noticed  in  \cite{Gurdogan:2015csr}, the bulk structure of a sufficiently large globe or fishnet graph is represented by the regular square lattice.  The ``defects'' for the globe graph appear only at the north and south poles where the ``curvature'' defects are inserted into the regular ``flat'' lattice. The 4D massless Feynman graphs of the shape of regular square lattice (``fishnet'' graphs) were long ago considered by A.Zamolodchikov \cite{Zamolodchikov:1980mb} who showed that, by virtue of star-triangle relations (a version of Yang-Baxter relations) they define an integrable lattice quantum spin model with the 
$\mathfrak{su}(2,2)$ conformal symmetry. So the anomalous dimensions are in principle computable as well by the conformal spin chain approach of 
\cite{Chicherin2013,Derkachov1999,Derkachov2003b,Derkachov2006a,Derkachov2006b,Derkachov2009,Derkachov2011}.\footnote{  N.Gromov, V.Kazakov, G.Korchemsky, S.Negro,  work in progress. }

\begin{figure}[t]
   \begin{center}
     \includegraphics[height=4cm]{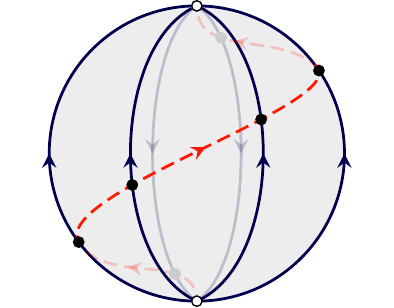}
     \hspace{2cm}
     \includegraphics[height=4cm]{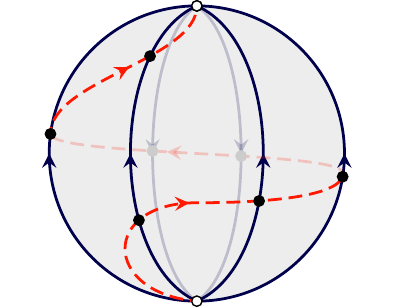}
\end{center}
\caption{One-magnon spiral graphs, without wrappings (on the left) and with a wrapping (on the right). The unwrapped configuration can be studied by ABA equations in the DS limit, whereas the wrapped one needs more sophisticated integrability tools (L\"uscher corrections or QSC).}
\label{one_m_spiral}
\end{figure}

\begin{figure}[t]
\begin{center}
  \includegraphics[height=4cm]{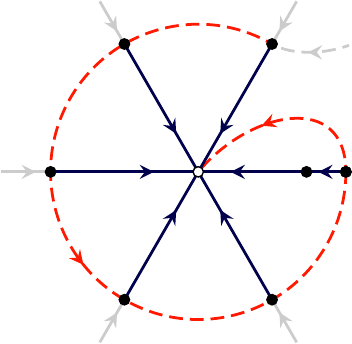}
  \hspace{2cm}
  \includegraphics[height=4cm]{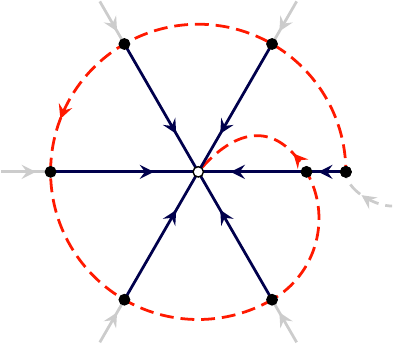}
\end{center}
\caption{Amputated one magnon  graphs: an unwrapped graph where each of the radial propagators of the \(\phi^1\) fields (shown in solid blue) are crossed by the propagators of \(\phi^2\) fields (shown in dashed red) at most once (left figure), and a wrapped one-magnon graph where there is a \(\phi^1\) propagator crossed twice by the same \(\phi^2\) field (right figure).    }
 \label{two_m_wrap}
\end{figure}

\begin{figure}[t]
\begin{center}
  \includegraphics[height=4cm]{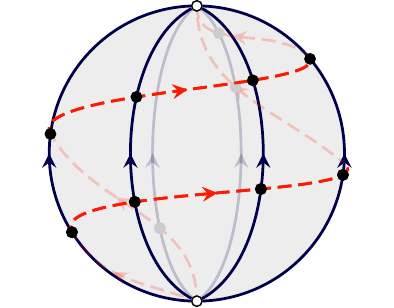}
  \hspace{2cm}
  \includegraphics[height=4cm]{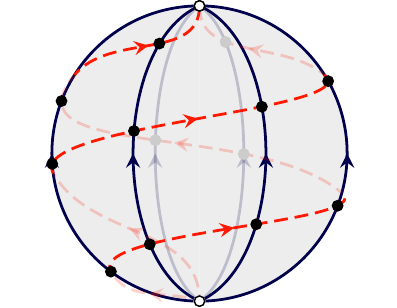}
\end{center}
\caption{Two-magnon spiral graphs, without wrappings (on the left) and with a wrapping (on the right).}
\label{two_m_spiral}
\end{figure}

\begin{figure}[t]
\begin{center}
  \includegraphics[height=4cm]{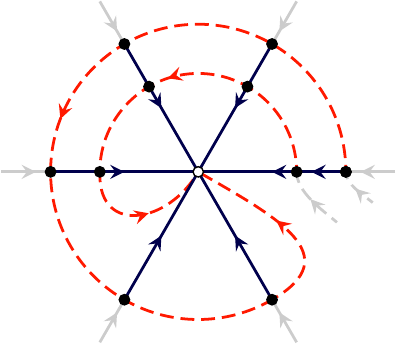}
\end{center}
\caption{Feynman graphs for two-magnon amputated graphs.}
\label{2magnon_wrap}
\end{figure}

\paragraph{Multi-magnon operators and spider-web graphs} Let us now describe the diagrammatics of more complex, multi-magnon operators of the type 
\begin{equation}\label{multi-magnons}
\cO_{L_1,L_2,\dots}(x)
= \Tr[(\phi^1)^{L_1}\,\phi^2\,(\phi^1)^{L_2}\,\phi^2\,(\phi^1)^{L_3}\,\phi^2\dots]\,,\qquad\qquad m+ \sum_{j=1}^m L_j=L\,,
\end{equation} where \(m\) is the number of magnons (insertions of \(\phi^2\) field). Notice that we take here only "chiral" operators, without mixing \(\phi^j\) with \(\phi_j^\dagger\) fields (this case will be discussed later).  
 The only possible Feynman graphs for two point functions of such operators for one magnon at a given loop   order are depicted on figure \ref{one_m_spiral} for two cases: an unwrapped and a wrapped magnon (dashed) lines.    For the purposes of easier computation of anomalous dimensions, we can again amputate the propagators converging at one of the operators.   For one magnon they are depicted on figure \ref{two_m_wrap}. For two magnons the corresponding wrapped and unwrapped graphs are presented on  figure \ref{two_m_spiral} and their amputated version -- on figure \ref{2magnon_wrap}. A natural name for the  amputated spiral graphs  is  ``spider-web" graphs. Amputation  appears to be not an innocent operation due to possible infrared divergences, especially for the multi-magnon operators, so  in the section~\ref{graphsABA} we prefer to do the direct computations with unwrapped graphs.

It is easy to convince oneself that, for a high loop order, the Feynman graphs look in the bulk, apart from the boundary effects close to operators, as integrable fishnet graphs. But the integrability should also persist for these multi-magnon configurations, thus enlarging the collection of integrable Feynman graphs from simplest fishnet graphs of \cite{Zamolodchikov:1980mb}  typical for the BMN vacuum operators, to more involved multi-magnon graphs.
Notice that replacing, say, \(\phi^2\) by \(\phi_2^\dagger\) does not change the magnon graph picture except changing the whole orientation of the planar graphs (one simply turns the sphere inside out). 

Let us note as an example, that for the abovementioned  particular case of the \(\beta\)-deformed \(\chi\)FT\(_4\) model \eqref{chiFT4beta}  we can extract one of the two anomalous dimensions for the  shortest two-magnon, scalar Konishi-like operator \({\cal O}_{K}=c_1\Tr(\phi^1\phi^2)^2+c_2\Tr(\phi^1\phi^1\phi^2\phi^2)\)  up to 4 loops (one wrapping) from  the results of the paper \cite{Ahn:2010yv} and even compare it to the direct perturbative expansion of \cite{Fiamberti2008a} given there. We will see later in section \ref{DSABA}, that our double scaled spectral equations correctly reproduces this result when the double scaling limit of these results is taken.

\paragraph{Comments on non-chiral operators and fermions}

More general scalar operators \eqref{general_operators}
of the bi-scalar model mixing the original and conjugated scalars
have even more complicated
graphs. An even more involved picture emerges in the full doubly scaled model \eqref{chiFT4}-\eqref{chiFT4int} if one mixes the scalar and fermionic fields within the same operator. We do not consider these operators in detail and 
limit ourselves only to a couple of comments:
\begin{itemize}
\item 
The operators of the form \begin{equation}\label{true_vacuum}
\cO_{L_1,L_2,\dots}(x)= \Tr[(\phi^1)^{L_1}\,\phi_1^\dagger\,(\phi^1)^{L_2}\,\phi_1^\dagger\,(\phi^1)^{L_3}\,\phi_1^\dagger\dots],\qquad\qquad m+ \sum_{j=1}^m L_j=L\,,\,
\end{equation}
  or the similar ones, when replacing all \(\phi^1,\phi_1^\dagger\to\phi^2,\phi_2^\dagger\), represent a ``true vacuum'': their anomalous dimension vanishes in all orders of perturbation theory, so that \(\Delta=L\). It is easy to convince oneself that no Feynman graphs correcting their tree order exist in the bi-scalar model.\footnote{Notice that we should exclude disconnected graphs, with the propagators  that connect \(\phi^1\) and \(\phi_1^\dagger\) fields of the same operator. }
\item    The situation becomes more involved when we include inside the same operators three types of fields, say \(\phi^1,\phi^2\) and \(\phi_1^\dagger\), or even all four fields  \(\phi^1,\phi^2,\phi_1^\dagger,\phi_2^\dagger\). The wrapped graphs seem then to be impossible, the loop expansion terminates at a finite order and  it should be in principle possible to study the spectra of such operators entirely within the ABA.

\end{itemize}

\section{$\chi\mathrm{FT}_3$ as DS limit of ABJM and its planar diagrammatics } \label{chift3diagrams}
We now explore a very analogous double scaling limit for the twisted ABJM theory. We first need to determine the form of the $\gamma$-deformed Lagrangian which to our knowledge has not been written explicitly in the literature. It can be obtained from the undeformed one by the standard procedure of replacing the ordered matrix product of $n$    fields  
\begin{equation}
A_1 * A_2 * \dots * A_n \equiv e^{-\frac{i}{2} \sum_{m>n }\epsilon_{ijk}  \gamma_i\, q_{A_m}^j q_{A_n}^k} A_1 A_2\dots A_n\,.
\end{equation}
Note that changing the order of fields will change the signs of corresponding terms in the exponent.
The  three $U(1)_i$ Cartan charges $q^{i}_{A}$ of the original $\mathfrak{su}(4)$ $R-$symmetry  associated to each field $A_i$ entering the Lagrangian are given in the table \ref{ABJMcharges}.

The resulting three-dimensional Lagrangian in our conventions reads

\begin{equation}
\begin{aligned}[y]
\mathcal{L}= N_c \,\Tr\biggl[
&\frac{1}{4\pi \lambda}\,\epsilon^{\mu \nu \lambda}\,\left(A_{\mu} \partial_{\nu} A_{\lambda} + \frac{2 i \sqrt{\lambda}}{3} A_{\mu} A_{\nu} A_{\lambda}\right) -\frac{1}{4\pi \lambda} \,\epsilon^{\mu \nu \lambda}\, \left(\hat{A}_{\mu} \partial_{\nu} \hat{A}_{\lambda} + \frac{2 i \sqrt{\lambda}}{3} \hat{A}_{\mu} \hat{A}_{\nu} \hat{A}_{\lambda}\right) -\\
& D_{\mu} Y^{\dagger}_{A}D^{\mu} Y^{A} + i\Psi^{ \dagger A} \slashed{D}\Psi_{A}\biggr] +  \mathcal{L}_{\mathrm{scalar}} + \mathcal{L}_{\mathrm{ferm}}
\end{aligned}
\end{equation}
with the  interacting part for scalars given by

\begin{equation} \label{twistedABJMlag}
\begin{aligned}
\mathcal{L}_{\mathrm{scalar}}={}N_c\frac{ (2 \pi \lambda)^2}{3}\, \Tr \biggl[ &
Y^{A} Y_{A}^{\dagger} Y^{B} Y_{B}^{\dagger}Y^{C} Y_{C}^{\dagger}+Y^{A} Y_{B}^{\dagger} Y^{B} Y_{C}^{\dagger}Y^{C} Y_{A}^{\dagger} \\
-\,\,& 6\, Y^{A} Y_{A}^{\dagger} Y^{B} Y_{C}^{\dagger}Y^{C} Y_{B}^{\dagger}-4\left(3-2 \delta_{AB}\right) Y^{A} Y_{A}^{\dagger} Y^{B} Y_{A}^{\dagger}Y^{A} Y_{B}^{\dagger}\\
-\sum_{A\neq B\neq C=1}^3  \biggl( e^{i\, \epsilon_{ABC}( \gamma_A+ \gamma_B- \gamma_C)}\, & Y^A  Y_{B}^{\dagger} Y^{4} Y_{A}^{\dagger}Y^{B} Y_{4}^{\dagger}+\frac{1}{3}e^{i\, \epsilon_{ABC}( \gamma_A+ \gamma_B+ \gamma_C)} Y^A  Y_{B}^{\dagger} Y^{C} Y_{A}^{\dagger}Y^{B} Y_{C}^{\dagger}\biggr)\biggr] 
\end{aligned} 
\end{equation}
where $\lambda = \frac{N_c}{k}$ is the 't Hooft coupling. The fermionic part $\mathcal{L}_{\text{ferm}}$ turns out to be rather lengthy and is left to the Appendix \ref{twstABJM}.
\begin{table}[t]
\centering
\begin{tabular}{@{}ccccccccc@{}}
\toprule 
$f$     & $Y^1$         & $Y^2$          & $Y^3$         & $Y^4$          & $\Psi^{ \dagger 2}$ & $\Psi^{ \dagger 2}$ & $\Psi^{ \dagger 3}$ & $\Psi^{ \dagger 4}$   \\ \midrule
$q^1_f$ & $-\frac{1}{2}$ & $\frac{1}{2}$ & $\frac{1}{2}$           & $-\frac{1}{2}$            & $-\frac{1}{2}$      & $\frac{1}{2}$     & $\frac{1}{2}$                & $-\frac{1}{2}$                \smallskip \\
$q^2_f$ & $\frac{1}{2}$           & $-\frac{1}{2}$            & $\frac{1}{2}$ & $-\frac{1}{2}$ & $\frac{1}{2}$                & $-\frac{1}{2}$                & $\frac{1}{2}$      & $-\frac{1}{2}$  \smallskip   \\
$q^3_f$ & $\frac{1}{2}$ & $\frac{1}{2}$  & $-\frac{1}{2}$ & $-\frac{1}{2}$  & $\frac{1}{2}$     & $\frac{1}{2}$     & $-\frac{1}{2}$     & $-\frac{1}{2}$    \smallskip \\ \bottomrule
\end{tabular}
\caption{ Cartan $U(1)$ charges of the $\mathfrak{su}(4)$ $R-$symmetry associated to the fields in the \(\gamma\)-deformed  ABJM model. The corresponding conjugate fields have the reversed charges.}
\label{ABJMcharges}
\end{table}

There is a plethora of models that one can generate by playing with different scalings of the parameters in the original twisted Lagrangian together with the coupling constant. We first consider the most general limit where we scale to infinity a single parameter while keeping the others fixed as follows
\begin{equation}
\begin{aligned}
&q_{3} \equiv e^{-i \gamma_{3}} \rightarrow \infty\,\,\,\text{and } \lambda \rightarrow 0 \\
&\xi_{i} \equiv e^{-i\gamma_{i}} \,\,\,\text{for }i=1,2\,\,\,\, \text{and}\,\,\,\, \xi_3 = q_3 \lambda^2\,\,\, \text{fixed}\,.
\end{aligned}
\end{equation} 
With this scaling we end up with a Lagrangian without gauge fields as they decouple in the weak coupling limit. It reads now
\begin{equation}\label{3paramDS_ABJM}
\mathcal{L} = N_c \,\Tr\biggl[- \sum_{i=1}^{4} \left(\partial_{\mu} Y^{\dagger}_{i}\partial^{\mu} Y^{i}  + i\Psi^{ \dagger i} \slashed{\partial}\Psi_{i}\right) \biggr] +  \mathcal{L}_{\mathrm{scalar}}+  \mathcal{L}_{\mathrm{ferm}}
\end{equation}
where the interaction term involving only scalars is given by
\begin{equation}
\begin{aligned}
\mathcal{L}_{\mathrm{scalar}} ={} (4\pi)^2 \, \Tr \Biggl[ \, &\frac{\xi _2 \xi_3 }{\xi_1} Y^2 Y^{\dagger}_4 Y^3 Y^{\dagger}_2 Y^4 Y^{\dagger}_3+\frac{ \xi _1 \xi_3 }{\xi_2} Y^1 Y^{\dagger}_3 Y^4 Y^{\dagger}_1 Y^3 Y^{\dagger}_4\\
 +&\frac{\xi_3}{  \xi _1 \xi _2  }Y^1 Y^{\dagger}_2 Y^4 Y^{\dagger}_1 Y^2 Y^{\dagger}_4+ \xi _1 \xi _2  \xi_3  Y^1 Y^{\dagger}_3 Y^2 Y^{\dagger}_1 Y^3 Y^{\dagger}_2 \Biggr]
\end{aligned}
\end{equation}
and the mixed fermionic and scalar interacting part is given by
\begin{equation}
\begin{aligned}
\mathcal{L}_{\mathrm{ferm}} ={}&  \frac{i}{4\pi \sqrt{\xi_3}}\,\Tr \Biggl[ {\frac{1}{\sqrt{\xi _1}}} Y^4\Psi^{\dagger 2}\Psi _4Y^{\dagger}_2-\frac{1}{\sqrt{\xi _2}} Y^4Y^{\dagger}_1\Psi _4\Psi^{\dagger 1}+\sqrt{\xi _1 } Y^3\Psi^{\dagger 1}\Psi _3Y^{\dagger}_1-\sqrt{\xi _2 } Y^3Y^{\dagger}_2\Psi _3\Psi^{\dagger 2} \\
&-\frac{1}{\sqrt{\xi _1}} Y^2Y^{\dagger}_4\Psi _2\Psi^{\dagger 4}+ \sqrt{\xi _2 } Y^2\Psi^{\dagger 3}\Psi _2Y^{\dagger}_3+\frac{1}{\sqrt{\xi _2}} Y^1\Psi^{\dagger 4}\Psi _1Y^{\dagger}_4-\sqrt{\xi _1 } Y^1Y^{\dagger}_3\Psi _1\Psi^{\dagger 3}\\
&-Y^3\Psi^{\dagger 1}Y^4\Psi^{\dagger 2}+Y^1\Psi^{\dagger 4}Y^2\Psi^{\dagger 3}+\Psi _1Y^{\dagger}_4\Psi _2Y^{\dagger}_3- \Psi _3Y^{\dagger}_1\Psi _4Y^{\dagger}_2 \Biggr]\, .
\end{aligned}
\end{equation}
We can generate simpler models by scaling more parameters. For instance, by considering the following limit
\begin{equation}
\begin{aligned}
&q_{i} \equiv e^{-i \gamma_{i}} \rightarrow \infty\,\,\,\mathrm{for}\,\,\, i=2,3\,,\,\,\,\text{and } \lambda \rightarrow 0 \\
&\xi_{1} \equiv e^{-i\gamma_{1}} \,\,\,\, \text{and}\,\,\,\, \xi_i = \frac{q_i}{\lambda}\,\,\, \text{fixed}\,\,\, \mathrm{for}\,\,\, i=2,3\,,
\end{aligned}
\end{equation} 
we generate a simpler Lagrangian
\begin{equation}
\begin{aligned}
\mathcal{L} = N_c \,\Tr\biggl[- \sum_{i=1}^{4} \partial_{\mu} Y^{\dagger}_{i}\partial^{\mu} Y^{i}  -\sum_{i=2}^{3} i\Psi^{ \dagger i} \slashed{\partial}\Psi_{i} \biggr] +  \mathcal{L}_{\mathrm{int}}
\end{aligned}
\end{equation}
with the interacting terms  given by
\begin{equation}
\mathcal{L}_{\mathrm{int}} = \frac{(4\pi)^2}{\xi _1 \xi _2 \xi _3} \left(\xi _1^2\, Y^2Y^{\dagger}_3Y^4 Y^{\dagger}_2 Y^3 Y^{\dagger}_4+Y^1Y^{\dagger}_2Y^3Y^{\dagger}_1Y^2Y^{\dagger}_3\right) -\frac{4 \pi i}{\sqrt{\xi _2 \xi _3}}  \left(Y^2 Y^{\dagger}_3 \Psi _2 \Psi^{\dagger 3}-Y^3 \Psi^{\dagger 2} \Psi _3 Y^{\dagger}_2\right)\,.
\end{equation}
This is the simplest Lagrangian we can generate containing both scalars and fermions.
Finally, of particular interest is the case where we take all the three \(\gamma\)-parameters to be imaginary and large with the following scaling with $\lambda$
\begin{equation}
\begin{aligned}
& q_i  \equiv e^{-i \gamma_i}\rightarrow \infty\,\,\,\text{for }i=1,2\,,\,\,\, q_3  \equiv e^{-i \gamma_3}\rightarrow 0\,\,\,\, \text{and}\,\,\,\, \lambda \rightarrow 0\,, \\
& \xi_i \equiv q_i \, \lambda^{2/3}\,\,\,\text{for }i=1,2\,,\,\,\,  \xi_3 \equiv \frac{q_3}{ \lambda^{2/3}} \,\,\,\,\, \text{fixed}\,.
\end{aligned}
\end{equation}
The resulting Lagrangian is the simplest one can get and in particular both the gauge fields and fermions decouple as a result of the weak coupling limit. We end up with a single sextic scalar interacting term from the action, namely
\begin{equation}
\mathcal{L} = N_c \,\Tr\biggl[- \partial_{\mu} Y^{\dagger}_{1}\partial^{\mu} Y^{1} - \partial_{\mu} Y^{\dagger}_{2}\partial^{\mu} Y^{2}- \partial_{\mu} Y^{\dagger}_{4}\partial^{\mu} Y^{4}\biggr]+ \mathcal{L}_{\text{int}}
\end{equation}
where the interaction term in the Lagrangian is given by
\begin{equation}
\begin{aligned} \label{finalabjm}
\mathcal{L}_{\text{int}} =\, & (4 \pi)^2 \, \xi \, Y^1 Y_4^{\dagger }Y^2 Y_1^{\dagger }Y^4 Y_2^{\dagger } \,,
\end{aligned}
\end{equation}
and we have defined $\xi \equiv \xi_3/( \xi_1 \, \xi_2) $\,. The resulting theory is again chiral and the complex conjugate term can be obtained by taking instead $\gamma_1,\gamma_2 \rightarrow -i\infty$ and $\gamma_3 \rightarrow +i\infty$.  We will study the spectrum of this model using the asymptotic Bethe ansatz in section \ref{ABAABJM}.

Another particularly interesting limit of the $\gamma$-deformed Lagrangian where some of the supersymmetry is preserved can be obtained by setting $\gamma_1 = \gamma_2 =0$ while keeping the remaining $\gamma_3$ parameter to be finite. This is the well known $\beta$-deformation\cite{Imeroni:2008cr,He:2013hxd} and the resulting theory is $\mathcal{N}=2$ supersymmetric.
We can then consider a double scaling limit in this $\beta$-deformed Lagrangian as follows
\begin{equation} \label{betaABJM}
\begin{aligned}
& q_3  \equiv e^{-i \gamma_3}\rightarrow \infty\,,\,\,\,\, \lambda \rightarrow 0\,, \\
& \xi_3 \equiv q_3\,\lambda^2 \,\,\,\,\, \text{fixed}\,.
\end{aligned}
\end{equation}
Unlike the $\mathcal{N}=4$ SYM case, the resulting Lagrangian keeps all fermions and scalars as in the original theory while the gauge fields decouple. Its explicit form is presented in Appendix \ref{ABJMbetalagrange}.
\subsection{Planar diagrams for correlation function of $\chi\mathrm{FT}_3$ }
\begin{figure}
  \centering
  \includegraphics{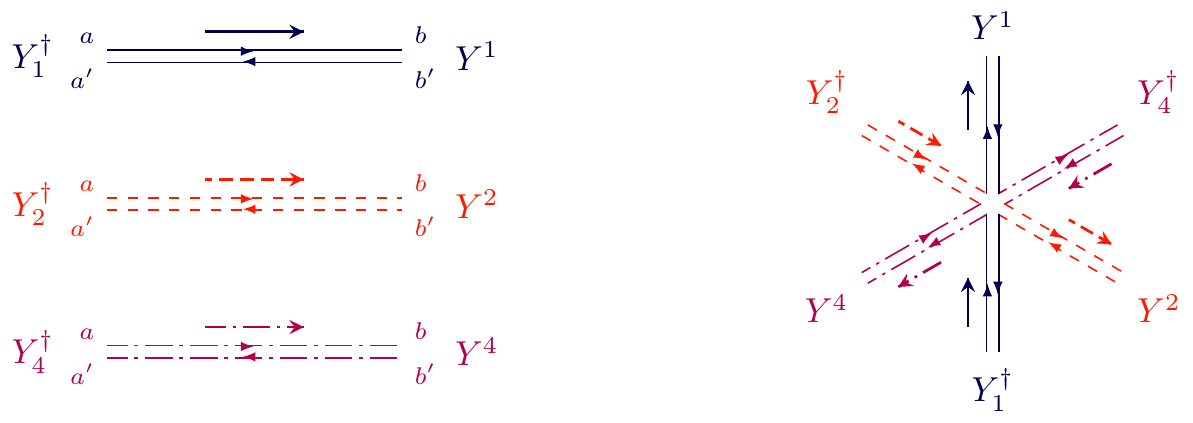}
  \caption{Feynman rules for $\chi\mathrm{FT}_3$ in double-line notation. The indices $a$, $a'$ and $b$, $b'$ belong to the bi-fundamental representation of the $SU(N_c)\times SU(N_c)$ gauge group. The arrows besides the double lines represent the flow of flavour.}
  \label{fig:chift3feynrules}
\end{figure}

\begin{figure}[t]
\begin{center}
  \includegraphics[height=6cm]{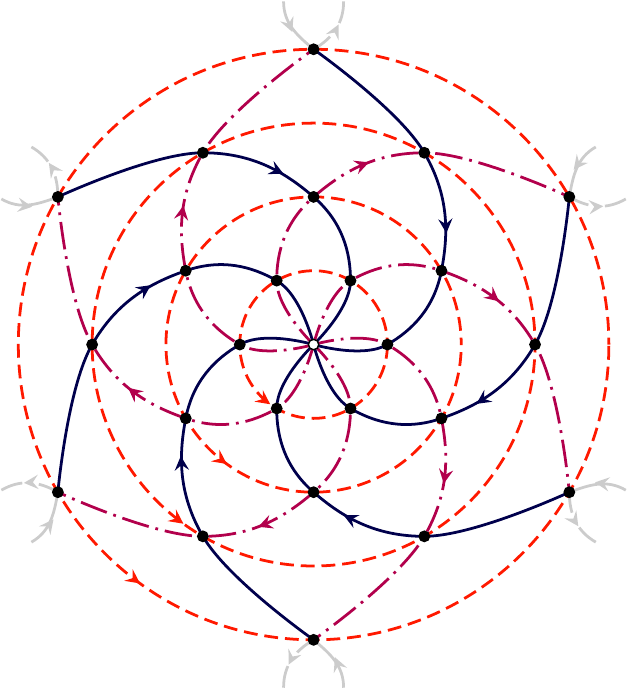}
\end{center}
\caption{The three-dimensional analogue of wheel graphs. These graphs
  are the only ones that renormalise the BMN vacuum in
  \(\chi\)FT\(_3\).}
\label{triangularwheel}
\end{figure}

We  now  describe the Feynman graphs contributing to the perturbative computations of the two point functions in the scalar model with the Lagrangian (\ref{finalabjm}) which turns out to be rather simple. The Feynman rules are presented in figure \ref{fig:chift3feynrules} and they simply contain three types of scalar propagators and a single interacting vertex.
As in the $\chi\mathrm{FT}_4$ model, we will be considering local operators made out of single-traces of the type
\begin{equation}
\mathcal{O}_{\chi_1 \chi_2 \dots \chi_L}(x) = \Tr [ \chi_1(x) \chi_2(x) \dots \chi_L (x) ]\,, \;\;\;\; \text{where}\;\;\; \chi_{j} \in \{ Y^{1}, Y_{1}^{\dag}, Y^{2}, Y_{2}^{\dag},Y^{4}, Y_{4}^{\dag}\}
\end{equation}
and consider two-point correlation functions of these operators.
\paragraph{Vacuum two point functions and wheels} The simplest state we are going to consider is the standard vacuum of the undeformed theory, 
\begin{equation}
\mathcal{O}(x) =   \Tr \left[ (Y^1 Y_4^{\dagger})^{L} \right]\,.
\end{equation}
The story here is similar to the previous findings of the 4D bi-scalar model.
Although this state is protected in the undeformed theory, it again receives quantum corrections in this model from wrapping graphs. They are simply the wheel diagrams depicted in figure \ref{triangularwheel}, which are obtained after amputation of the north pole (or equivalently the south pole) of the globe. 

\begin{figure}[t]
    \centering
    \begin{subfigure}{0.5\textwidth}
        \centering
        \includegraphics[height=4.5cm]{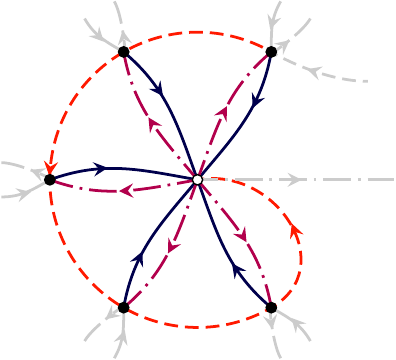}
        \caption{A one-magnon graph without wrapping}
        \label{xiftthreewrappingnot}
    \end{subfigure}%
    \begin{subfigure}{0.5\textwidth}
        \centering
        \includegraphics[height=4.5cm]{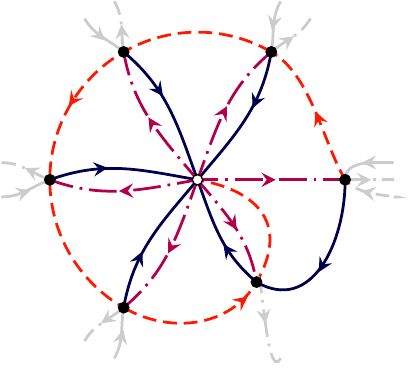}
        \caption{A one-magnon graph with wrapping}
                \label{xiftthreewrapping}
      \end{subfigure}\\
          \begin{subfigure}{0.5\textwidth}
        \centering
        \includegraphics[height=4.5cm]{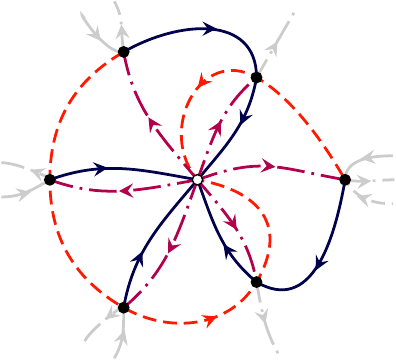}
        \caption{A two-magnon graph without wrapping}
        \label{xiftthreetwomagwrappingnot}
    \end{subfigure}%
    \begin{subfigure}{0.5\textwidth}
        \centering
        \includegraphics[height=4.5cm]{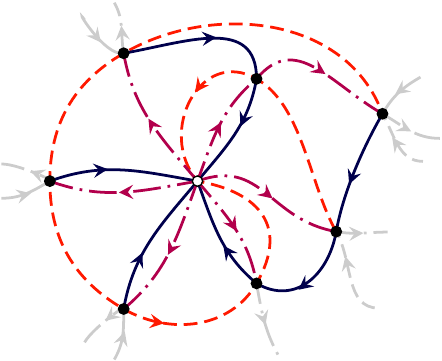}
        \caption{A two-magnon graph with wrapping}
        \label{xiftthreetwomagwrapping}
    \end{subfigure}
    \caption{Types of graphs for the two-point functions of single-magnon and two-magnon operators in $\chi$FT$_3$.}
\end{figure}

For a globe with a sufficiently large number of frames (red lines corresponding to \(Y^2\)
propagators in figure \ref{triangularwheel}), a regular triangular lattice structure emerges in the bulk. If we ignore the lattice  ``curvature defects'' in the north and south poles of the globe due to the insertion of the local operators, we can regard this as a   ``flat'', regular triangular lattice. In the same work~\cite{Zamolodchikov:1980mb}   of Zamolodchikov  such a regular triangular flat lattice formed by massless scalar propagators was also shown to form an integrable statistical-mechanical system, whose free energy per spin was computed there in the limit of an infinite lattice. 
\paragraph{Multi-magnon states and 3D ladders}
Let us now consider certain excited states in this model. We take them to be made out of linear combinations of single-traces of the form
\begin{equation}
\mathcal{O}(x) =   \Tr \left[ (Y^1 Y_4^{\dagger})^{L-N}(Y^2 Y_4^{\dagger})^{N} \right]\,,
\end{equation}
which falls into the so-called $\mathfrak{su}(2)$ subsector of the undeformed theory. For a single magnon ($N=1$), the typical $l$-loop graph in an asymptotically long spin chain ($L>l$) is represented in figure \ref{xiftthreewrappingnot} and consists of  ladder-type graphs. This class of 3D ladders was already considered in~\cite{Bak:2009tq} and the results for the (amputated) graphs were computed there at arbitrary loop order. When we include wrapping we get again the spiral graphs depicted in figure \ref{xiftthreewrapping}.

For two magnons ($N=2$), the resulting graphs before wrapping are represented in figure~\ref{xiftthreetwomagwrappingnot} and at wrapping order in figure \ref{xiftthreetwomagwrapping}. As we will see, integrability can be used to provide information about this class of graphs in three dimensions. 

\section{Asymptotic Bethe ansatz for the spectrum of  $\chi\mathrm{FT}_4$ and $\chi\mathrm{FT}_3$} \label{DSABA}
In this section, we derive equations describing the spectrum of asymptotically
long operators in the $\chi\mathrm{FT}_4$ and $\chi\mathrm{FT}_3$ from the double scaling limit of the twisted asymptotic Bethe ansatz equations of the corresponding ``mother'' theories:  $\gamma$-deformed $\mathcal{N}=4$ SYM and ABJM models, respectively.  

\subsection{Spectrum of $\chi\mathrm{FT}_4$}

\subsubsection{Dispersion relation}
In the $\gamma$-deformed $\mathcal{N}=4$ SYM the form of the dispersion relation
in the bulk of effective asymptotically long spin chain is not modified with respect to the original theory \cite{Beisert:2005if}. The dependence of the anomalous dimensions on the deformation parameters comes solely from the twisted periodic boundary
conditions, through the solutions of the twisted Bethe equations.
As a consequence, in order to determine the effective dispersion relation for any of the double scaled models we first need to consider the one-magnon solution of the corresponding Bethe equations, plug it in the original dispersion relation and then take there the double scaling limit.

Let us consider the simplest single-magnon operator in the bi-scalar model (\ref{bi-scalarL}) with one excitation $\phi^2$ on top of the vacuum made out of a sea of $L-1$ scalars  $\phi^1$.  At the level of the twisted asymptotic Bethe ansatz equations for the $\gamma$-deformed $\mathcal{N}=4$ SYM written in Appendix \ref{appbetheSYM}, such an operator is achieved by setting  $K_4=1$, $K_i=0$ for all other $i$ and take the charges $J_1=L-1,\, J_2=1$ and $J_3=0$. The Bethe equations reduce to a single equation that is equivalent to the total momentum conservation condition and reads
\begin{equation}\label{periodicity}
q_3^{2L} e^{iL p} \,=1.
\end{equation} 
Using the parameterization of momentum \begin{equation}\label{momentumx+x-}
e^{ ip}= \frac{x_{4}^{+}}{x_{4}^{-}}
\end{equation}   in terms of  the Zhukovski variables  obeying the identity 
\begin{equation}
x^+ + \frac{1}{x^+}-x^{-} - \frac{1}{x^-}=\frac{i}{g}
\end{equation}
we can find the solutions for $x_{4}^+$ and $x_{4}^-$ in terms of $g$ and $q_3$. There will be two such solutions. Then we can plug them in the original dispersion relation expressing the anomalous dimension \(\gamma(g,q_3)=\Delta(g,q_3)-L\)
\begin{equation} \label{disp}
\gamma =2 i g \left( \frac{1}{x_4^+}-\frac{1}{x_4^-}\right)\,
\end{equation} 
through the momentum \(p\)  and take the double scaling limit. We notice that in the double scaling limit \(q_3\to\infty,\quad g\to 0,\quad \xi_3\equiv \xi=gq_3\)-fixed, the only way to satisfy equation \eqref{periodicity}, \(e^{-ip}=q_3^2\),  is to send  \(p\to i\infty\).

 Then in the exact  dispersion relation  
 \eqref{disp} 
 \begin{equation} \label{displim}
\gamma = -1\pm \sqrt{1+16 \,g^2\sin^2\frac{p}{2}}\simeq -1 \pm \sqrt{1-4 \,g^2q_3^2} \,\,
\end{equation} 
we have to drop one of the modes (left-moving for \(q_3\to\infty\) or right-moving for \(q_3\to 0\)), which means that we favor one chirality along the ``spin chain'' w.r.t. another one. We finally  get the effective double-scaled dispersion relation
\begin{equation} \label{finaldisp4}
\gamma = -1 \pm \sqrt{1-4 \,\xi_3^2}\,.
\end{equation}
We choose the solution with $+$ sign since it is the only one that has a sensitive expansion at weak coupling, namely it starts at order $\xi_3^2$ for small $\xi_3$.  


\subsubsection{Multi-magnon states}
We consider now operators with an arbitrary number of magnons over the vacuum in both models (\ref{chiFT4}) and (\ref{bi-scalarL}).  Starting again from the twisted asymptotic Bethe ansatz equations written in the Appendix \ref{appbetheSYM}, we consider for simplicity the $\mathfrak{su}(2|3)$ closed sector of the original theory.  It is well known that in $\mathcal{N}=4$ SYM there will be a mixing of scalars with the gluino $\psi_4$ already at two loops. In our double scaled model (\ref{chiFT4}) though, the gluino $\psi_4$ dropped out. 
Therefore, the field content is narrowed down to simply three chiral scalars. In particular, the global $AdS_5$ charges $S_1$ and $S_2$ are zero while the remaining $S^5$ charges satisfy $J_1=L-J_2-J_3>0$ and $J_2, J_3>0$. The length $L$ is conserved and the hypercharge $B$ is also zero.
 We refer to such sector made out of operators of the form $\Tr\left((\phi^1)^{J_1} (\phi^2)^{J_2} (\phi^3)^{J_3} \right)$ as the broken $\mathfrak{su}(3)$ sector even though remnant symmetry is simply $\mathfrak{u}(1)^3$. 
In the bi-scalar model (\ref{bi-scalarL}), it boils down to a broken $\mathfrak{su}(2)$ sector where the operators are of the form $\Tr \left((\phi^1)^{J_1} (\phi^2)^{J_2}  \right)$.

The above constraints on the global charges of the operators greatly simplify the Bethe equations. In particular, the right wing of the corresponding Dynkin diagram trivializes (i.e., $K_5=K_6=K_7=0$) and the number of roots of the left wing is fixed to be $K_1=K_2=K_3=J_3$. In the middle node equation we have $K_4 = J_2+J_3$\,.
The set of ABA  equations we get in this sector is the following \cite{Beisert2005,Beisert:2006ez},
\begin{equation}
\begin{aligned}
q_1^{-J_2-J_3}q_2^{J_1+J_3}q_3^{-J_1+J_2}&=\prod_{i=1}^{J_2+J_3}\frac{1-\frac{1}{x^{+}_{4,i} x_{1,j}}}{1-\frac{1}{x^{-}_{4,i} x_{1,j}}}  \prod_{l=1}^{J_3}\frac{u_{1,j} -u_{2,l} +i/2}{u_{1,j} -u_{2,l} -i/2}
\\
1&= \underset{k\neq l}{\underset{k=1}{\prod^{J_3}}}\frac{u_{2,l} -u_{2,k} -i}{u_{2,l} -u_{2,k} +i}\prod_{j=1}^{J_3}\frac{u_{2,l} -u_{1,j} +i/2}{u_{2,l} -u_{1,j} -i/2}\prod_{j=1}^{J_3}\frac{u_{2,l} -u_{3,j} +i/2}{u_{2,l} -u_{3,j} -i/2}   
\\
q_1^{-J_2-J_3}q_2^{J_1-J_3}q_3^{-J_1-J_2} & =  \prod_{i=1}^{J_2+J_3}\frac{x^{+}_{4,i}-x_{3,j} }{x^{-}_{4,i}-x_{3,j}}\prod_{l=1}^{J_3}\frac{u_{3,j} -u_{2,l} +i/2}{u_{3,j} -u_{2,l} -i/2}\nonumber
\\
q_1^{2J_3}q_2^{2J_3}q_3^{2(J_1+J_2)} & = \left( \frac{x_{4,k}^{-}}{x_{4,k}^{+}}\right)^L\, \underset{i\neq l}{\underset{i=1}{\prod^{J_2+J_3}}}\frac{x^{+}_{4,k}-x^{-}_{4,i} }{x^{-}_{4,k}-x^{+}_{4,i} } \frac{1-\frac{1}{x^{+}_{4,k} x^{-}_{4,i}}}{1-\frac{1}{x^{-}_{4,k} x^{+}_{4,i}}}\, \sigma(x_{4,k},x_{4,i})^2 \\
&\quad \, \times \prod_{j=1}^{J_3}\frac{x^{-}_{4,k}-x_{3,j} }{x^{+}_{4,k}-x_{3,j} }\prod_{j=1}^{J_3}\frac{1-\frac{1}{x^{-}_{4,k} x_{1,j}}}{1-\frac{1}{x^{+}_{4,k} x_{1,j}}} \,,
\end{aligned}
\label{1357AhnConventions2}
\end{equation}
where the rapidities  \(u_{a,j}\) are related to the momenta of magnons \eqref{momentumx+x-}  through the Zhukovsky map  \(u=\frac{1}{g}(x+\frac{1}{x})\). We would like to have them written in terms of the double scaled parameters $\xi_j$. In order to achieve the scaled version of these ABA equations
we first observe that simply doing a series expansion of the Zhukovski variable for small coupling constant does not produce the correct powers of $g$ to form the desired combination  $\xi_i =g q_i$ with the twist parameters. For example take the third equation. The expansion  in $g$ of the first factor of the last line gives 
\begin{equation}
 \prod_{i=1}^{J_2+J_3}\frac{x^{+}_{4,i}-x_{3,j} }{x^{-}_{4,i}-x_{3,j}} \simeq  \prod_{i=1}^{J_2+J_3}\frac{u_{4,i}-u_{3,j}+i/2}{u_{4,i}-u_{3,j}-i/2}+\mathcal{O}(g)\,,
\end{equation}
which does not balance the large factors of $q_i$ in the left hand side of that equation. To overcome this apparent problem we have to look for the
solutions on other sheets of Zhukovsky variables. We note that by first performing crossing transformations \(x\to 1/x\) of some of the Zhukovski variables we can then generate extra powers of $g$ upon expanding the resulting equations for small $g$. With a judicious choice of the variables to cross, we can produce the precise powers of the coupling constant that recombine nicely with the twist parameters. Indeed, the following analytic continuations lead to the correct double scaled equations
\begin{equation} \label{crossvars}
x_{3} \rightarrow 1/x_3\quad \text{and}\quad x^{+}_4 \rightarrow 1/x_{4}^{+}\,,\,\, x^{-}_4 \rightarrow x_{4}^{-}\,
\end{equation}
keeping all other variables untouched.
For example, consider again the third equation above. Once we transform the variables according to (\ref{crossvars}), the first factor on the last line will become
\begin{equation}
 \prod_{i=1}^{J_2+J_3}\frac{1/x^{+}_{4,i}-1/x_{3,j} }{x^{-}_{4,i}-1/x_{3,j}} \simeq (g^2)^{J_2+J_3}\prod_{i=1}^{J_2+J_3}\frac{u_{3,j}-u_{4,i}-i/2 }{u_{3,j} \left(u_{4,i}^2+1/4 \right)}+\mathcal{O}(g^{2(J_2+J_3)+2})\,,
\end{equation}
and this power of $g$ of the leading term is precisely what we need to form the double scaled combination $\xi_j$ with the twist parameters in the left-hand side.

More generally, after performing the replacements (\ref{crossvars}) in the equations (\ref{1357AhnConventions2}) and series expanding them for small $g$, we get the properly doubly scaled spectral equations
\begin{equation}
\begin{aligned}
\xi_1^{-J_2-J_3}\xi_2^{J_1+J_3}\xi_3^{-J_1+J_2}&=\prod_{j=1}^{J_2+J_3}\frac{u_{1,i}-u_{4,j}-i/2}{u_{1,i}} \prod_{l=1}^{J_3}\frac{u_{1,j} -u_{2,l} +i/2}{u_{1,j} -u_{2,l} -i/2}
\\
1&= \underset{k\neq l}{\underset{k=1}{\prod^{J_3}}}\frac{u_{2,l} -u_{2,k} -i}{u_{2,l} -u_{2,k} +i}\prod_{j=1}^{J_3}\frac{u_{2,l} -u_{1,j} +i/2}{u_{2,l} -u_{1,j} -i/2}\prod_{j=1}^{J_3}\frac{u_{2,l} -u_{3,j} +i/2}{u_{2,l} -u_{3,j} -i/2}   
\\
\xi_1^{-J_2-J_3}\xi_2^{J_1-J_3}\xi_3^{-J_1-J_2} & =  \prod_{i=1}^{J_2+J_3}\frac{u_{3,j}-u_{4,i}-i/2 }{u_{3,j} \left(u_{4,i}^2+1/4 \right)}\prod_{l=1}^{J_3}\frac{u_{3,j} -u_{2,l} +i/2}{u_{3,j} -u_{2,l} -i/2}
\\
\xi_1^{2J_3} \xi_2^{2J_3} \xi_3^{2(L-J_3)} & = \left(u_{4,k}^2 +1/4\right)^L\, \underset{i\neq l}{\underset{i=1}{\prod^{J_2+J_3}}}\frac{u_{4,k} -u_{4,i}+i}{u_{4,k}-u_{4,i}-i} \, \sigma^{m, m}_{0}(u_{4,k},u_{4,i})^2 \\
&\quad \, \times \prod_{j=1}^{J_3}\frac{u_{3,j} \left(u_{4,k}^2+1/4 \right)}{u_{3,j}-u_{4,k} -i/2}\prod_{j=1}^{J_3}\frac{u_{1,j}}{u_{1,j} - u_{4,k} -i/2} \,.
\end{aligned}
\label{dsspectraleqs}
\end{equation}
In the middle node equation, $\sigma^{m, m}_{0}(u_{4,k},u_{4,i})$ stands for the leading order term of the dressing phase in the mirror-mirror kinematics. Its expression can be derived from \cite{Gromov:2009bc,Arutyunov:2009ur,Gromov2010d}.   We find it to be equal to
\begin{equation}\label{dressing-f}
\sigma^{m, m}_{0}(u,v)^2 =\frac{\left(4 v^2+1\right) \Gamma \left(i u+\frac{1}{2}\right) \Gamma \left(i u+\frac{3}{2}\right) \Gamma \left(\frac{1}{2}-i v\right) \Gamma \left(\frac{3}{2}-i v\right) \Gamma (-i u+i v+1)^2}{\left(4 u^2+1\right) \Gamma \left(\frac{1}{2}-i u\right) \Gamma \left(\frac{3}{2}-i u\right) \Gamma \left(i v+\frac{1}{2}\right) \Gamma \left(i v+\frac{3}{2}\right) \Gamma (i u-i v+1)^2} \,.
\end{equation}
In particular, in order to get the spectrum for chiral operators of the type \eqref{multi-magnons} of the bi-scalar model (\ref{bi-scalarL}) we simply set $J_3=0$ above and limit ourselves to the broken $\mathfrak{su}(2)$ sector. Then only the middle equation survives and it reads
\begin{equation} \label{brokensu2}
 \xi_3^{2L} = \left(u_{4,k}^2 +1/4\right)^L\, \underset{i\neq l}{\underset{i=1}{\prod^{J_2}}}\frac{u_{4,k} -u_{4,i}+i}{u_{4,k}-u_{4,i}-i} \, \sigma^{m, m}_{0}(u_{4,k},u_{4,i})^2\,. 
 \end{equation}
We can finally compute the energy of a state starting from the  dispersion relation  (\ref{disp}). After doing the crossing transformation (\ref{crossvars}) in the standard ABA dispersion relation \begin{equation}\gamma =2 i g\sum_{j} \left( \frac{1}{x^+(u_{4,j})}-\frac{1}{x^-(u_{4,j})}\right)\,  \end{equation}
 we series expand for small $g$ and keep only the leading term
\begin{equation}
\gamma = \sum_{k=1}^{J_2+J_3} 2i \left(u_{4,k}+\frac{i}{2} \right) +\mathcal{O}(g)\,,
\end{equation}
The leading term provides the correct expression for the energy of  multi-magnon state in the models (\ref{chiFT4}) and (\ref{bi-scalarL}) (with $J_3=0$ for the latter). The dependence on the effective couplings $\xi_i$ comes already through the solution of the double scaled spectral equations (\ref{dsspectraleqs}).

\paragraph{A simple check: Konishi in the strongly $\beta$-deformed twisted theory}
As mentioned earlier, the anomalous dimension of Konishi  in the $\beta$-deformed theory was computed previously both from Feynman perturbation theory \cite{Fiamberti2009,Fiamberti2008a}
and by integrability methods \cite{Ahn2011} using  the asymptotic Bethe ansatz with the first L\"uscher correction and TBA/Y-system \cite{Gromov2011}. 
After taking the double scaling limit, the four loop result is given by the asymptotic part and the wrapping correction $\gamma^{\mathrm{wrap}}$ which starts at four loop
\begin{equation} \label{konishi}
\gamma_{\mathcal{K}} = 2 \sqrt{2}\, \xi^2-2\, \xi^4 +\frac{\xi^6}{\sqrt{2}}+(4-4 \zeta_3)\,\xi^8 +\gamma^{\mathrm{wrap}}\,\xi^{8}+\mathcal{O}(\xi^{10})\,,
\end{equation}
with $\xi =g\, e^{-i \beta/2}$\, and
\begin{equation}
\gamma^{\mathrm{wrap}} = 4-8\,\zeta_{3}\,.
\end{equation}
The asymptotic result (\ref{konishi}), apart from the wrapping term, matches precisely with one of the solutions of equation (\ref{brokensu2}) for $L=4$ and $J_2=2$.
 The wrapping contribution 
 can be obtained either by computing the L\"uscher corrections or from 
 the TBA/Y-system, as in the abovementioned papers.
The wrapping correction to the one-magnon state and the calculation of the corresponding wrapped Feynman graph will be discussed in section~\ref{sec_wrap}.

\subsection{Spectrum of $\chi\mathrm{FT}_3$}

\subsubsection{Dispersion relation}
 As a starting point to determine the dispersion relation of the $\chi\mathrm{FT}_3$ model, we consider the twisted Bethe equations for the $\mathfrak{su}(2)\times\mathfrak{su}(2)$ sector of the ABJM model written in  Appendix \ref{appbetheABJM}. The simplest excited state we will be studying contains a single magnon
 \begin{equation}
 \Tr \left[ (Y^1 Y_4^{\dagger})^{L-1} (Y^2 Y_4^{\dagger}) \right]
 \end{equation}
 which corresponds to the excitation of one $\mathfrak{su}(2)$ wing of the $\mathfrak{su}(2) \times \mathfrak{su}(2)$ subsector. At the level of Bethe roots, this is equivalent to setting $K_{\bar{4}} = K_3 =0$ and excite a single root $K_{4}=1$. The following discussion then parallels the one from the previous section. The resulting Bethe equation for this state is simply given by
\begin{equation}\label{periodicityabjm}
q_1^{-L}q_2^{-L}q_3^{+L} e^{iL p} \,=1\,,
\end{equation} 
which fixes the momentum $p$ to be given solely in terms of the twists $q_1,\, q_2$ and $q_3$.
We consider now the original dispersion relation of ABJM, namely
\begin{equation}
\gamma= \sqrt{\frac{1}{4} + 4 h(\lambda)^2 \sin(p/2)^2}-\frac{1}{2}\,.
\end{equation}
In the double scaling limit, we use that at weak coupling $h(\lambda) \simeq \lambda$ which leads to the following effective dispersion relation
\begin{equation} \label{anomeff}
\gamma =  - \frac{1}{2}+ \sqrt{\frac{1}{4}- \,\lambda^2\, q_1 q_2 q_3^{-1}}= - \frac{1}{2}+\frac{1}{2}\sqrt{1- 4\,\xi} \,.
\end{equation}
Notice now that the effective coupling $\lambda^2 \, q_1 q_2 q_3^{-1}=  \xi_3 /(\xi_1 \xi_2) \equiv \xi$ is consistent with the Lagrangian coupling defined in (\ref{finalabjm}).
This simple dispersion relation is exactly the same as in the doubled scaled model from $\mathcal{N}=4$ SYM (see (\ref{displim})). However, it now arises from the re-summation of a series of 3D ladder diagrams discussed in section \ref{chift3diagrams}. Such ladder diagrams were previously computed in \cite{Bak:2009tq} and they produce precisely the anomalous dimension given in (\ref{anomeff}).

\subsubsection{Multi-magnon states} \label{ABAABJM} 
We consider now multi-magnon states in the  $\chi\mathrm{FT}_3$ of the form
 \begin{equation}
 \Tr \left[ (Y^1 Y_4^{\dagger})^{L-N_1} (Y^2 Y_4^{\dagger})^{N_1} \right]\,+\text{permutations}.
 \end{equation}
At the level of the twisted Bethe equations (\ref{twistedabjm}), this type of operators is contained in the closed $\mathfrak{su}(2)$ sector and it amounts to setting the number of roots to $K_{\bar{4}}=K_3 =0$ and $K_4=N_1$. 
The equations we will be considering are then the following
\begin{equation} \label{twistedabjm1}
q_1^{-L}q_2^{-L}q_3^{+L}\left(\frac{x^+_{4,k}}{x^-_{4,k}}\right)^{
L} =\,
\prod_{j\neq k}^{K_4}
\left[ \frac{u_{4,k}-u_{4,j}+i}{u_{4,k}-u_{4,j}-i}\, \sigma_{\rm BES}(u_{ 4,k},u_{ 4,j}) \right]  \,.
\end{equation}
We now proceed in  complete analogy  to the case of the twisted $\mathcal{N}=4$ SYM. Namely, we perform the following crossing transformation on the Zhukovski variables which allows us to get the properly double scaled equations
\begin{equation}
x_{4,k}^{-} \rightarrow  1/x_{4,k}^{-} \,,\,x_{4,k}^{+} \rightarrow  x_{4,k}^{+}\,.
\end{equation}
Upon expanding for small $\lambda$ we are led to the double scaled equations,
\begin{equation} \label{twistedabjm2}
\xi^{-L}\,(u_{4,j}^2+1/4)^{L} =
\prod_{j\neq k}^{K_4}
\left[ \frac{u_{4,k}-u_{4,j}+i}{u_{4,k}-u_{4,j}-i} \,\sigma_{\rm 0}^{m,m}(u_{ 4,k},u_{ 4,j})  \right] \,
\end{equation}
where $\xi = \frac{\lambda^2 q_1  q_2}{q_3} = \frac{\xi_1 \xi_2}{\xi_3} $ is precisely the coupling constant appearing in the Lagrangian (\ref{finalabjm}).
The dressing factor is given here by the same formula \eqref{dressing-f} and the energy has the following expression
\begin{equation}
\gamma = \sum_{k=1}^{K_4} i \left(u_{4,k}+\frac{i}{2} \right)\,.
\end{equation}

This set of equations is analogous to the  equations (\ref{brokensu2}) for the broken \(\mathfrak{su}(2)\) sector of $\mathcal{N}=4$ SYM. However, despite of the similarities 
they can be used to compute UV divergences of very different, three-dimensional Feynman graphs, as discussed in section \ref{chift3diagrams}.

\section{Computing multi-loop graphs from the ABA spectrum }
\label{graphsABA}

The goal of this section is to provide a method of the study, and sometimes of explicit calculation, of multi-loop Feynman integrals entering the two-point functions and the dilatation operator of the bi-scalar model described above, by exploiting a combination of direct graph computation and of the integrability of  the dilatation operator of the model. The specific feature of our model where each order of perturbation theory is defined by at most a single Feynman graph, makes possible the study of certain individual graphs   unachievable by conventional methods.

\subsection{The dilatation operator}
Let us begin by introducing  generic flavour structures  needed for the construction of the dilatation operator. In the bi-scalar model, the dilatation operator will depend on a very limited number of such structures as there is a single interaction vertex. Consider the standard basis of states where the reference state $\Tr\left[(\phi^{1})^L\right]$ is represented by $|\uparrow \dots \uparrow\rangle$  and the excitations $\phi^2$ are placed on top of it  as spins pointing down e.g., \begin{equation}\Tr\left[\phi^{1} \phi^2 \phi^1 \phi^1\dots \right] \equiv |\uparrow \downarrow \uparrow \uparrow \dots \rangle\,.\end{equation} We consider here for simplicity the closed ``broken \(\mathfrak{su}(2\))'' sector of the model formed by the operators of the type \eqref{multi-magnons}.  Then the typical flavour structures in this basis are defined by
\begin{equation}
\mathcal{X} ( \dots,a,b,c,\dots )
 \equiv 
\sum_{i=1}^L 
\dots 
\bigl(
  \sigma^{+}_{i+a} \sigma^{-}_{i+a+1}
\bigr)
 \bigl(
  \sigma^{+}_{i+b} \sigma^{-}_{i+b+1}
\bigr)
\bigl(
  \sigma^{+}_{i+c} \sigma^{-}_{i+c+1}
\bigr) 
\dots\,.
\end{equation}
where \(\s_j^+=\begin{pmatrix}0 & 1 \\
0 & 0 \\
\end{pmatrix}\)
and  \(\s_j^-=\begin{pmatrix}0 & 0 \\
1 & 0 \\
\end{pmatrix}\) are acting on the spin at the site \(j\).
Pictorially, each factor in every term corresponds to  exchanging the positions of $\phi^2$ and $\phi^1$ provided $\phi^2$ is to the left of $\phi^1$ (the term $\sigma_{j}^{-}\sigma_{j+1}^{+}$ is absent). Equipped with this structure, we can easily write down the expected form of the dilatation operator as a linear combination of the  structures allowed by the Lagrangian, with arbitrary coefficients. These coefficients can then be in principle fixed from the computation of corresponding Feynman integrals. The peculiarity of this  bi-scalar chiral model is that for a given structure at a given loop order there is a single Feynman integral contributing to it at that loop order.\footnote{Of course, due to the renormalization, products  of graphs  giving rise to that structure at lower loop orders will also contribute to the coefficients  of structures at  a given loop order.}

Accounting for the symmetries of the dilatation operator, we are led to the linear combination of the following structures up to four loops\footnote{This form of the dilatation operator is similar to the one for the original $\mathcal{N}=4$ SYM  employed, for instance, in \cite{Beisert2007}. The differences are in the structures defined by \(\mathcal{X}\)s which contrasts with the full standard permutation operator, and in that we only allow for the so-called maximal range interactions (that is the interactions that reshuffle the spins in a maximal way). 
}
\begin{equation}
\begin{aligned} \label{dilatationansatz}
\delta\mathcal{D} = \,\,  & \xi^2 \, \mathcal{C}_{11}\, \mathcal{X} (1)\\
+\, &\xi^4\, \mathcal{C}_{21} \left( \mathcal{X} (1,2) +\mathcal{X} (2,1) \right) \\
 +\, & \xi^6  \bigl( \mathcal{C}_{31} \, (\mathcal{X} (1,3,2) + \mathcal{X}(2,1,3) )+\mathcal{C}_{32} \, \mathcal{X}(3,2,1) +\mathcal{C}_{33}\, \mathcal{X}(1,2,3) \bigr)  \\
+ \, & \xi^8  \bigl[ \mathcal{C}_{41}  \mathcal{X} (2,1,3,2) +\mathcal{C}_{42} \, (\mathcal{X}(2,1,4,3) + \mathcal{X}(1,3,2,4))  \\
+ \,  &\mathcal{C}_{43} (\mathcal{X}(1,4,3,2) + \mathcal{X}(1,2,4,3)) +\mathcal{C}_{44} (\mathcal{X}(3,2,1,4)+ \mathcal{X}(2,1,3,4)  )\\
+\,  &\mathcal{C}_{45} (\mathcal{X}(1,2,3,4) + \mathcal{X}(4,3,2,1)) \bigr]\,.
\end{aligned}
\end{equation}
 Some of these coefficients depend on the renormalization scheme for subtraction of the divergences. In the next section, we will restrict ourselves to the minimal subtraction scheme in 
 dimensional regularization  and will fix completely the dilatation operator within this scheme.
This also allows to fix, in the same scheme, the eigenvectors representing operators with particular dimensions. Potentially, this might be useful for computing more complicated quantities such as multi-point correlators.
\subsection{Two-point functions in dimensional regularization}

One way of extracting the dilatation operator is through the perturbative computation of two-point correlation functions of local operators.
Let us very briefly review how to do it in the dimensional regularization scheme, fixing some useful notations along the way.
Consider a bare two point function
\begin{equation}
\mathcal{G}_{\alpha \beta}^{\text{bare}}\equiv\langle \mathcal{O}^{\text{bare}\, \dagger}_{\alpha}(x) \, \mathcal{O}^{\text{bare}}_{\beta}(0) \rangle
\end{equation}
in dimensional regularization, where the dimension $D$ is set to be $4-2\epsilon$ with the parameter $\epsilon$ serving as regulator. To avoid the cluttering of formulas we suppress the arguments of $\mathcal{G}_{\alpha \beta}^{\text{bare}}$. In perturbation theory, the structure of $\mathcal{G}_{\alpha \beta}^{\text{bare}}$ is given by
\begin{equation} \label{bare}
\mathcal{G}_{\alpha \beta}^{\text{bare}} = \frac{1}{x^{2 \Delta_0}} \left( \mathcal{N}_{\alpha \beta} +\sum_{n=1}^{\infty} \xi^{2n} \, (x^{-2} \mu^{2})^{n \epsilon}\, \mathcal{I}^{(n)}_{\alpha \beta}(\epsilon)  \right)\,,
\end{equation}
where $\mu$ is a scale introduced to keep the coupling constant dimensionless, $\mathcal{N}$ is a tree level normalization factor and the $n-$loop Feynman integrals $\mathcal{I}^{(n)}_{\alpha \beta}(\epsilon) $ generically admit a Laurent series in $\epsilon$ of the form
\begin{equation}
  \mathcal{I}^{(n)}_{\alpha \beta}(\epsilon)  = \frac{c^{(n)}_n}{\epsilon^n}+\frac{c^{(n)}_{n-1}}{\epsilon^{n-1}} +\dots +\frac{c^{(n)}_1}{\epsilon} + c^{(n)}_{0} + \mathcal{O}(\epsilon)\,.
\end{equation}

Note that we are considering a slightly different regularization as compared to the most standard dimensional regularization, where the scalar propagators  appear raised to the power $1-\epsilon$. Therefore, in our scheme each $n-$loop integral spits out a factor of $x^{-2 n \epsilon}$ instead of $x^{+2 n \epsilon}$. This scheme is convenient for computing the loop integrals using IBP identities with FIRE \cite{Smirnov:2014hma}, which requires the propagators to have integer powers.

In order to extract the anomalous dimension of the local operators, we renormalize them by constructing the wave-function $\mathcal{Z}_{\alpha \beta}$ also as a Laurent series in $\epsilon$, in such a way that the renormalized two point function is finite in the limit  $\epsilon\to 0$  and it is given by the standard conformal invariant form. 
More concretely, the renormalized operators are defined through
\begin{equation}
\mathcal{O}_{\alpha} =\mathcal{Z}_{\alpha \beta}\, \mathcal{O}^{\text{bare}}_{\beta}\,.
\end{equation}
where the wave-function $\mathcal{Z}_{\alpha \beta}$ is a function of $\epsilon$ and of the coupling $\xi$ (and of the corresponding scale $\mu$) only. The $n^{\text{th}}$-loop term of the wave-function is generically given by the follolwing expansion
\begin{equation}
 \mathcal{Z}_{\alpha \beta} \bigr|_{n-\text{loop}} = \frac{z^{(n)}_n}{\epsilon^n}+\frac{z^{(n)}_{n-1}}{\epsilon^{n-1}} +\dots +\frac{z^{(n)}_1}{\epsilon}\,,
 \end{equation}
where we have truncated the series in $\epsilon$ up to the term $\epsilon^{-1}$. This simply corresponds to the choice of a minimal subtraction renormalization scheme. The coefficients of the wave-function and the corresponding anomalous dimension are simultaneously fixed by solving the following set of equations\footnote{The scale $\mu$ does not play any role in this discussion and hence we have set it to one.} 
\begin{equation} \label{renotwopt}
  \lim_{\epsilon \rightarrow 0}
  \bigl[ \mathcal{Z}^{*}_{\alpha \rho} \,  \mathcal{G}^{\text{bare}}_{\rho \gamma} \, \mathcal{Z}_{\beta\gamma} \bigr]
  =
  \bigl[
  \mathcal{C} \,(x^2)^{-\Delta_{0} - \left(\delta\mathcal{D}\right)}
  \bigr]_{\alpha \beta}\,,
\end{equation}
where $\left(\delta\mathcal{D}\right)_{\alpha \beta}$ is the (anomalous part of) the dilatation operator which provides the anomalous dimensions upon diagonalization and $\mathcal{C}_{\alpha \beta}$ is a normalization constant.
 In the next section we will use this set of equations to first confirm the prediction of the asymptotic Bethe ansatz up to four loops (where all integrals are well known) and then   make a prediction for a particular combination of five-loop integrals as an illustration\footnote{We have not found these particular five-loop integrals in the literature.}.

\subsection{Two magnons at four loops}
In this section we consider the anomalous dimensions for $N=2$ up to
four loops. The length of the operators $L$ is taken to be
sufficiently large such that no wrapping diagrams contribute at this
loop order, i.e. $L>4$. This ensures that no $\phi_1$ particle interacts
with any $\phi_2$ particle more than once.

The following diagrams contribute to the bare two-point function up to
four loops for $L=7$, 
\begin{multline}
  \mathcal{G}_{\alpha\beta}
  =
    \mathbbm{1}_{3\times 3}
    + \left[
      \begin{array}{ccc}
        0& \dgab & 0\\
        \dgba& 0 & \dgbc\\
        0& \dgcb & \dgcc\\
      \end{array}
    \right] 
    + \left[
      \begin{array}{ccc}
        \dggaa & 0                & \dggac\\
        0      &\dggbb  & \dggbc     \\
        \dggca & \dggcb                & \dgcc
      \end{array}
    \right] \\
    + \left[
      \begin{array}{ccc}
        0& \dgggab &\dgggac       \\
        \dgggba &\dgggbb                     & \dgggbc\\
        \dgggca & \dgggcb              &\dgggcc
      \end{array}
    \right] 
    + \left[
      \begin{array}{ccc}
        \dggggaa & \dggggab & \dggggac\\
        \dggggba & \dggggbb & \dggggbc\\
        \dggggca & \dggggcb & \dggggcc 
      \end{array}
    \right]  + \mathcal{O}(\xi^{10})\, .
\end{multline}

Although almost each entry of these diagrams is a different graph,
some of the entries share the spacetime integral and the bare mixing
matrix of the two point functions is given in terms of the following
integrals:

\begin{multline} \label{bare7}
  \mathcal{G}_{\alpha\beta}
  =
  \mathbbm{1}_{3\times 3}
  +
  \left[
    \begin{array}{ccc}
      0,& I_{[;1]} & 0\\
      I_{[;1]}& 0 & I_{[;1]}\\
      0 & I_{[;1]} & I_{[;1]}
    \end{array}
  \right] \xi^2
  +
  \left[
    \begin{array}{ccc}
      I_{[;2]}& 0& I_{[;2]}\\
      0& I_{[;1]}^2 & I_{[;2]}\\
      I_{[;2]}& I_{[;2]} &I_{[;1]}^2 
    \end{array}
  \right] \xi^4\\
  \hfill
    +
  \left[
    \begin{array}{ccc}
      0&I_{[0;1,2]}&     I_{[;3]}\\
      I_{[0;1,2]}&      I_{[;3]}&      I_{[;1]}I_{[;2]}\\
      I_{[;3]}&I_{[;1]}I_{[;2]}&     I_{[;1]}I_{[;2]}
    \end{array}
  \right] \xi^6
  +
  \left[
    \begin{array}{ccc}
      I_{[0;2,2]}&I_{[;4]}&          I_{[0;1,3]}\\
      I_{[;4]}         &I_{[1;2,2]} & I_{[;1]}I_{[;3]}\\
      I_{[2;3,1]}&I_{[;1]}I_{[;3]}& I_{[;2]}I_{[;2]}
    \end{array}
  \right] \xi^8
  +
  \mathcal{O}(\xi^{10})\, ,
\end{multline}
where $I_{[a_1,\dotsc,a_m; b_1,\dotsc, b_m]}$ denotes an $m$-magnon
graph, for which the $i^{\text{th}}$ magnon is $a_{i-1}$ sites to the
right of the $(i-1)$'th one, and crosses $b_i$ vacua. They can be read
off from the graphical representation in equation (\ref{bare7}). Note
that some of these integrals are equivalent because two Feynman graphs
that are related by a 180$^\circ$ rotation give rise to the same
kinematical factor, such as $I_{[0;1,3]}$ and $I_{[2;3,1]}$.

These integrals have UV divergences for each integration vertex that
is attached to one of the external vertices by two powers of a scalar
propagator. To regulate these UV divergences, we consider these
integrals in $4-2\epsilon$ dimensions but keep the propagators as
$1/{x^2}$ .


Being logarithmically divergent, these integrals have a Laurent expansion in
$\epsilon$ as follows:
\begin{equation}
  \label{eq:4}
  I_{[a_1,\dotsc,a_m; b_1,\dotsc, b_m]}
  =
  \sum_{k=-h}^{\infty}
  I_{[a_1,\dotsc,a_m; b_1,\dotsc, b_m]}^{(k)} \epsilon^{k}\, ,
\end{equation}
where $h = \sum_{i=1}^m b_i$ is the number of loop integrations, i.e.
the total number of interactions undergone by all magnons.

The values of the integrals considered in this section can be found in
Appendix \ref{appendixintegrals}.

We now plug (\ref{bare7}) into the defining formula of the dilatation
operator (\ref{renotwopt}) and allow for arbitrary entries for the
wave-function $\mathcal{Z}$. Once we solve these equations we obtain
the following dilatation operator
\begin{equation}
\delta\mathcal{D} = \left[
\begin{array}{ccc}
 -2 \xi ^4 +4 (1-2 \zeta_3) \xi ^8& -2 \xi ^2-2 \xi ^6 -10 \xi ^8&-2 \xi ^4- 4 \xi ^6-\frac{8 \xi ^8}{3} \\
-2 \xi ^2+2 \xi ^6- 10 \xi ^8 &- 4 \xi ^6+2 \xi ^8 &-2 \xi ^2 - 2 \xi ^4\\
-2 \xi ^4-4 \xi ^6  +\frac{8 \xi ^8}{3}& -2 \xi ^2 -2 \xi ^4&- 2 \xi ^2 \\
\end{array}
\right]\,,
\end{equation}
which fixes the constants $\mathcal{C}_{ij}$ of the ansatz (\ref{dilatationansatz}) to be
\begin{equation}
\begin{aligned}
&\mathcal{C}_{11}=-2,\,\,\,\,\mathcal{C}_{21}=-2,\,\,\,\,\mathcal{C}_{31}=2,\,\,\,\,\mathcal{C}_{32}=-2,\,\,\,\,\mathcal{C}_{33}=-4,\\
&\mathcal{C}_{41}=-4 (2 \zeta (3)-1),\,\,\,\,\mathcal{C}_{42}=2,\,\,\,\, \mathcal{C}_{43}=\frac{8}{3},\,\,\,\, \mathcal{C}_{44}=-\frac{8}{3},\,\,\,\, \mathcal{C}_{45}=-10\,.
\end{aligned}
\end{equation}
We can now check that the eigenvalues of this dilatation operator match exactly with the solutions of the Bethe equations (\ref{brokensu2}).
%


\subsection{Predictions at five loops}

In this section, we illustrate how to make use of the spectrum to compute the UV divergences of multi-loop Feynman integrals of the $\phi^4$ interaction type. We will focus on five-loop integrals but one can equally well generate information about even higher loops. 

Essentially we reverse the sequence  of steps described in the previous section. We start by writing the five-loop integrals as a series in $\epsilon$ with arbitrary coefficients and solve the set of equations (\ref{renotwopt}) in terms of these coefficients. We then diagonalize the resulting dilatation operator and equate its eigenvalues to the integrability prediction from the solutions of (\ref{brokensu2}). The structure of the five loop contribution to the bare two point function for $L=7$ operators with two magnons is given by the following graphs 
\begin{equation}
\mathcal{G}_{\alpha \beta} \bigr|_{5\text{ loops}}= 
\left[
      \begin{array}{ccc}
        \dgggggaa & \dgggggab & \dgggggac\\
        \dgggggba & \dgggggbb & \dgggggbc\\
        \dgggggca & \dgggggcb & \dgggggcc 
      \end{array}
    \right] = 
     \left[
    \begin{array}{ccc}
      I_{[;5]}&I_{[0;2,3]}&          I_{[0;1,4]}\\
      I_{[1;3,2]}         &I_{[;4]}I_{[;1]} & I_{[1;2,3]}\\
      I_{[3;4,1]}&I_{[2;3,2]}& I_{[;3]}I_{[;2]}
    \end{array}
  \right]
\end{equation}
where  in the second equality we have explicitly written the graphs in terms of the corresponding integrals in our notation.
The information provided by the spectrum is going to be used to fix the $1/\epsilon$ terms of the integrals. The remaining higher order poles are almost completely fixed as a condition for the exponentiation of the result as given in $(\ref{renotwopt})$. There is some freedom here related to the choice of subtraction scheme. This manifests itself in the fact that the $1/\eps^2$ terms of the integrals are fixed up to a constant which we denote by $d_1$ below.  As a solution of the equations  $(\ref{renotwopt})$ we find
\begin{equation}
\begin{aligned}
  \mathcal{I}_{[0;1,4]}& =-\frac{16}{15 \epsilon ^5}-\frac{13}{6 \epsilon ^4}+\frac{\frac{37}{10}+\frac{4 \pi ^2}{9}}{\epsilon ^3}+ \frac{-144 \, d_{1}+5440 \zeta_3+325 \pi ^2+612}{360 \epsilon ^2}+\frac{p_1}{\epsilon } \\ 
    \mathcal{I}_{[1;3,2]}& =-\frac{12}{5 \epsilon ^5}-\frac{43}{6 \epsilon ^4}+\frac{\frac{13}{10}+\pi ^2}{\epsilon ^3}
+\frac{144\,d_{1}+10080 \zeta_3+1075 \pi ^2+4548}{360 \epsilon ^2}+\frac{p_2}{\epsilon }  \\ 
 \mathcal{I}_{[2;3,2]}& =-\frac{4}{3 \epsilon ^5}-\frac{7}{2 \epsilon ^4}+\frac{\frac{19}{6}+\frac{5 \pi ^2}{9}}{\epsilon ^3}  
+\frac{-144\,d_1+5360 \zeta_3+525 \pi ^2+3588}{360 \epsilon ^2}+\frac{p_3}{\epsilon }\\
\mathcal{I}_{[;5]} &=-\frac{4}{15 \epsilon ^5}+\frac{15+\pi ^2}{9 \epsilon ^3}
+\frac{\frac{52 \zeta_3}{9}-\frac{14}{3}}{\epsilon ^2} + \frac{p_4}{\epsilon } \end{aligned} 
\end{equation}
Now we inject the spectrum information from the solution of the Bethe equations. This fixes
$p_4$ and imposes a nontrivial relation between the constants $p_1, p_2$ and $p_3$. They read
\begin{equation}
\begin{aligned}
p_2&=-p_1-\frac{196 \zeta_3}{9}+\frac{133 \pi ^4}{360}-\frac{25 \pi ^2}{12}-\frac{401}{15}\\
p_3&=p_1+\frac{238 \zeta_3}{45}+\frac{2 \pi ^2}{9}-\frac{176}{15}-\frac{11 \pi ^4}{360}\\
p_4&=-\frac{56 \zeta_3}{3}+\frac{181}{15}-\frac{17 \pi ^2}{36}-\frac{31 \pi ^4}{360}
\end{aligned}
\end{equation}
The conclusion is that the computation up to the order \(1/\epsilon\) of a single five-loop integral, say $  I_{[0;1,4]}$ which looks simpler than the other two, would fix the unknowns $d_1$ and $p_1$ and therefore constrain the other two integrals completely. Therefore, we profit substantially from using the spectrum\footnote{We could have also applied this method to the previous four loop example. In this case, we have three two-magnon integrals at four loops. We have checked that by computing one of these integrals one recovers the two other integrals already known in the literature.}.

This clearly carries over to the higher loops and one can raise the question of how efficient is this method for constraining higher loop integrals. In order to answer this we observe that what we did here is equivalent to fixing as much as possible the Hamiltonian from the spectrum constraints. This has been extensively done for both $\mathcal{N}=4$ SYM\cite{Beisert2007} and ABJM \cite{Minahan:2009wg, Leoni:2010tb}.
In our context the difference is that we can relate a particular, connected or disconnected graph to each unknown coefficient of the Hamiltonian. From this point of view it is clear that the spectrum is not enough due to similarity transformations one can perform  without changing the spectrum. As the loop order increases, the number of unknowns will also increase and in order to fix it completely we have to supplement the spectrum with explicit results for particular integrals. Therefore, when complemented with other methods for determining some higher loop integrals, this has the potential of providing valuable information about them.





\section{Wrapping effects in the bi-scalar chiral model}
\label{sec_wrap}

We will discuss here the wrapping effects for the simplest single-magnon operator  \begin{equation}\label{single-magnon}
\cO_{L}(x)= \Tr[ (\phi^1)^{L-1}\,\phi^2 ]\,,
\end{equation} in the bi-scalar model \eqref{bi-scalarL}. As seen before, for asymptotically long such operators their anomalous dimension is described by the expression (\ref{finaldisp4}). This dispersion relation turns out to be the same also for the DS limit stemming from the $\beta-$deformed SYM, so that the asymptotic anomalous dimensions of this operator coincide for these two distinct models. 
Perhaps more non-trivial is the fact that this equality of the asymptotic anomalous dimensions of both models actually persists when the wrapping correction are included. This follows from the diagrammatic analysis of the perturbative two point functions for these operators. It turns out that in both models the only allowed vertex is the one from the single interacting term of the bi-scalar model.  
The leading wrapping correction \( \gamma_L^\mathrm{wrap}\) is of order $\xi^{2L}$ and we will focus on it from now on. In order to extract it, we follow the observation above and use the results of the computation of  \(\gamma^{\text{wrap}}_L\) in the \(\beta\)-deformed $\mathcal{N}=4$ SYM known in the literature, both from the direct Feynman graph computation \cite{Fiamberti2009} and from integrability \cite{Gromov2011} (with a perfect match between the results of these two methods) and take its DS limit.

In conjunction with the dimensionally regularized expression for unwrapped graphs at any \(L\) computed in Appendix \ref{appendixintegrals}, integrability of $\chi$FT$_4$ allows us to determine the spiral graphs in figure \ref{two_m_wrap} for any $L$.
 
The result of  \cite{Fiamberti2009} states that the contribution of the wrapping diagrams to the anomalous dimension is given by the following expression
\begin{equation}\label{deltaGammaExact}
\gamma^{\mathrm{wrap}}_L(g,q)=-2 L (4 \pi  g)^{2 L} \left(P_L \left(c_L^{(0)}-c_L^{(L-1)}\right)-2 \sum _{j=0}^{\left\lfloor \frac{L}{2}\right\rfloor -1} \left(c_L^{(j)}-c_L^{(-j+L-1)}\right) I_L^{(j+1)}\right)+\gamma_{L-\mathrm{loop}}^{\mathrm{ABA}}(g,q)\,,
\end{equation}  where  
\begin{equation}
c_L^{(j)}\text{=}\left(q-\frac{1}{q}\right)^2 \left(q^{-2 j+2 L-2}+\left(\frac{1}{q}\right)^{-2 j+2 L-2}\right)\,.
\end{equation} 
Here $\gamma^{\mathrm{ABA}}_{L-\mathrm{loop}}$ is the \(L\)-loop
asymptotic Bethe ansatz contribution\footnote{The addition of the term
  $\gamma^{\mathrm{ABA}}_{L-\mathrm{loop}}$ is justified by the fact
  that in \cite{Fiamberti2009} the authors consider the full $L$-loop
  contribution on top of the asymptotic result for this type of
  operators. That requires the subtraction of the $L-$loop graphs with
  $(L+1)$-range of interaction. In our model, this last contribution
  is precisely given by the $L-$loop asymptotic Bethe ansatz
  contribution.} to the anomalous dimension for an operator with
length greater than $L$ (which in our model arises precisely from the
single-magnon $L$-loop ladder graph).
Additionally we have that 
\begin{equation}
P_L=\frac{2}{(4\pi)^{2L}L}\binom{2L-3}{L-1}\zeta_{2L-3}  
\end{equation} is the leading term in a wheel-type graph and \(I_L^{(j)}\) is explicitly given for some $j$ in \cite{Fiamberti2009}.

In the DS limit, only the term with \(j=0\) contributes to \eqref{deltaGammaExact} since  \(c_L^{(j)}= q^{-2 j+2 L}(1+{\cal O}(q^{-2}))\) and the corresponding integral is given by (see \cite{Fiamberti2009} for explicit expressions)   \begin{eqnarray}\label{Iint}
I_L^{(1)}=\frac{1}{2}P_L +\frac{1}{L} \sum _{k=L-1-\left\lfloor \frac{L-1}{2}\right\rfloor }^{L-3} \binom{2k+1}{2k+3-L}\zeta_{2k+1}+\frac{1}{2L}[1+(-1)^L](L-2)\zeta_{ L-1}\,. \end{eqnarray}  Hence,  keeping the leading \(q^{2L} \) term in  \eqref{deltaGammaExact} we obtain for the one-wrapping contribution to the  dimension of one-magnon operator in the bi-scalar model   \begin{equation}\label{wrapG}
\gamma_L^{\text{wrap}}(g,q)-\gamma_{L-\mathrm{loop}}^{\mathrm{ABA}}(g,q)=\xi^{2L}f_L\,,
\end{equation}
where
\begin{equation}\label{wrapf}
f_L\text{=}-2  (4 \pi  )^{2 L} \left(2 \sum _{k=L-1-\left\lfloor \frac{L-1}{2}\right\rfloor }^{L-3} \binom{2k+1}{2k+3-L}\zeta_{2k+1}+[1+(-1)^L](L-2)\zeta_{ L-1}\right)\,.\end{equation}
 We will now proceed to compute the first wrapping graph drawn in
 figure \ref{two_m_wrap}.
     In perturbation theory, up $L$ loops, the
 length-$L$ operators are renormalised by the following graphs:
 \begin{equation}
   \label{wrappeddimension}
  \sum_{k=1}^{L-1}
  \raisebox{0mm}{
    \begin{minipage}{3.5cm}
      \begin{center}
              \includegraphics{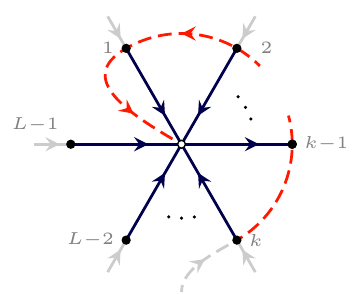}
    \end{center}
  \end{minipage}
  }
  \qquad 
  +
  \quad
  \raisebox{0mm}{
  \begin{minipage}{3.5cm}
    \begin{center}
                    \includegraphics{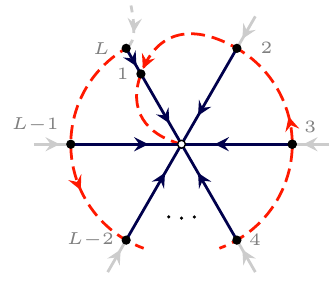}
    \end{center}
  \end{minipage}
  }
\end{equation}
We will use for that the knowledge of the asymptotic one-magnon (amputated) graphs at any \(L\) and at any \(n<L\)  loops, i.e. before the wrapping order determined in Appendix \ref{appendixintegrals}  \footnote{The momentum-space versions of these integrals were computed in \cite{Gross2002}, which would correspond to position space integrals with propagators of the form \((x^2_{ab})^{-1+\epsilon}\).} in dimensional regularization,
\begin{equation}\label{Gneps}
G_{n\geq 2}(x,\epsilon)= (x^2)^{-n \epsilon}
  \prod_{k=1}^{n}
  \frac{  \Gamma(1-\epsilon)\Gamma(-k\epsilon) \Gamma(1+k \epsilon)}{\Gamma(1-(k+1)\epsilon) \Gamma(2+(k-1)\epsilon)}=(x^2)^{-n \epsilon }\sum_{k=-n}^{\infty}\e^{k}G_n^{(k)}\, .
\end{equation}  
\begin{table}[t] 
\centering
\begin{tabular}{| c | c | }
\hline
$L $ & $t_L$ \\ \hline \hline
 \begin{minipage}[c][1.2cm][c]{1cm}\centering  4\end{minipage} & \begin{minipage}[c][1.2cm][c]{14cm}
   \begin{equation}\nonumber 
     \frac{1365}{36}
     -\frac{7}{36}\pi^2
     -\frac{41 \zeta_3}{9} 
   \end{equation}\end{minipage}\\  \hline 
 \begin{minipage}[c][1.2cm][c]{1cm}\centering  5\end{minipage} & \begin{minipage}[c][1.2cm][c]{14cm}
   \begin{equation}\nonumber
     \frac{6727}{40}
     -\frac{245\pi^2}{288}
     -\frac{335 \zeta_3}{18}
     -\frac{169\pi^4}{11520}
     -8 \zeta_5
   \end{equation}\end{minipage}\\  \hline   
 \begin{minipage}[c][1.2cm][c]{1cm}\centering  6\end{minipage} & \begin{minipage}[c][1.2cm][c]{14cm}
   \begin{equation}\nonumber
     \frac{62601}{80}
     -\frac{301 \pi ^2}{80}
     -\frac{4991 \zeta_3}{45}
     -\frac{351 \pi ^4}{3200} 
     -\frac{712 \zeta_5}{75}
     +\frac{23 \pi^2 \zeta_3}{180}
     -\frac{70 \zeta_7}{3}
   \end{equation}\end{minipage}\\  \hline 
\begin{minipage}[c][1.2cm][c]{1cm}\centering   7\end{minipage} &\begin{minipage}[c][2.6cm][c]{14cm}\begin{equation}\nonumber 
    \begin{aligned}[t]
      &\frac{4753177}{1260}
      -\frac{49049 \pi
        ^2}{2880}
      -\frac{44891\zeta_3}{72}
      -\frac{110551 \pi ^4}{172800}
      +\frac{5929 \pi^2\zeta_3}{5184}\\
      &
      -\frac{49567 \zeta_5}{720}
      -\frac{393677 \pi  ^6}{87091200}
      +\frac{102487 \zeta_3^2}{12960}
      -12 \zeta_7
      -72\zeta_9
    \end{aligned}
\end{equation}\end{minipage} \\ \hline 
\end{tabular}
{\caption{ We present a few examples of the coefficient of the $1/\epsilon$ pole of the first wrapping integral at order $L$. The higher order poles coincide with those for the unwrapped single-magnon  integrals given in (\ref{Gneps})} \label{spiraltable}}
 \end{table} 
Let us now define the wrapped Feynman integral in equation (\ref{wrappeddimension})  as a Laurent series in $\epsilon$\begin{equation}\label{wrappedK}
K_L(\e)=  {\cal G}_{L}(\epsilon)+\frac{t_L}{\e}{\cal +O}(\e^0)
\end{equation}  where \({\cal G}_L(\epsilon)=\frac{G_L^{(L)}}{\e^{L}}+\frac{G_L^{(L-1)}}{\e^{L-1}}+\dots+\frac{G_L^{(2)}}{\e^{2}}\) with the coefficients
 defined through
\eqref{Gneps}. All the coefficients of the higher order poles in $\epsilon$ 
ought to be the same as for the unwrapped graphs at lengths greater than \(L+1\). Given the uniqueness of the graphs at each loop order this is the only possibility for them to exponentiate into the scheme independent anomalous dimension, which is finite and unambigous in the \(\epsilon\) limit. The first order pole  \(t_L /\e \) represents the only correction relevant to wrapping. We can fix it by the knowledge of one-wrapping contribution to the anomalous dimension \eqref{wrapG} (which is known also from integrability computations of \cite{Gromov2011} - a potentially more powerful method than the direct Feynman graph calculus). 
In order to relate the first order pole coefficient $t_L$ to the anomalous dimension we follow the standard renormalization of the operator $\cO_{L}(x)$. Since there is no mixing for this particular type of operators, it renormalizes multiplicatively by
\begin{equation} \label{anomazed}
\mathcal{O}_L^{\mathrm{ren}} = \mathcal{Z} \,  \cO_{L}
\end{equation}
where $\mathcal{Z}$ is such that the poles in $\epsilon$ coming from the quantum corrections to the bare operator $\cO_{L}$ will be cancelled. Once we determine it the corresponding anomalous dimension is given by the standard form
\begin{equation}
\gamma = \lim_{\epsilon \rightarrow 0} \left[ - \epsilon \, \xi\,  \frac{\partial \log \mathcal{Z} }{\partial \xi} \right]\,.
\end{equation}

Following this procedure, we compute $\mathcal{Z}$ at $L$-loop order from the above one-magnon unwrapped graphs $(\ref{Gneps})$ at $n<L$ loops and at order $L$ we use the wrapped diagram given in (\ref{wrappedK}). Equating the result for the anomalous dimension as given by (\ref{anomazed}) to the wrapping result (\ref{deltaGammaExact}) we fix the $t_L$ at any desired loop order. As illustration we have some results presented in table \ref{spiraltable} which can be easily generated for any loop $L$.


\section{Conclusions}

In this paper, we studied the properties of chiral field theories - \(\chi\)FTs emerging from the \(\gamma\)-deformed \(\mathcal{N}=4\) SYM theory and ABJM model in the  double scaling (DS) limit which combines a strong \(\gamma\)-twist with the weak coupling limit. While such a  DS limit was already proposed in \cite{Gurdogan:2015csr} by two of the authors for the \(\mathcal{N}=4\) SYM, a similar DS limit for ABJM is a new result of this paper. On the one hand, these  theories do not have any supersymmetry (apart from a very special choice of effective couplings for the \(\beta\)-deformation where some supersymmetry survives: \(\mathcal{N}=1\) in the case of $\mathcal{N}=4$ SYM or \(\mathcal{N}=2\) for ABJM) and they do not contain the gauge fields anymore. On the other hand they possess a much simpler Feynman graph expansion. Say, for the simplest such \(\chi\)FT  model considered here -- the four dimensional bi-scalar \(\chi\)FT\(_4\) -- for most of the interesting physical quantities, such as multi-point correlation functions, there is at most a single planar graph per loop order.     

A remarkable property of both \(\chi\)FT\(_4\) and \(\chi\)FT\(_3\)  theories is their quantum integrability in the 't~Hooft approximation. Unlike their ``mother''-theories, the integrability in these \(\chi\)FTs  is visible explicitly, say, for various two-point correlation functions, on the level of a single integrable planar graph at each order of perturbation theory.  

 The integrability  holds for all local single-trace
operators which can be constructed in these \(\chi\)FTs, provided that
their length is greater than two. This can be the BMN vacuum operator, which is unprotected for the twisted
theories, as well as more complicated multi-magnon operators. For the BMN state, integrability
becomes transparent since the corresponding wheel graphs, which make
up the only contribution in perturbation theory, appear to have the
same bulk structure as a ``fishnet'' graph with a square lattice
structure:  such fishnet graphs are shown in
\cite{Zamolodchikov:1980mb} to represent an integrable statistical
mechanical system.  The integrability for the wheel graphs can be also
explicitly demonstrated from the integrable \(\mathfrak{su}(2,2) \) spin chain
construction \cite{Gromov:2017cja}.  However, for  more
complicated states, such as multi-magnon operators, we have to rely on
more hypothetical but nevertheless well-established integrability
methods for the spectral problem of the twisted ``mother''-theories of
our \(\chi\)FTs -- \(\mathcal{N}=4\) SYM and ABJM. On the other hand, we are
able to establish the single  Feynman graph at each loop
order corresponding to a given local single-trace operator and pose the
question: what is the precise spin-chain picture behind its
integrability in the planar limit. This question is yet to be answered.

In this paper we studied  operators with magnons,
i.e. with insertions of new fields into the BMN vacuum. We clarified the Feynman graph picture for the renormalization of such operators. We explored
the integrability of spectral equations for the asymptotic limit of
very long operators in the form of Beisert-Staudacher ABA
equations. We managed to reduce these equations in the DS limit for
both $\mathcal{N}=4$ SYM and ABJM, at least in specific interesting sectors: for the
operators in the sector with (broken) \(\mathfrak{su}(3)\) symmetry for \({\cal N}=4\) SYM, and in the sector with (broken) \(\mathfrak{su}(2)\) for ABJM. It would be interesting to double-scale the full system
of ABA equations and  to  get the corresponding full  set of DS ABA
equations for the complete 4D \(\chi\)FT with 3-coupling action
\eqref{chiFT4int}, as well as for its 3D analog
\eqref{3paramDS_ABJM}.  Even for the bi-scalar \(\chi\)FT \eqref{bi-scalarL}
we do not yet understand how to get the double scaled ABA for the
operators mixing both chiralities (any single-trace product of
\(\phi^1,\phi^2,\phi_1^\dagger,\phi_2^\dagger\)).
 
A very simple Feynman graph picture of our \(\chi\)FTs suggests a
possibility to use integrability as a new, powerful tool of exact
computations of a large variety of multi-loop Feynman integrals. Some
classes of graphs, such as one-magnon graphs on figures \ref{one_m_spiral} or single- and double-wheel graphs
of the type drawn on figures \ref{doublewheel}, admit computations at arbitrary loop
order. We demonstrated here the efficiency of ABA equations that, when combined
with the direct computations of some multi-loop graphs in dimensional
regularization, 
fixes completely 
some unknown 
five loop two-magnon graphs of the length \(L\ge6\) for the bi-scalar
\(\chi\)FT.

The ultimate method for studying the spectrum of anomalous dimensions
in these integrable \(\chi\)FTs should be the twisted QSC equations of
\cite{Kazakov:2015efa}.\footnote{See also \cite{Gromov:2015dfa} for a
  similar twisted QSC for a different problem -- of a cusped Wilson
  line.  Recently, the QSC equations have been established for the
  cusped Wilson line~\cite{Gromov:2016rrp} in a similar limit for a
  simpler problem of computation of quark-antiquark potental in planar
  $\mathcal{N}=4$ SYM, where the DS limit of QSC sums up a ladder of
  Feynman graphs.}  But the appropriate doubly scaled version of this
QSC is yet to be found.  We hope that the ABA equations established in
the current paper are a significant step in this direction.

\begin{figure}
  \centering
  \includegraphics{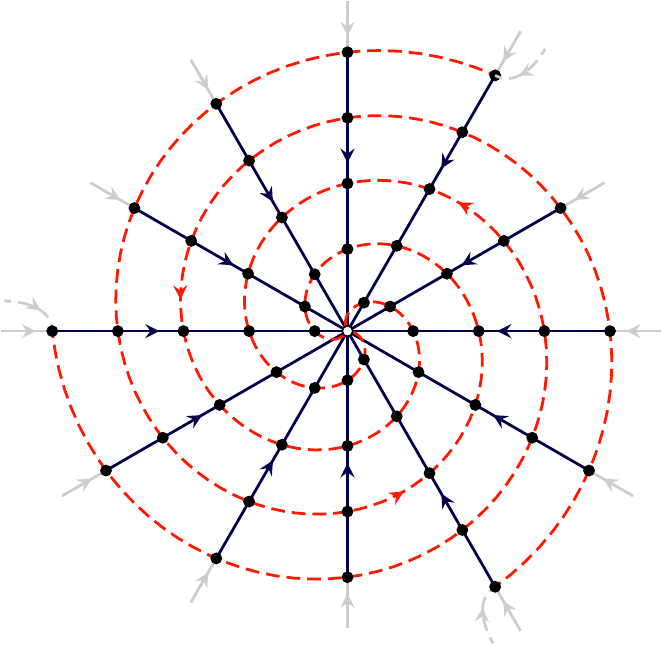}
  \caption{Even though the presence of magnons in single-trace
    operators imposes different boundary conditions, the bulk of these ``spider-web''
    diagrams that renormalise these operators is still a square lattice.}
  \label{fig:multimagnonspiderweb}
\end{figure}
The  perturbation theory computing the anomalous dimension of an   operator contains, in sufficiently high order of perturbation theory,  the ``wrapped'' Feynman graphs, where the magnon lines make at least one full circle around the operator. ABA is not enough for studying the wrapping contribution  and the Y-system/TBA/QSC equations are for the moment the only method to compute them.  For the \(\chi\)FTs studied here, the typical graphs (say, having sufficiently many loops but not too many external legs at the boundary)  are of a ``fishnet'' type, i.e., containing large pieces of regular rectangular lattice. For a particular complicated operator,  the graphs  computing  its anomalous dimension could be quite involved, such as for example ``spider-web'' graphs of figure \ref{fig:multimagnonspiderweb}. The double scaled QSC approach should be able to compute, at least numerically or analytically, to a high  order of perturbation theory, the corresponding anomalous dimensions which are given in the case of planar \(\chi\)FTs by a limited combination of graphs. This provides many scheme independent relations among these graphs. If completed by some direct graph calculus in a given scheme, these relations provide the method of direct computation of those graphs.  We demonstrated here the method by computing, using the integrability of mixing matrix, the unwrapped graphs of bi-scalar \(\chi\)FT\(_4\) at four and (partially) at five loops in the minimal dimensional regularization scheme.

The \(\chi\)FTs studied here should be a good, and still non-trivial testing ground for an even  more complicated important problem:  computation of structure constants in planar $\mathcal{N}=4$ SYM. There has been  recently progress in this problem\cite{Basso:2015zoa,Basso:2015eqa,Fleury:2016ykk,Eden:2015ija,Eden:2016xvg}, but the current methods allow only   for the ABA analogue of this quantity and some wrapping corrections. The  emerging difficulties concern the wrapping effects, both in direct diagrammatic approach, due to the large   number  and complexity  of diagrams, as well as in the hexagon approach where important analyticity ingredients are still missing. The \(\chi\)FTs, thanks to the drastic simplification of Feynman diagramatics on the one hand, and to a more direct integrability approach to ``fishnet'' graphs, has good chances to clarify the emerging problems and to help moving forward towards the complete construction of the OPE for local operators in planar $\mathcal{N}=4$ SYM. Moreover, the study of the running of the double-trace couplings, or even of some non-planar, \({\cal O}(1/N^2)\) effects might be accessible for computations in all orders of perturbation theory due to the simplicity of diagrammatics and spin chain type integrability of the related Feynman graphs for which the  fishnet structure will be still the typical bulk element.  

Another interesting set of physical quantities to explore in \(\chi\)FT are the scattering amplitudes.  For example, in the bi-scalar \(\chi\)FT the scattering amplitude of two species of bosons is given by a single finite Feynman graph depending on the ordering of external particles, i.e. a single loop-order gives the full contribution. The computation of such objects and the  observation of its Yangian symmetry claimed to be present in the amplitudes of the ``mother''-theories, is another curious direction of research.
 
 An important question to answer is: do these  \(\chi\)FTs have any gravity duals? On the one hand, from the standard twisted AdS/CFT point of view, we are at the  weak coupling on the CFT side which means the strong coupling, i.e. strong quantum fluctuations on the string side. That is why the standard classical gamma-twisted AdS picture \cite{Frolov:2005dj}  is not directly applicable. On the other hand, the planar graph expansion, especially in the strong coupling regime, suggests a world-sheet picture for sufficiently big graphs.   The \(\chi\)FTs suggest the existence of only chiral, or only anti-chiral excitations on this hypothetical world-sheet. It might  be also possible to combine both chiralities within the same string model, order by order in the string coupling, as one does in the standard critical string theory in the flat background. It would  be good to understand the fate of the tachyon in the gamma-deformed string theory on \(AdS_5\times S_5\) background.


\section*{Acknowledgments}
\label{sec:acknowledgments}

We thank B. Basso, M. Costa, T. Fleury, V. Gon\c{c}alves, N. Gromov, R. Janik, 
S. Komatsu, G. Korchemsky, E. Panzer, G. Sizov, S. van Tongeren,
P. Vieira, D. Volin, P. Wiegmann, M. Wilhelm, K. Zarembo for
discussions. We also thank B. Basso, R. Janik, S. Komatsu, I. Kostov, G. Korchemsky,
K. Larsen, K. Zarembo for reading the manuscript and their
comments. The work of J.C., \"O.G. and V.K. was supported by the
People Programme (Marie Curie Actions) of the European Union's Seventh
Framework Programme FP7/2007-2013/ under REA Grant Agreement No 317089
(GATIS), by the European Research Council (Programme ``Ideas''
ERC-2012-AdG 320769 AdS-CFT-solvable), from the ANR grant StrongInt
(BLANC- SIMI- 4-2011). The work of J.C. is supported by the research
grant CERN/FIS-NUC/0045/2015.  The work of \"O.G. is supported by the
ERC grant ERC-2015-CoG 648630 IQFT. J.C. and V.K. are grateful to
Humboldt University (Berlin) for the hospitality and financial support
of this work in the framework of the “Kosmos” programme.

\appendix


\section{Action of  $\g$-twisted $\mathcal{N}=4$ SYM}
\label{app:actionssym}
The lagrangian of \(\gamma\)-deformed \({\cal\ N}=4\) 
SYM reads  (see e.g.\cite{Fokken:2013aea})
\begin{equation*}
\label{N=4SYMlagrangian}
  {\cal L}=N_c\Tr\biggl[
  -\frac{1}{4} F_{\mu\nu}F^{\mu\nu}
  -\frac{1}{2}D^\mu\phi^\dagger_iD_\mu\phi^i
  +i\bar\psi^{\dot\alpha}_{ A}D^\alpha_{\dot\alpha}\psi^A_{\alpha }\biggr]
+{\cal L}_{\rm int}
\end{equation*}
where \(i=1,2,3\)\,\, \(A =1,2,3,4\),
\(D^{\alpha}_{\dot\alpha}= D_\mu
(\tilde\sigma^{\mu})^\alpha_{\dot\alpha}\) with
\((\tilde\sigma^{\mu})^\alpha_{\dot\alpha}=(-i\sigma_2,i\sigma_3,
\mathbb{I},-i\sigma_1)^\alpha_{\dot\alpha}\,\) and 
 \begin{equation*}
   \begin{aligned}[y]
     &\mathcal{L}_{\rm int} =N_cg\,\,\Tr\bigl[\frac{g}{4} \{\phi^\dagger_i,\phi^i\}
     \{\phi^\dagger_j,\phi^j\}-g\,e^{-i\epsilon^{ijk}\gamma_k}
     \phi^\dagger_i\phi^\dagger_j\phi^i\phi^j\\
     &-e^{-\frac{i}{2}\gamma^-_{j}}\bar\psi^{}_{ j}\phi^j\bar\psi_{ 4}
     +e^{+\frac{i}{2}\gamma^-_{j}}\bar\psi^{}_{ 4}\phi^j\bar\psi_{ j}
     + i\epsilon_{ijk} e^{\frac{i}{2} \epsilon_{jkm} \gamma^+_m} \psi^k \phi^i \psi^{ j}\\
     &-e^{+\frac{i}{2}\gamma^-_{j}}\psi^{}_{ 4}\phi^\dagger_j\psi_{
       j}
     +e^{-\frac{i}{2}\gamma^-_{j}}\psi^{}_{j}\phi^\dagger_j\psi_{
       4}
     + i\epsilon^{ijk} e^{\frac{i}{2} \epsilon_{jkm} \gamma^+_m} \bar\psi_{ k} \phi^\dagger_i \bar\psi_j\bigr]\hfill\, .
   \end{aligned}
 \end{equation*}
where the summation is assumed w.r.t. doubly and triply repeating indices. We suppress the  Lorentz indices of fermions, assuming the contractions
  \(\psi_i^\alpha \psi_{j,\alpha}\) 
 and  \(\bar\psi_{i,\dot\alpha} \bar\psi_j^{\dot\alpha}\). We also use the notations
 \begin{equation*}
\gamma_1^{\pm}=-\frac{\gamma_3\pm\gamma_2}{2},\quad\!\!
\gamma_2^{\pm}=-\frac{\gamma_1\pm\gamma_3}{2},\quad\!\!
\gamma_3^{\pm}=-\frac{\gamma_2\pm\gamma_1}{2}\, .
\end{equation*}
The parameters of the $\gamma$-deformation 
{$q_j=e^{-\frac{i}{2}\gamma_{j}}$ $j=1,2,3$} are related to
the Cartan subalgebra:  \(\mathfrak{u}(1)^3\subset \mathfrak{su}(4) \cong \mathfrak{so}(6)\).

\section{$\g$-twisted ABJM} \label{twstABJM}
We present here explicit form of the part of  the  $\gamma$-twisted ABJM action involving the fermion-scalar interactions. It reads 
\begin{equation}
\begin{aligned}
  &\mathcal{L}_{\text{ferm}}=\\
  &-  N_c \lambda \,( 2 \pi i)\,
  \Tr\biggl[
  - 4 \, Y^{i} Y_{j}^{\dagger}  \Psi_i \Psi^{ \dagger j}
  + 4 \, Y_{i}^{\dagger} Y^{j} \Psi^{ \dagger i} \Psi_j 
  - Y^{i} Y_{i}^{\dagger}  \Psi_j \Psi^{ \dagger j}
  + Y_{i}^{\dagger} Y^{i} \Psi^{ \dagger j} \Psi_j \\
  &
  - Y^{4} Y_{4}^{\dagger} \Psi_i \Psi^{ \dagger i}
  + Y_{4}^{\dagger} Y^{4} \Psi^{ \dagger i} \Psi_i
  - Y^{i} Y_{i}^{\dagger} \Psi_4 \Psi^{ \dagger 4}
  + Y_{i}^{\dagger} Y^{i} \Psi^{ \dagger 4} \Psi_4
  + Y^{4} Y_{4}^{\dagger} \Psi_4 \Psi^{ \dagger 4}
  - Y_{4}^{\dagger} Y^{4} \Psi^{ \dagger 4} \Psi_4
\\
  &+
  2 \, \epsilon_{ijk} \, e^{-i  \epsilon_{ijk} \gamma^-_j} 
  Y^\dagger_i  \Psi_j  Y^\dagger_k \Psi_4  
  -2 \, \epsilon_{ijk} \, e^{-i  \epsilon_{ijk}\gamma^-_j} 
  \Psi_i  Y^\dagger_j \Psi_i  Y^\dagger_4 
  -2 \, e^{i \gamma^-_i} Y^\dagger_4 Y^i \Psi^{\dagger 4} \Psi_i 
  +2 \, e^{-i \gamma^-_i} Y^i Y^\dagger_4 \Psi_i \Psi^{\dagger 4}\\
  &
  -2 \, \epsilon_{ijk} \, e^{i  \epsilon_{ijk} \gamma^-_j}
  Y^i  \Psi^{\dagger 4}  Y^k  \Psi^{\dagger j}
  +2 \, \epsilon_{ijk} \, e^{i  \epsilon_{ijk}\gamma^-_j}
  \Psi^{\dagger i}  Y^4 \Psi^{\dagger i}  Y_j 
  +2 \, e^{-i \gamma^-_i} Y^4 \Psi^{\dagger i} \Psi_4 Y^\dagger_i 
  -2 \, e^{i \gamma^-_i} Y^\dagger_i \Psi_4 \Psi^{\dagger i} Y^4
\\
  &+\frac{2}{3}\sum_{l=1}^3\biggl(3 \, e^{-i  \epsilon_{ijl}\gamma^+_l} -2\biggr)
  \biggl(
  Y^i  Y^\dagger_j \Psi_i   \Psi^{\dagger j}
  -
  Y^\dagger_i Y^j \Psi^{\dagger i} \Psi_j
  \biggr)\biggr]\, ,
\end{aligned}
\end{equation}
where the indices $i,j,k$ are summed from 1 to 3.

\subsection{Strongly twisted $\beta$-deformed ABJM Lagrangian} \label{ABJMbetalagrange}
In the double scaling limit of the $\beta$-deformed ABJM which was described in the main text (see (\ref{betaABJM})), we obtain the following $\mathcal{N}=2$ supersymmetric \(\chi\)FT\(_3\) model 
\begin{equation}
\mathcal{L} = N_c \,\Tr\biggl[- \partial_{\mu} Y^{\dagger}_{A}\partial^{\mu} Y^{A} + i\Psi^{ \dagger A} \slashed{\partial}\Psi_{A}\biggr]+ \mathcal{L}_{\text{int}}
\end{equation}
with the interacting term given by 
\begin{equation}
\begin{aligned}
\mathcal{L}_{int} =\,&\xi _3\,   \Tr\biggl[ Y^2 {Y}^{\dagger}_4 Y^3 {Y}^{\dagger}_2 Y^4 {Y}^{\dagger}_3+Y^1 {Y}^{\dagger}_4 Y^3 {Y}^{\dagger}_1 Y^4 {Y}^{\dagger}_3+Y^1 {Y}^{\dagger}_2 Y^4 {Y}^{\dagger}_1 Y^2 {Y}^{\dagger}_4+Y^1 {Y}^{\dagger}_2 Y^3 {Y}^{\dagger}_1 Y^2 {Y}^{\dagger}_3\biggr]-\\
&\frac{i \, \xi_3}{2}   \Tr\biggl[ 2 Y^4 {Y}^{\dagger}_1 \Psi _4 {\Psi }^{1\dagger}-2 {Y}^{\dagger}_2Y^4{\Psi }^{2\dagger}\Psi _4+2 Y^3{Y}^{\dagger}_2\Psi _3{\Psi }^{2\dagger}-2 {Y}^{\dagger}_1Y^3{\Psi }^{1\dagger}\Psi _3+\\
&2 Y^2{Y}^{\dagger}_4\Psi _2{\Psi }^{4\dagger}-2 {Y}^{\dagger}_3Y^2{\Psi }^{3\dagger}\Psi _2+2 Y^1{Y}^{\dagger}_3\Psi _1{\Psi }^{3\dagger}-2 {Y}^{\dagger}_4Y^1{\Psi }^{4\dagger}\Psi _1-\\
&Y^3{\Psi }^{1\dagger}Y^4{\Psi }^{2 \dagger}-Y^4 {\Psi }^{2\dagger}Y^3 {\Psi }^{1\dagger}+Y^1 {\Psi }^{4\dagger}Y^2{\Psi }^{3\dagger}+Y^2{\Psi }^{3\dagger}Y^1{\Psi }^{4\dagger}+\\
&{Y}^{\dagger}_1\Psi _4{Y}^{\dagger}_2\Psi _3+{Y}^{\dagger}_2\Psi _3{Y}^{\dagger}_1\Psi _4-{Y}^{\dagger}_3\Psi _1{Y}^{\dagger}_4\Psi _2-{Y}^{\dagger}_4\Psi _2{Y}^{\dagger}_3\Psi _1 \biggr]\,.
\end{aligned}
\end{equation}

\section{Twisted Asymptotic Bethe equations for $\gamma$-deformed $\mathcal{N}=4$ SYM}\label{appbetheSYM}
In this appendix we remind the full twisted Asymptotic Bethe equations for the $\gamma$-deformed $\mathcal{N}=4$ SYM. The full twisted version was first discussed in  \cite{Beisert:2005if,Gromov:2007ky}. In the conventions of  \cite{Beisert:2005if,Ahn2011} they read
{
  \allowdisplaybreaks
  \begin{equation}
    \allowdisplaybreaks
  \begin{aligned}
q_1^{-J_2-J_3}q_2^{J_1+J_3}q_3^{-J_1+J_2}&=\prod_{i=1}^{K_4}\frac{1-\frac{1}{x^{+}_{4,i} x_{1,j}}}{1-\frac{1}{x^{-}_{4,i} x_{1,j}}}  \prod_{l=1}^{K_2}\frac{u_{1,j} -u_{2,l} +i/2}{u_{1,j} -u_{2,l} -i/2}
\\
1&= \underset{k\neq l}{\underset{k=1}{\prod^{K_2}}}\frac{u_{2,l} -u_{2,k} -i}{u_{2,l} -u_{2,k} +i}\prod_{j=1}^{K_1}\frac{u_{2,l} -u_{1,j} +i/2}{u_{2,l} -u_{1,j} -i/2}\prod_{j=1}^{K_3}\frac{u_{2,l} -u_{3,j} +i/2}{u_{2,l} -u_{3,j} -i/2}   
\\
q_1^{-J_2-J_3}q_2^{J_1-J_3}q_3^{-J_1-J_2} & =  \prod_{i=1}^{K_4}\frac{x^{+}_{4,i}-x_{3,j} }{x^{-}_{4,i}-x_{3,j}}\prod_{l=1}^{K_2}\frac{u_{3,j} -u_{2,l} +i/2}{u_{3,j} -u_{2,l} -i/2}\nonumber
\\
q_1^{2J_3}q_2^{2J_3}q_3^{2(J_1+J_2)} & = \left( \frac{x_{4,k}^{-}}{x_{4,k}^{+}}\right)^L\, \underset{i\neq l}{\underset{i=1}{\prod^{K_4}}}\frac{x^{+}_{4,k}-x^{-}_{4,i} }{x^{-}_{4,k}-x^{+}_{4,i} } \frac{1-\frac{1}{x^{+}_{4,k} x^{-}_{4,i}}}{1-\frac{1}{x^{-}_{4,k} x^{+}_{4,i}}} \sigma(p_k,p_i)^2 \prod_{j=1}^{K_3}\frac{x^{-}_{4,k}-x_{3,j} }{x^{+}_{4,k}-x_{3,j} }\\
&\quad \times \,  \prod_{j=1}^{K_1}\frac{1-\frac{1}{x^{-}_{4,k} x_{1,j}}}{1-\frac{1}{x^{+}_{4,k} x_{1,j}}} \prod_{j=1}^{K_5}\frac{x^{-}_{4,k}-x_{5,j} }{x^{+}_{4,k}-x_{5,j} }  \prod_{j=1}^{K_7}\frac{1-\frac{1}{x^{-}_{4,k} x_{7,j}}}{1-\frac{1}{x^{+}_{4,k} x_{7,j}}}
\\
q_1^{J_2-J_3}q_2^{-J_1-J_3}q_3^{-J_1-J_2} & = \prod_{i=1}^{K_4}\frac{x^{+}_{4,i}-x_{5,j} }{x^{-}_{4,i}-x_{5,j}} \prod_{l=1}^{K_6}\frac{u_{5,j} -u_{6,l} +i/2}{u_{5,j} -u_{6,l} -i/2} \nonumber
\\
1 &= \underset{k\neq l}{\underset{k=1}{\prod^{K_6}}}\frac{u_{6,l} -u_{6,k} -i}{u_{6,l} -u_{6,k} +i}     \prod_{j=1}^{K_5}\frac{u_{6,l} -u_{5,j} +i/2}{u_{6,l} -u_{5,j} -i/2}\prod_{j=1}^{K_7}\frac{u_{6,l} -u_{7,j} +i/2}{u_{6,l} -u_{7,j} -i/2}
\\
q_1^{J_2-J_3}q_2^{-J_1+J_3}q_3^{-J_1+J_2}&=\prod_{i=1}^{K_4}\frac{1-\frac{1}{x^{+}_{4,i} x_{7,j}}}{1-\frac{1}{x^{-}_{4,i} x_{7,j}}}  \prod_{l=1}^{K_6}\frac{u_{7,j} -u_{6,l} +i/2}{u_{7,j} -u_{6,l} -i/2} \,.
\end{aligned}
\label{1357AhnConventions}
\end{equation}
}

where \(q_j=e^{-\frac{i}{2}\gamma_j},\,\,\, j=1,2,3\) are the gamma-twist parameters. The total momentum condition is now given by
\begin{equation}
\prod_{k=1}^{K_4}\frac{x_{4,k}^{+}}{x_{4,k}^{-}} = q_2^{-2 J_3}q_3^{-2J_2}\,.
\end{equation}
The number of roots at seven different nodes of the ABA Dynkin diagram is given in terms of the charges by
\begin{equation}
\begin{aligned}
K_1&=\frac{1}{2}\left( L-B -J_1-J_2+J_3\right)\,,\\
K_2& = \frac{1}{2}\left( \Delta_0 -J_1-J_2+J_3-S_1-S_2 \right)\,,\\
K_3& = \frac{1}{2}\left(B-L+ 2\Delta_0 -J_1-J_2+J_3 \right)\,\\
K_4& =  \Delta_0 -J_1\,,\\
K_5& = \frac{1}{2}\left(2\Delta_0-B-L -J_1-J_2-J_3 \right)\,\\
K_6& = \frac{1}{2}\left(\Delta_0 -J_1-J_2-J_3-S_1+S_2 \right)\,\\
K_7& = \frac{1}{2}\left(B+L -J_1-J_2-J_3\right)\,.
\end{aligned}
\end{equation}
Finally, the anomalous dimension is given by
\begin{equation}
\gamma = 2 ig \sum_{k=1}^{K_4}\left( \frac{1}{x_{4,k}^{+}} - \frac{1}{x_{4,k}^{-}}\right)\,.
\end{equation}
\section{Twisted Asymptotic Bethe equations for $\gamma$-deformed ABJM}\label{appbetheABJM} \footnote{The untwisted ABJM  Bethe equations were first proposed in \cite{Gromov:2008qe}.}
We present here the twisted Bethe equations for the closed sector 
$\mathfrak{su}(2)\times \mathfrak{su}(2)$ of ABJM
\begin{eqnarray} \label{twistedabjm}q_1^{- L} q_2^{ - L} q_3^{ -2 K_{\bar{4}}+L}\left(\frac{x^+_{4,k}}{x^-_{4,k}}\right)^{
L} \label{BAE3}&=&
\prod_{j\neq k}^{K_4}
\frac{u_{4,k}-u_{4,j}+i}{u_{4,k}-u_{4,j}-i}  \, \prod_{j\neq k}^{K_4}
\sigma_{\rm BES}(u_{ 4,k},u_{ 4,j}) \prod_{j=1}^{K_{\bar 4}} \sigma_{\rm BES}(u_{ 4,k},u_{ \bar 4,j}) \nonumber \,, 
\\
q_1^{- L} q_2^{ L } q_3^{2 K_4  -L}\left(\frac{x^+_{\bar 4,k}}{x^-_{\bar  4,k}}\right)^{
L} &=&
\prod_{j=1}^{K_{\bar 4}}
\frac{u_{\bar  4,k}-u_{\bar 4,j}+i}{u_{\bar 4,k}-u_{\bar 4,j}-i}  \,
 \prod_{j\neq k}^{K_{\bar 4}}
\sigma_{\rm BES}(u_{ \bar 4,k},u_{ \bar 4,j}) \prod_{j=1}^{K_{  4}} \sigma_{\rm BES}(u_{ \bar 4,k},u_{ 4,j}) \,,
\end{eqnarray}
where $q_i = e^{-i\gamma_i}$. 

\section{Feynman integrals}
\label{appendixintegrals}
In this appendix we list the values of the integrals that enter the
two point functions considered in the main text. These integrals are UV
divergent and we regulate these divergences by integrating loop
momenta in $4-2\epsilon$ dimensions but keeping the position-space
propagators four dimensional, such that generic one and two magnon integrals are defined as:
\begin{equation}
  \label{eq:onemagnonintdef}
  I_{[;h_1]} \equiv
  \left[\prod_{i=1}^{h_1}\int \frac{{\mathrm d}^{D}x_{b_{1,i}}}{\pi^D}\right]
  \frac{1}{x^2_{1\,b_{1,1}}}
  \left[\prod_{i=1}^{h_1} 
  \frac{1}{x^2_{1\,b_{1,i}} x^2_{b_{1,i}\,b_{1,(i+1)}} x_{b_{1,i}2} }\right]
\end{equation}
\begin{multline}
  I_{[d_1;h_1,h_2]} \equiv
  \left[\prod_{i=1}^{h_1}\int \frac{{\mathrm d}^{D}x_{b_{1,i}}}{\pi^D}\right]
  \left[\prod_{i=1}^{h_2}\int \frac{{\mathrm d}^{D}x_{b_{2,i}}}{\pi^D}\right]
  \frac{1}{x^2_{1\,b_{1,1}}}  \frac{1}{x^2_{1\,b_{2,1}}}
  \left[\prod_{i=1}^{d_1} 
  \frac{1}{x^2_{1\,b_{1,i}} x^2_{b_{1,i}\,b_{1,(i+1)}} x^2_{b_{1,i}2} }\right]\\
  \times  \left[\prod_{i=d_1+1}^{h_1}
    \frac{1}{x^2_{1\,b_{2,i}} x^2_{b_{1,i}\,b_{1,(i+1)}} x^2_{b_{1,i}\,b_{2,i}}  x^2_{b_{2,i}\,b_{2,(i+1)}} x^2_{b_{1,i}\,2}}\right]
  \left[\prod_{i=h_1-d_1+1}^{h_2} \frac{1}{x^2_{1\,b_{2,i}} x^2_{b_{2,i}\,b_{2,(i+1)}} x^2_{b_{2,i},2} }\right]
\end{multline}
where we identified $x_{b_{1\,h_1+1}} \equiv x_2$ and $x_{b_{2\,h_2+1}} \equiv x_2$. If a standard dimensionful expression is factored out, then they take the form
\begin{equation}
  I_{[;h_1]} \equiv (x_{12}^2)^{-h_1(1+\epsilon)-1} e^{-(h_1) \gamma_E \epsilon }\,\mathcal{I}_{[;h_1]}
  \qquad
  I_{[d_1;h_1,h_2]} \equiv (x_{12}^2)^{-h_2-d_1-2-(h_1+h_2) \epsilon} e^{-(h_1+h_2) \gamma_E \epsilon }\,\mathcal{I}_{[d_1;h_1,h_2]}
\end{equation}
where the factors $\mathcal{I}$ have a Laurent series near $\epsilon
\to 0$ with coefficients that are rational numbers or zeta values.

These integrals can be efficiently related through IBP identities
using FIRE \cite{Smirnov:2014hma} to master integrals provided in
\cite{Baikov:2010hf} to a sufficiently large order in $\epsilon$, namely,\allowdisplaybreaks
\begin{subequations}
  \begin{align}
  \mathcal{I}_{[;1]}
  &=
  \frac{-2}{\epsilon}
  -2
  +\zeta_2 \epsilon
  + \biggl(\zeta_2 + \frac{14\zeta_3}{3}\biggr)\epsilon^2
  + {\cal O}(\epsilon^3)\\
  \mathcal{I}_{[;2]}
  =
  \mathcal{I}_{[0;1,1]}
   &=
  \begin{aligned}[t]
    & \frac{2}{\epsilon^2} +\frac{3}{\epsilon} -(2\zeta_2 +1) -
    \biggl(1+3\zeta_2 + \frac{28\zeta_3}{3}\biggr)\epsilon \\
    & -\biggl(3+3\zeta_2 +4\zeta_3- \frac{21\zeta_4}{2}\biggr)\epsilon^2 + {\cal O}(\epsilon^3) \end{aligned}\\
  \mathcal{I}_{[;3]}
  &=
    \begin{aligned}[t]
      &  - \frac{4}{3\epsilon^3}
      - \frac{2}{\epsilon^2}
      + \biggl(\frac{8}{3}+2\zeta_2\biggr)\frac{1}{\epsilon}
      + \biggl(3\zeta_2 + \frac{32}{3}\zeta_3\biggr)\\
      & + \biggl(-\frac{16}{3} - \frac{2\pi^2}{3} + \frac{\pi^4}{8} - 18 \zeta_3\biggr)\epsilon\\
      &+ \biggl(8- \frac{91 \pi^4}{240} -\frac{568 \zeta_3}{3} - \frac{8\pi^2\zeta_3}{3}+\frac{504\zeta_5}{5}\biggr)\epsilon^2
      + {\cal O}(\epsilon^3)
    \end{aligned}\\
  \mathcal{I}_{[0;1,2]}
  =
  \mathcal{I}_{[1;2,1]}
  &
    =\begin{aligned}[t]
      &- \frac{8}{3\epsilon^3}
      - \frac{6}{\epsilon^2}
      + \frac{4 \zeta_2}{\epsilon}
      +\biggl(\frac{16}{3} + 9 \zeta_2 + \frac{52}{3} \zeta_3\biggr) 
      -\biggl(\frac{16}{3} - \frac{33 \zeta_4}{2} - 2 \zeta_3 \biggr)\epsilon \\
      &-\biggl(\frac{40}{3} + 8 \zeta_2 + 256 \zeta_3 + \frac{147 \zeta_4}{8} - 26 \zeta_2\zeta_3 + \frac{668\zeta_5}{5}  \biggr)\epsilon^2
      + \mathcal{O}(\epsilon^3)
    \end{aligned}\\
    \mathcal{I}_{[;4]}
    &
  =
  \begin{aligned}[t]
  &\frac{2}{3\epsilon^4}
  +
  \frac{2}{3\epsilon^3}
  -
  \frac{17+8\zeta_2}{6 \epsilon^2}
  +
  \biggl(\frac{19}{6} - \frac{4\zeta_2}{3} - \frac{80\zeta_3}{9 }\bigg)\frac{1}{\epsilon}\\
  &+ \biggl(\frac{2}{3} + \frac{17\zeta_2}{3}  + \frac{208 \zeta_3}{9} - \frac{28 \zeta_4}{3} \biggr)
  + \mathcal{O}(\epsilon)    
  \end{aligned}\\
  \mathcal{I}_{[0;2,2]}
  &=
  \begin{aligned}[t]
    &\frac{4}{3\epsilon^4} + \frac{3}{\epsilon^3} -
    \biggl(\frac{8}{3} + \frac{\zeta_2}{3}\biggr)\frac{1}{\epsilon^2} -
    \biggl(7+6\zeta_2+\frac{106\zeta_3}{9}\biggr)\frac{1}{\epsilon}\\
    &+ \biggl(\frac{46}{3} + \frac{16 \zeta_2}{3}  + 8 \zeta_3 - \frac{29 \zeta_4}{3}\biggr)
    + \mathcal{O}(\epsilon)
  \end{aligned}
\\
  \mathcal{I}_{[1;2,2]}
  &=
  \begin{aligned}[t]
  &\frac{10}{3\epsilon^4}
  +
  \frac{29}{3\epsilon^3}
  +
  \biggl(\frac{13}{6}+\frac{20 \zeta_2}{3} \biggr)\frac{1}{\epsilon^2}
  -\biggl(\frac{213}{18}+\frac{58\zeta_2}{3}+\frac{256\zeta_3}{9}\biggr)\frac{1}{\epsilon}\\
  &
  + \biggl(\frac{87}{9}-\frac{13}{3}\zeta_2-\frac{80}{9}\zeta_3-\frac{68}{3}\zeta_4\biggr)
  + \mathcal{O}(\epsilon)
\end{aligned}
\\
  \mathcal{I}_{[0;1,3]}
  =
  \mathcal{I}_{[2,3,1]}
  &=
  \begin{aligned}[t]
  &\frac{2}{\epsilon^4}
  + \frac{29}{6\epsilon^3}
  - \frac{2+4\zeta_2}{\epsilon^2}
  - \biggl(\frac{20}{3}+ \frac{29 \zeta_2}{3} + \frac{56 \zeta_3}{3}\biggr) \frac{1}{\epsilon}\\
  &+ \frac{19}{2} + 4 \zeta_2 + \frac{68}{9} - 16 \zeta_3
  + \mathcal{O}(\epsilon)
  \end{aligned}
\end{align}
\end{subequations}  

A different but related class of integrals we have considered in this paper
are the amputated versions of the one-magnon integrals~(\ref{eq:onemagnonintdef}):
\begin{equation}
  \label{eq:onemagnonintdefamp}
  I^{\mathrm{amp}}_{[;h_1]} \equiv
  \left[\prod_{i=1}^{h_1}\int \frac{{\mathrm d}^{D}x_{b_{1,i}}}{\pi^D}\right]
  \frac{1}{x^2_{1\,b_{1,1}}}
  \left[\prod_{i=1}^{h_1-1} 
  \frac{1}{x^2_{1\,b_{1,i}} x^2_{b_{1,i}\,b_{1,(i+1)}} }\right]\, \frac{1}{x^2_{b_{1,h_1}2}}\, ,
\end{equation}
These integrals are obtained from the one-magnon integrals
(\ref{eq:onemagnonintdef}) by removing all propagators that reach the
external vertex \(x_2\)
apart from \(1/{x^2_{b_{1,h_1}2}}\).
If this propagator was not present, the integral would have a sub
bubble with an IR divergence and therefore it acts as an IR regulator.
 
The integrals (\ref{eq:onemagnonintdefamp}) are nothing but nested bubble integrals and by noting the recursive structure
\begin{equation}
  I^{\mathrm{amp}}_{[;h_1]}
  =
  \frac{1}{\bigl(x_{12}^2\bigr)^{\epsilon}}
  \frac{\Gamma(1-\epsilon)\Gamma(-h_1 \epsilon)\Gamma(1+h_1\epsilon)}
  {\Gamma(1-(h_1+1)\epsilon)\Gamma(2+(h_1-1)\epsilon)}
  \,
  I^{\mathrm{amp}}_{[;h_1-1]} \, ,
\end{equation}
it is easy to find their value in an arbibtrary dimension and for any $h_1$ as:
\begin{equation}
  I^{\mathrm{amp}}_{[;h_1]}(0) 
  =
  \frac{  \Gamma^{h_1}(1-\epsilon)}{(x_{12}^2)^{1+h_1 \epsilon}}
  \prod_{k=1}^{h_1}
  \frac{\Gamma(-k\epsilon) \Gamma(1+k \epsilon)}{\Gamma(1-(k+1)\epsilon) \Gamma(2+(k-1)\epsilon)} \, .
\end{equation}


\bibliographystyle{JHEP}
\bibliography{biblio_magnon_paper}

\end{document}